\documentclass[a4paper,11pt]{article}
\pdfoutput=1 

\usepackage{jheppub} 

\usepackage[T1]{fontenc} 
\usepackage{xspace}
\newcommand{\Py}{\textsc{Pythia}\xspace}

\title{\boldmath String Junctions Revisited}

\author[a,b]{J. Altmann,} 
\author[a,b]{P. Skands}

\affiliation[a]{Rudolf Peierls Centre for Theoretical Physics, University of Oxford, OX1 3PU, UK}
\affiliation[b]{School of Physics and Astronomy, Monash University, VIC-3800, Australia}

\emailAdd{javira.altmann@monash.edu}
\emailAdd{peter.skands@monash.edu}

\abstract{Recent measurements at the LHC have revealed heavy-flavour baryon fractions much larger than those observed at LEP, with e.g., $\Lambda_c^+/\mathrm{D}^0$ and $\Lambda_b^0/\mathrm{B}^0$ reaching $\sim 0.5$ at low $p_\perp$. One scenario that has been at least partly successful in predicting observed trends is QCD colour reconnections with string junctions. In previous work, however, the limit of a low-$p_\perp$ heavy quark was not well defined. We reconsider the string equations of motion for junction systems in this limit, and find that the junction effectively becomes bound to the heavy quark, a scenario  we refer to as a ``pearl on a string''. We extend string-junction fragmentation in \Py with a dedicated modelling of this limit for both light- and heavy-quark ``pearls''.
}

\begin{document} 
\setcounter{tocdepth}{2}
\maketitle
\flushbottom

\section{Introduction}
\label{sec:intro}

The  dynamical process by which high-energy quarks and gluons become confined inside hadrons --- \emph{hadronization} --- remains among the most challenging problems in particle physics. In the  context of theoretical models, one typically constrains a set of non-perturbative parameters in a reference process, like hadronic $Z$ decays which can be studied cleanly in $e^+e^-$ collisions, and then assumes that those same parameters can be reused in different environments, like $pp$ collisions. This assumption --- referred to as \emph{jet universality} --- is rooted in the factorization of long-distance non-perturbative physics from short-distance perturbative physics. It underpins, e.g., the formalism of fragmentation functions, and is also the starting point for the modelling of hadronization in Monte Carlo (MC) event generators.


It has become clear that there are interesting breakdowns of jet universality between $e^+e^-$ and $pp$ collisions and that these breakdowns tend to become more pronounced with the charged-particle multiplicity of the latter. Many of the observations, such as enhanced baryon and strange hadron production \cite{CMS:2011fsn,ATLAS:2011xhu,CMS:2013zgf,CDF:2013kip,ALICE:2016fzo,ALICE:2018pal,ALICE:2023egx}, appear reminiscent of phenomena that are also observed in heavy-ion collisions. Although it is not yet clear what precise physical conclusions to draw from this, it certainly motivates a reassessment of the baryon and strange-hadron production mechanisms in theoretical models, particularly for high-multiplicity $pp$ collisions.  

In this work we focus in particular on the Lund string model of hadronization~\cite{ANDERSSON198331, Andersson:2001yu, ARTRU1983147, Andersson:1998tv}, as implemented in the \Py~8 event generator~\cite{Bierlich:2022pfr}. 

Two modelling aspects that are known to be especially important at high multiplicities in $pp$ collisions are: multi-parton interactions (MPI) and colour-space ambiguities / colour reconnections (CR). 

A comparatively simple textbook example of the latter, which has been studied extensively, is hadronic $e^+e^- \rightarrow WW$ events, in particular in the context of studying effects of corrections to the so-called Leading-Colour (LC) string configurations on precision measurements of the $W$ mass~\cite{Sjostrand:1993hi,L3:2003ohc, L3:2003oci, OPAL:2003njc, Siebel:2005uw, OPAL:2005cdr, ALEPH:2006cdc, DELPHI:2006tie, ALEPH:2006jcu}. However in  $e^+e^-$ collisions, there is a strong expectation that beyond-LC string configurations are suppressed, due to several factors including the  $1/N_C^2$ suppression, the relative boost and space-time separation of the two $W$ decay systems, and dynamical effects as QCD coherence suppresses non-planar (``zig-zagging'') colour flows within each $W$ decay system. Nevertheless, the verdict from a combination of results from all four LEP collaborations was that the no-CR hypothesis was excluded at 99.5\% CL~\cite{ALEPH:2013dgf}. 

The picture becomes more complex in $pp$ collision systems (or more generally hadron-hadron collisions), where one must consider the initial-state coloured partons, coloured beam remnants, and the contributions of multi-parton interactions (MPI). As the scattering centres of the MPI sit within $\sim$ a proton radius of each other, the effect of colour-space ambiguities is presumably not (significantly) suppressed by space-time separations, nor are there kinematic or (LC) coherence suppressions. Moreover, there is a combinatorial enhancement of the ambiguities in events with a large number of MPI $\sim$ high particle multiplicities, which counteracts the naive $1/N_C^2$ suppression. 

The very first MC model of MPI for $pp$ collisions already incorporated a simple CR model~\cite{Sjostrand:1987su}. This was essential to describe the observed growth of the average charged-particle $p_\perp$ with charged-particle multiplicity, $\left< p_\perp \right>(N_\mathrm{ch})$, in minimum-bias collisions. CR also turned out to be crucial for a good description of the underlying event (UE) at the Tevatron~\cite{Field:2005sa,Skands:2010ak}. Since then, many further CR toy models for $pp$ have been proposed \cite{Rathsman:1998tp, Sandhoff:2005jh,Buttar:2006zd,Skands:2007zg,Gieseke:2012ft,Argyropoulos:2014zoa,Christiansen:2015yqa,Bellm:2019wrh}, often with the additional motivation to study CR effects on precision determinations of the top quark mass~\cite{Skands:2007zg,Argyropoulos:2014zoa,Christiansen:2015yqa}. The models that are applied in the $pp$ context are often technically somewhat simpler than their $ee$ counterparts were, mainly due to the necessity of addressing more complicated parton topologies. They tend to reconfigure colour connections based on global potential-energy minimisation arguments. The QCD-based CR model of Ref.~\cite{Christiansen:2015yqa} was the first to combine this with reintroducing the colour-space ambiguities stochastically using SU(3) colour algebra. It builds  randomized SU(3)-weighted colour indices onto the LC partons from MPI + showers. Thus, full-colour event structures are restored --- if only in an approximate statistical sense. This introduces multiple ways to achieve colour neutralization, and along with string-length minimization can allow for alternative string configurations to be selected instead of the LC ones.  

A key feature of the SU(3) structure of QCD is the existence of the antisymmetric red-green-blue colour-singlet state. As Lund strings form between colour-connected partons, there naturally must also be a string type that connects red-green-blue partons in a colour singlet. These strings are modelled by a Y-shaped structure which we call a ``string junction’’. The baryon number of such a junction is conserved such that when junction strings fragment, baryon formation occurs around the junction itself. Thus the fragmentation of junction systems provides an additional baryon production mechanism in $pp$ that is not present in $e^+e^-$ collision events. Notably, junction formation scales with the prevalence of CR effects; they are particularly important for higher-multiplicity events. 

In the context of the Lund model, string junctions were first designed to study baryon-number violating SUSY processes~\cite{Sjostrand:2002ip}, which correspond to high-$p_\perp$ explosions with typically ultra-relativistic endpoints. The resulting junction-fragmentation formalism was then recycled for beam remnants, which also involve high-energy endpoints. In neither of these cases were soft (i.e., low-energy) junction endpoints encountered often and there was thus no incentive to develop a dedicated treatment of this limit. However, the same junction modelling was recycled again for QCD CR~\cite{Christiansen:2015yqa}, but this time due to the explicit string-length minimization used in the CR procedure, it is fairly frequently the that topologies are produced in which one or more junction legs are soft. This is particularly relevant for heavy-flavour baryons at low $p_\perp$, which have received significant experimental interest recently~\cite{LHCb:2017bdt,ALICE:2020wla,ALICE:2021bli,ALICE:2021rzj,LHCb:2021xyh,ALICE:2022cop,ALICE:2022exq,ALICE:2023jgm,ALICE:2023xiu,ALICE:2023wbx,ALICE:2023sgl,LHCb:2023wbo}. The aim of this work is therefore to consider this limit more carefully than was done in the past, enabling us to make firmer predictions. 

We proceed systematically, and consider low-energy excitations around the junction in general, including also how soft gluon kinks affect the effective juntion motion and the resulting junction baryon spectra. This paper is organised as follows. In sec.~\ref{sec:background} we summarize briefly the Lund String Model with particular focus on string junctions. In sec.~\ref{sec:Theory} we elaborate on the modelling of junction motion, exploring both the existing modelling and previously unconsidered so-called soft-leg cases. We also provide a model of gluon kinks passing through junctions, however this is limited to a theoretical description and is not implemented in \Py. In sec.~\ref{sec:implementation} we describe the updated  implementation in \Py, addressing the treatment of soft gluon kinks on junction motion. Finally, sec.~\ref{sec:results} outlines some comparison of the updated implementation to both the old junction modelling and to data.

\section{Background}
\label{sec:background}
\subsection{Lund String Model}

The Lund String Model is a semi-classical phenomenological hadronization model, based on confinement in Quantum Chromodynamics (QCD). The model collapses the colour confinement field into an infinitely narrow flux tube spanned between coloured particles, modelled as a 1+1 dimensional relativistic worldsheet which we call a string, and is modelled using relativistic string dynamics \cite{ANDERSSON198331, Andersson:2001yu, ARTRU1983147}. Modelling the confinement field as a string is largely motivated by the Cornell potential \cite{Eichten:1978tg, Bali:1992ab}, a lattice QCD result that demonstrates a linear potential between static colour charges. Such a linear potential corresponds to a constant tension, and hence a string characterised by the constant string tension $\kappa \sim 1$ GeV/fm. 

\begin{figure}[ht!]
    \begin{center}
    \includegraphics[width = 0.4\textwidth]{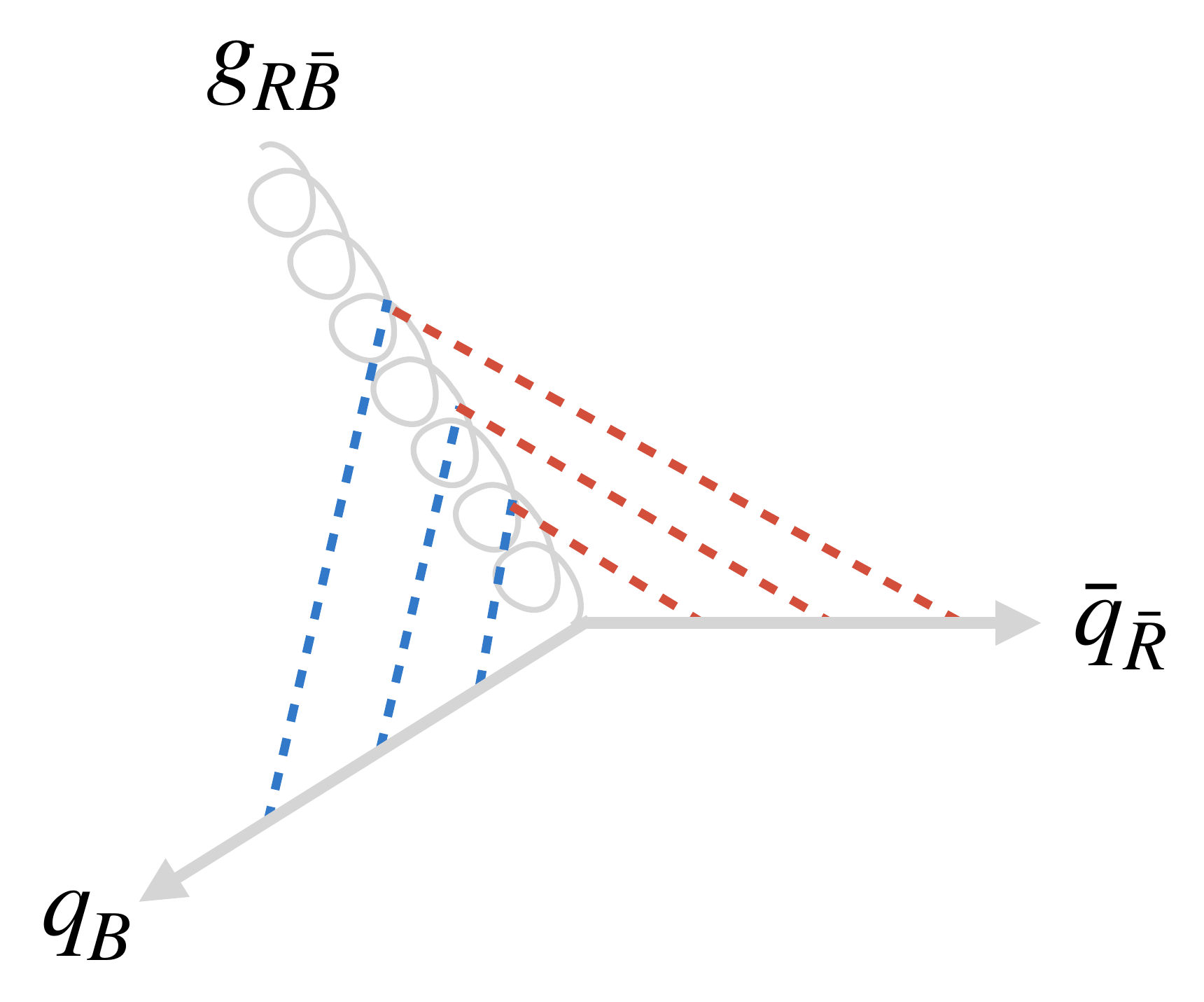}
    \end{center}
    \caption{Sketch of a string configuration with a single gluon-kink on a dipole string. Here the strings are represented by the dashed coloured lines and are shown at several different times. Here one can see that a gluon-kink results in two separate string segments, one connecting $g_{R\bar{B}}$ to $q_{B}$ and another connecting $g_{R\bar{B}}$ to $\bar{q}_{\bar{R}}$.}
    \label{fig:dip}
\end{figure}

These confinement fields form between colour-connected partons, such that the overall string configuration is in a singlet state according to the SU(3) colour structure of QCD. This means that quarks/antiquarks (i.e. triplet/antitriplet colour charges) are colour connected to a single other parton, and gluons (i.e. octet colour charges) will be colour connected to two other partons via two string pieces. The simplest string form to consider is the so-called dipole string, which makes use of the colour-anticolour colour neutral state and has a quark and antiqaurk endpoint, with arbitrarily many intermediate gluons that form transverse ``kinks'' on the string. An example of such a dipole string with one gluon kink is shown in fig.~\ref{fig:dip}. 

\subsubsection{Beyond Leading Colour}
\label{sec:BLC}

In event generators like \Py, the mapping of colour flow is simplified by considering the Leading-Colour (LC) limit, in which QCD is modelled as an $SU(N_C)$ gauge theory with the number of colours taken to infinity (i.e. $N_C \rightarrow \infty$). In this limit, gluon (adjoint) colour states can be reduced to simple direct products of fundamental and antifundamental ones, since the singlet can be neglected in the group relation 
\begin{equation}
N_C \otimes \bar{N}_C ~=~ \overbrace{(N_C^2 - 1)}^{\mathrm{adjoint}} ~\oplus~ 1~.
\end{equation} 
This vastly simplifies the problem of colour flow in parton cascades and also eliminates colour interference effects beyond the dipole level; each colour in an event is unique and is matched to a single other anticolour. At the perturbative level, these colour connections are used to set up the system of radiating LC ``dipoles'', while at the non-perturbative level the LC colour connections dictate between which partons confining potentials should arise: in the Lund string model, each LC dipole is dual to a non-perturbative string piece~\cite{Gustafson:1986db}. Thus, the LC limit provides for unambiguous string topologies, with each string piece having a colour and an anticolour charged endpoint, i.e.\ dipole strings. 

When generalising from a single parton system to multi-parton interactions, the LC limit does not address if or how different MPI systems should be correlated in colour space. In the limit of high $Q^2$, one may assume that each MPI is independent of any others in colour space. In an LC picture, this would imply that there would be no string connections directly between outgoing partons from different (high-$Q^2$) multi-parton interactions (MPIs). (At lower $Q$ of order the inverse proton size, coherence and/or saturation effects presumably modify this picture, but not much is known about the details.)

What about effects beyond LC? In hadronic $Z$ decays, effects beyond LC are expected to be very small, as they are suppressed both by $1/N_C^2 \sim 10\%$ and further by a combination of kinematics and coherence (no overlapping jets and dominance of angular ordered colour structures within each jet). However, in systems with several independent colour sources, such as in hadronic $e^+e^-\to WW$ events (in the limit of vanishing $W$ lifetime\footnote{ Presumably a reasonable starting approximation since $\Gamma_W \sim 2\,\mathrm{GeV}\,\gg\,\Lambda_\mathrm{QCD}$.}) the kinematic suppression can be lifted in parts of phase space, and in dense string environments such as (high-multiplicity) $pp$ collisions, combinatorics can further counteract the naive $1/N_C^2$ suppression. That is, while effects beyond LC between two dipoles, say, are suppressed by $1/N_C^2$, the probability that there are no beyond-LC effects in a system of $n$ dipoles will scale like $(1-1/N_C^2)^{n-1}$, which becomes asymptotically small for large $n$. 

Focusing now on QCD, with $N_C=3$, the finite number of colours allows for a potentially large number of \emph{possible} string configurations to produce overall singlet states, resulting in ambiguity in where the confining fields form. The QCD-based colour reconnection (CR) model of Ref.~\cite{Christiansen:2015yqa} stochastically reintroduces such colour-space ambiguities by assigning colour/anticolour indices from 0 to 8 to partons in order to reproduce probabilities defined by the $\mathrm{SU}(3)$ algebra,

\begin{subequations}
\label{eqn:SU3}
    \begin{equation}
        3\otimes\bar{3}=8\oplus 1~,
        \label{eqn:33bar}
    \end{equation}
    \begin{equation}
        3\otimes 3=6\oplus\bar{3}~,
        \label{eqn:33}
    \end{equation}
\end{subequations}
where the octet and sextet states are interpreted as non-confining, while the singlet and antitriplet are interpreted as confining and partially confining respectively.

To determine which among the many resulting possible string configurations is realised in terms of being the one selected for hadronization, the configuration that gives the overall ``shortest string lengths'' is chosen. Here the ``length'' of the string actually refers to a momentum-space Lorentz-invariant measure of the integrated energy density per unit length of the string, which we call the $\lambda$-measure. In the context of the QCD CR model, the default form for the $\lambda-$measure for a $q\bar{q}$ dipole string was previously
\begin{equation}
    \lambda^{q\bar{q}} = \ln{\left( 1+\frac{\sqrt{2}E_1}{m_0} \right)} + \ln{\left( 1+\frac{\sqrt{2}E_2}{m_0} \right)}~,
    \label{eqn:lambdaOld}
\end{equation}
where $E_i$ is the energy of the parton in the dipole rest frame, and $m_0$ is a regularisation parameter of order $\Lambda_\mathrm{QCD}$. For massive endpoint partons, however, this form overestimates the physical string length, since the endpoint masses are allowed to contribute to the energies in the numerators. 

To ensure a more sensible treatment of QCD CR involving strings with heavy-quark endpoints, we introduce the following generalisation (made default from \Py 8.311 onwards),
\begin{equation}
    \lambda^{q\bar{q}} = \text{max} \left[
        \ln{\left(  \frac{E_1+|\vec{p}|}{m_1+m_0} \right)}, 0 \right] 
        + \text{max} \left[\ln{\left( \frac{ E_2+|\vec{p}|}{m_2+m_0}\right)}, 0 \right] ~,
    \label{eqn:lambdaNew}
\end{equation}
with  $m_i$ the mass and $(E_i,\pm\vec{p})$ the energies and momenta of  the partons in the rest frame of the pair. This measure ensures the limit of  $|\vec{p}|\rightarrow 0$ produces $\lambda\rightarrow 0$. Here $m_0$ is used to protect against massless partons, and the $\text{max}$ functions impose that no endpoint can be associated with a negative contribution to the effective $\lambda$ measure. 
 
We note that an alternative $\lambda$-measure  for string pieces involving heavy-quark endpoints was already proposed in \cite{Bierlich:2023okq}, based on the rapidity of a heavy-quark endpoint as a measure of its associated string length. But that measure only applied to heavy quarks and could not be used for light or massless partons. The measure defined in eq.~\eqref{eqn:lambdaNew} is similar in spirit to the one proposed in \cite{Bierlich:2023okq} but with modifications so that the same form of the $\lambda$-measure can be used for all partons regardless of mass, allowing consistency across all types of string lengths being compared in the CR treatment. This same string-length measure is then also used to calculate lengths for $qg$, $gg$ and $g\bar{q}$ string connections as well. 

\begin{figure}[b!]
\centering\includegraphics*[width=0.96\textwidth]{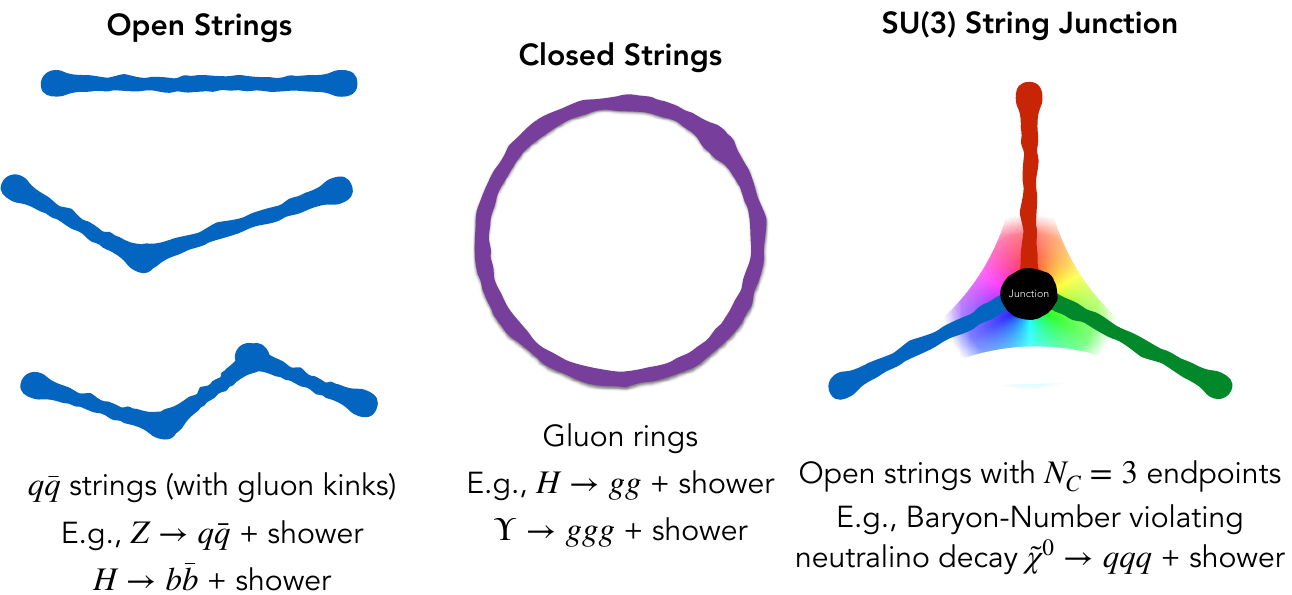}
\caption{Types of string topologies, with ``textbook examples'' of each kind. \label{fig:stringTopologies}}
\end{figure}

By allowing colour reconnections, string connections between different MPI can be formed. Moreover, in addition to strings spanned between triplet and antitriplet endpoints, two other string topologies can arise: gluon loops \cite{Andersson:1984af} and junctions \cite{Sjostrand:2002ip}, illustrated in fig.~\ref{fig:stringTopologies}.

The latter will be the focus of this paper and revisited in sec.~\ref{sec:juncIntro}, including details on the $\lambda-$measure for junction topologies and the fragmentation of such strings.

\subsubsection{Fragmentation}
\label{sec:frag}

\begin{figure}[tp]
    \begin{center}
    \includegraphics[width = 0.9\textwidth]{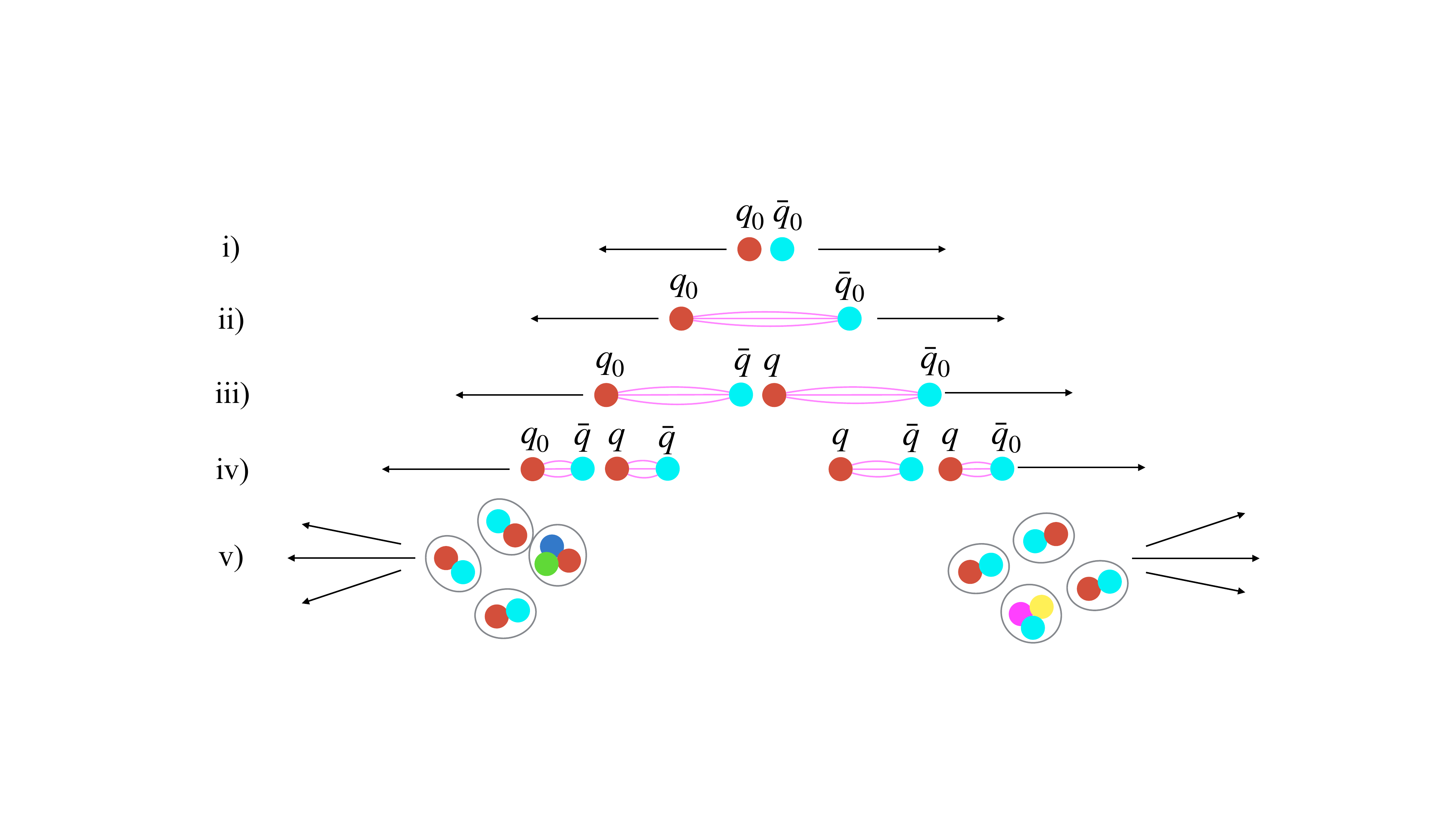}
    \end{center}
    \caption{A basic schematic showing the hadronization process at different times for a simple dipole string. i) The initial colour-anticolour pair begin to move apart at high energies. ii) The colour flux tube forms between the colour-anticolour charge pair. iii)-iv) New $q\bar{q}$ pairs are formed from string breaks. v) After hadronization is complete, we are left with colourless hadrons which form jets. }
    \label{fig:frag}
\end{figure}

In high-energy collisions such as those at the Large Hadron Collider (LHC), colour-connected partons move apart at high energies, resulting in high-invariant-mass string systems that stretch the confining potentials to the point of breaking. At the site of these string breaks, a quark-antiquark (or diquark-antidiquark) pair is created. This string-breaking process is called fragmentation, and occurs along the string until there is no longer sufficient energy to keep fragmenting the string, resulting in final-state ``primary'' hadrons (some of which may be unstable and undergo further decays, creating ``secondary'' hadrons). A basic schematic of this process is shown in Fig ~\ref{fig:frag}. 

The modelling of string fragmentation in \Py (the Lund model) relies on two main components; the quantum tunnelling process for spontaneous pair creation, which dictates the flavour and transverse components of the fragmentation process, and the fragmentation function that governs the longitudinal component. To model the spontaneous pair creation that occurs in a string break, a QCD analogy of the QED Schwinger mechanism~\cite{PhysRev.82.664} is used. The Schwinger mechanism was originally derived in the context of spontaneous electron-positron pair creation in the presence of a strong electric field. In the Lund model, an analogous QCD formulation is used for $q\bar{q}$ creation from the confinement field. The leading term of the Schwinger mechanism implies a Gaussian form with respect to the transverse mass $m_\perp$,
\begin{equation}
    \exp \left( \frac{-\pi m_\perp^2}{\kappa} \right) = 
    \exp \left( \frac{-\pi p_\perp^2}{\kappa} \right)
    \exp \left( \frac{-\pi m^2}{\kappa} \right),
    \label{eqn:schwinger}
\end{equation}
with parton mass $m$ and transverse momentum $p_\perp$ relative to the local string axis. This provides a suppression of both heavy-particle production and constrains the $p_\perp$ distribution. Alternative string-breaking models have also been proposed (e.g., the thermal model of Ref.~\cite{Fischer:2016zzs}), however so far the standard has remained Schwinger-type string breaks. This is what will be used throughout the remainder of this paper. An important consequence of the Gaussian suppression and the string tension $\kappa\sim 1$ GeV/fm is that only light-flavoured quarks (i.e. up, down and strange quarks) can be created via fragmentation. Thus all charm and beauty quarks must come from perturbative processes (which can include MPI), a distinction of particular importance as it allows heavy-flavour hadrons to serve as  interesting probes of fragmentation modelling. We will return to this point in sections~\ref{sec:juncIntro} and \ref{sec:Theory}.

The longitudinal component of fragmentation is governed by what is known as the Lund symmetric fragmentation function (or also the left-right symmetric fragmentation function). In the Lund string model, string breaks are treated as causally disconnected, and hence the time ordering of string breaks holds no physical significance. This allows us to perform string breaks in any order we choose. The simplest is to fragment hadrons off from either string endpoint, which makes it particularly easy to impose  on-shell hadron mass constraints on each produced hadron. Formulated with a simple dipole string in mind, the symmetric fragmentation function takes the form
\begin{equation}
	f(z) = N \frac{1}{z} (1-z)^a \exp{\left( \frac{-bm_\perp^2}{z} \right) },
    \label{symFrag}
\end{equation}
which is a probability distribution for the hadron to take fraction $z$ of the string momentum, with tuneable free parameters $a$ and $b$, and $N$ a normalisation constant. The string end from which to fragment is normally chosen randomly for each break. 

\subsection{Junction String Topologies}
\label{sec:juncIntro}

\begin{figure}[ht!]
    \begin{center}
    \includegraphics[width = 0.9\textwidth]{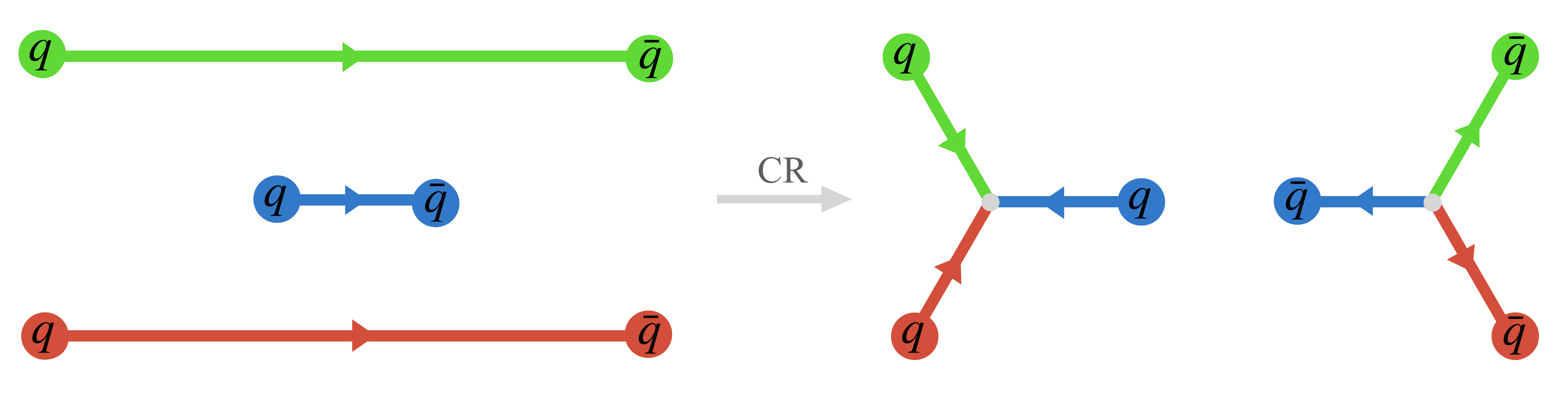}
    \end{center}
    \caption{The left image shows an LC string configuration consisting of three dipole strings. The coloured lines here represent the strings, with arrows indicating the direction of the colour flow (conventionally flowing from colour to anticolour). The right side image shows a possible alternative string configuration given a junction-type colour reconnection, resulting in the formation of a junction and an antijunction string system. }
    \label{fig:CR}
\end{figure}

String junctions arise naturally from the SU(3) colour structure of QCD and represent the confinement field spanned between three colour-triplet partons (or three colour-antitriplet ones) in an overall colour-neutral state. These states only exist for finite $N_C$ and hence intrinsically go beyond the LC picture. 
In the context of the Lund model, with $N_C=3$, they are represented by Y-shaped string topologies~\cite{Sjostrand:2002ip} as illustrated in the right-hand side of fig.~\ref{fig:CR}. 
Here the ``junction'' itself refers to the topological feature depicted by the central vertex point of the Y-shape, with the ``junction legs'' being the three string pieces extending out from the junction. 

In order for string junctions to be formed via colour reconnections, one needs a generalised string-length measure which allows to compare the effective string lengths between dipole- and junction-type configurations, such as those on the left and right of fig.~\ref{fig:CR} respectively. With the measure for dipole-type string pieces defined by eq.~\eqref{eqn:lambdaNew} above, we need an equivalent measure for parton-to-junction string pieces. This is constructed by reinterpreting each junction leg as half of a dipole string, with the midpoint of the dipole sitting at the junction itself (in a frame in which the junction is at rest). In a straightforward generalisation of eq.~\eqref{eqn:lambdaNew}, we define the $\lambda$-measure associated with each of the parton--junction string segments in a three-parton junction configuration, as
\begin{equation}
    \lambda^{qqq}_i = 
       \text{max}  \left[ \ln{  \left(\frac{E_i+|\vec{p}_i|}{j_0(m_i+m_0)} \right)}, 0 \right]~,
    \label{eqn:lambdaJunc}
\end{equation}
Importantly, the energies $E_i$ and momenta $\vec{p}_i$ of each of the three partons are defined in the rest frame of the junction, which generalises the notion of the dipole rest frame that was used in eq.~\eqref{eqn:lambdaNew}. What this junction rest frame (JRF) looks like and how to find it will be explored in the detail in sec.~\ref{sec:Theory}. 

The $\lambda$-measure defined in eq.~\eqref{eqn:lambdaJunc} also introduces a free parameter, $j_0$, which allows the user to modify how easily junction systems form. Larger values of $j_0$ make junction formation more likely (by decreasing the effective $\lambda$-measures for junction topologies relative to dipole ones), and vice versa. In the code, $j_0$ is set by the tuneable parameter \texttt{ColourReconnection:junctionCorrection}. As an option mainly intended for comparisons, we also retain the possibility to choose to use the old form of the string-length measure instead, as per eq.~\eqref{eqn:lambdaOld}, where this parameter $j_0$ multiplies the $m_0$ term in the denominator. 

\subsection{Junction Fragmentation}

\begin{figure}[t]
    \begin{center}
    \includegraphics[width = 0.98\textwidth]{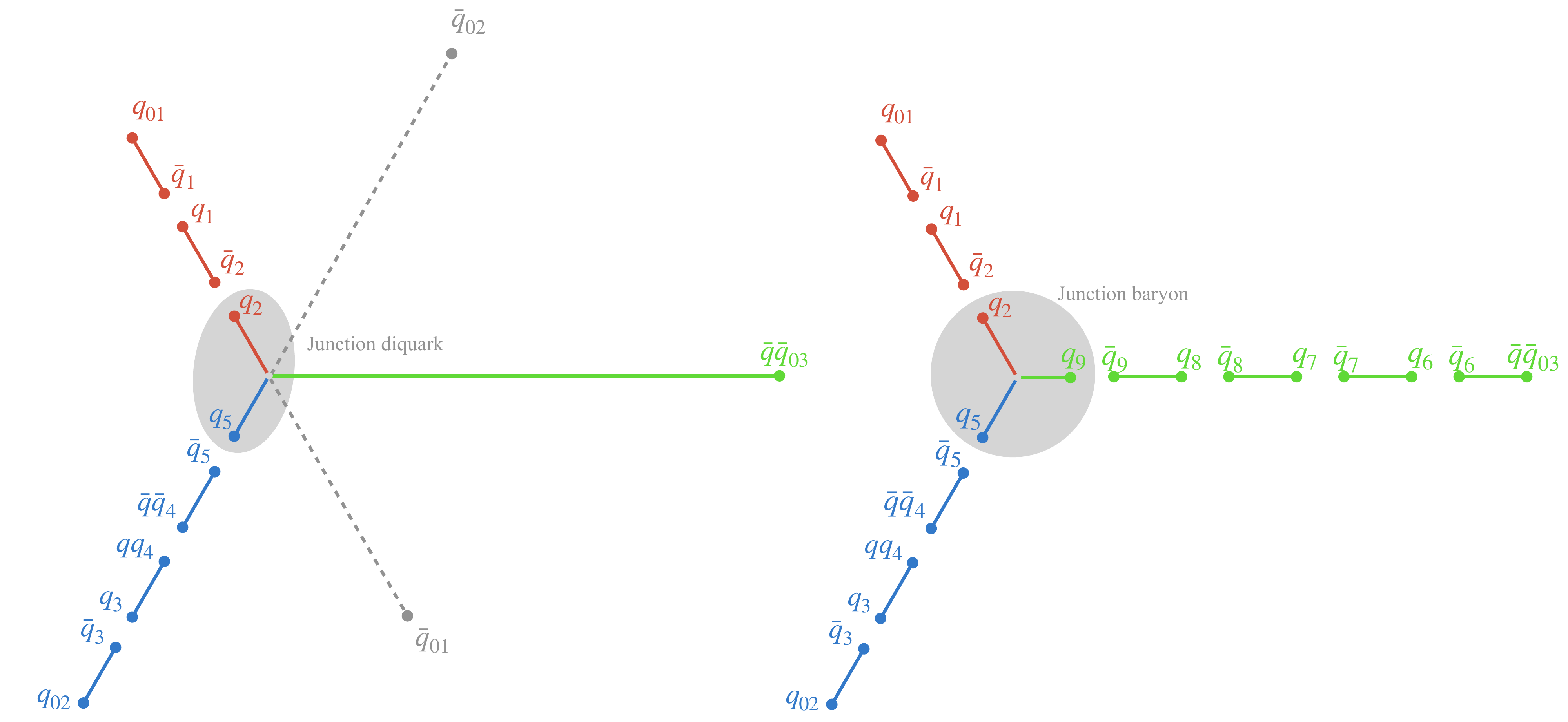}
    \end{center}
    \caption{Fragmentation of a junction topology using the standard junction fragmentation procedure in the Mercedes frame. The first frame shows the fragmentation of the two legs with the lowest momenta in the JRF (which here is the Mercedes frame). For these junction legs, fictitious legs are constructed on the opposite side of the junction. These are shown by the dashed lines with endpoints $\bar{q}_{01}$ and $\bar{q}_{02}$. These two legs have been fragmented from their real endpoints inwards towards the junction, after which we have the formation of the junction diquark, $q_3q_5$, from the string breaks next to the junction. The right image shows fragmentation of the final junction leg, treated as a dipole with endpoints $q_2q_5$ and $q_{03}$. Here the strings breaks next to the junction, $q_2q_5q_9$ combine to form the so-called junction baryon.}
    \label{fig:juncFrag}
\end{figure}

The string-length measures defined above govern whether and how easily junction topologies are formed in CR. The next question is how does one fragment a junction string system? 
The standard approach to junction fragmentation~\cite{Sjostrand:2002ip} is to use the concept of reinterpreting junction legs as equivalent to half of a dipole string, analogously to how the string-length measure was defined above. This allows us to simply recycle the dipole string-fragmentation procedure from sec.~\ref{sec:frag}. Construction of the full dipole string is done by boosting the junction system to the junction rest frame, and creating what we call a fictitious leg. This fictitious leg is a mirror image of a junction leg that extends on the opposite side of the junction, a depiction of which can be seen in fig.~\ref{fig:juncFrag}. This fictitious leg acts as the combined effective pull of the other two junction legs, providing a reservoir of oppositely oriented momentum. 

Of course we generally encounter more complicated junction systems than those shown in fig.~\ref{fig:juncFrag}, which can include one or more gluon kinks on each of the junction legs. In such cases, the rest frame of the junction will change over time as different gluon-kinks dominate the junction motion. 
Rather than fragmenting the junction legs with a dynamically changing junction velocity, an average JRF is determined and used in order to construct the fictitious legs. However, the original procedure for determining the average JRF proposed in Ref.~\cite{Sjostrand:2002ip} often failed to converge, especially when (slow) heavy quarks were involved. In this work, we develop a new and more stable procedure for determining the average JRF, described in sec.~\ref{sec:Theory}. However note that this average JRF is only used for fragmentation and not when deciding where colour reconnections are formed. (For CR string-length calculation purposes, only the ``first frame'', defined by the partons nearest to the junction, is considered.)

Once the average JRF has been found and the fictitious endpoints constructed, the two softest junction legs are fragmented in towards the junction, cf.~the left side of fig.~\ref{fig:juncFrag}. In \Py 8.310 and prior, the two softest junction legs are determined by the legs with the lowest JRF energies. In this work and from \Py 8.311 onwards, the lowest absolute momentum is used instead. This change was made as the energy of a junction leg does not reflect how soft the leg is for heavy-quark endpoints. 

In the example depicted in fig.~\ref{fig:juncFrag}, the two softest junction legs are defined by endpoints $q_{01}$ and $q_{02}$. Once these junction legs are identified and their fictitious endpoints constructed, each of these junction legs is fragmented from the real endpoint in towards the junction, using standard string-fragmentation methods (except for not alternating between the two ends), with fragmentation stopping once the junction is reached. To measure when to stop fragmentation of a given leg, the energy of the leg in the JRF is compared to the summed energies of the produced hadrons. 
There are also parameters \texttt{StringFragmentation:eBothLeftJunction} and \texttt{StringFragmentation:eMaxLeftJunction} which regulate the remaining energy of the two softest junction legs in the JRF so that it is not too large.

Additional to this energy measure, as of \Py 8.311 we have also included a further constraint which ensures that the mass of the so-called ``junction diquark'' remains positive. This junction diquark is comprised of the quarks from the string breaks closest to the junction from the fragmentation of the two softest junction legs. 
In fig.~\ref{fig:juncFrag}, this junction diquark is defined by $q_2q_5$. For the momentum of this junction diquark, energy-momentum conservation yields
\begin{equation}
    p_{\text{diq}} = p_{\text{min}} + p_{\text{mid}} - p_{\text{had}},
    \label{eqn:pDiq}
\end{equation}
where $p_{\text{min}}$ and $p_{\text{mid}}$ are the momenta of the softest and next-to-softest junction legs respectively, and $p_{\text{had}}$ is the momentum sum of the produced hadrons from the fragmentation of these two legs. Once the junction diquark is formed, the last junction leg can be hadronized as a standard dipole string, which in the example in fig.~\ref{fig:juncFrag} would have endpoints $q_3q_5$ and $\bar{q}\bar{q}_{03}$. Note that the choice of fragmenting the softest two junction legs first is somewhat arbitrary, however allows easier handling of hadron mass constraints as it avoids the dipole constructed from the junction diquark and final junction endpoint from being too small. The impact of making  alternative ordering choices was studied in Ref.~\cite{Sjostrand:2002ip} and found to be small.

Notably one can see in fig.~\ref{fig:juncFrag} that the string breaks surrounding the junction necessarily form a baryon, which here we call the ``junction baryon''. This provides an additional baryon production mechanism unique to beyond-LC modelling of colour reconnections plus string fragmentation (with standard baryon production resulting from diquark-antidiquark string breaks or beam remnants). The significant effect of junction baryon production has already been seen in increased baryon-to-meson ratios in $pp$ collisions. An early example of this was the $\Lambda/K$ ratio measured by CMS~\cite{CMS:2011jlm}, which --- as shown in Ref.~\cite{Christiansen:2015yqa} --- default \Py (which relies on LEP-LHC universality~\cite{Skands:2014pea}) underpredicts by over 20\% while the CR model with junctions fits the data well. 

Though junction modelling is relevant for all baryon production, it can be particularly relevant for heavy-flavour baryon production. Heavy-flavour (charm and beauty) quarks act as interesting probes into hadronization as heavy flavours cannot themselves be created via string breaks. Thus in order to form a heavy-flavour junction baryon, the junction leg containing the heavy-quark endpoint must be sufficiently soft that it does not fragment but remains part of the junction baryon. This leaves the heavy-flavour baryon sensitive to the modelling of the junction motion and junction fragmentation. 

Large effects of junctions on baryon production in the heavy-flavour sector have already been seen, particularly at low $p_\perp$. This effect can be seen in the increased $\Lambda_c/D^0$ ratio at low $p_\perp$ \cite{ALICE:2021rzj}, where the default \Py tune (Monash 2013~\cite{Skands:2014pea}) predicts an approximately flat distribution. Given that the probability of diquark-antidiquark pair production via string breaks has some fixed probability (in the absence of collective effects such as ropes \cite{Bierlich:2016faw, Bierlich:2017sxk} or close-packing \cite{Fischer:2016zzs}), the universality-limit flat distribution is unsurprising. The rise at low $p_\perp$ exhibited by the QCD CR models can be explained by junction baryons as junctions predominantly sit at low $p_\perp$. This is a natural consequence of the string-length minimisation in CR, which results in junction reconnections largely occurring between jets the tips of which fragment into hard mesons (and baryons with the normal --- universal --- ratio) while the junction baryon is the most subleading and hence typically softest hadron produced (in the JRF), hence it generally also sits at low $p_\perp$ (in the LAB).

Another interesting consequence of including junctions in a string model is the decrease of the $\Lambda_b$-baryon asymmetry at low $p_\perp$ as seen by LHCb~\cite{LHCb:2021xyh}. For $pp$ collisions,  models without junction CR predict a large asymmetry. This arises from $\Lambda_b$ production via combination of a $b$ quark with the proton beam remnant, whereas there is no equivalent mechanism for $\bar{\Lambda}_b$ due to the absence of an antiproton beam remnant for a $\bar{b}$ quark to combine with. However in CR models with junctions, junctions are always created in pairs (total baryon number is conserved), cf., e.g., fig.~\ref{fig:CR}. The contributions from junction topologies to $\Lambda_b$ and $\bar{\Lambda}_b$ production are therefore approximately equal, and as previously discussed particularly occurs at low $p_\perp$ which in turn dilutes the asymmetry in that part of phase space. 

The original formulation of junction fragmentation, however, focused on high-energy junction legs~\cite{Sjostrand:2002ip}. Special consideration of the limit of soft endpoints was not built into the model, nor were mass effects arising from gluon kinks and endpoint masses under good control. Consequently, this initial construction, which remained the implementation for \Py 8.310 and earlier, was not stable against such cases, with the procedure in \Py 8.310 unable to construct an average JRF for around 10\% of  minimum-bias events at LHC energies. The interesting predictions seen for low-$p_\perp$ heavy-baryon production with the QCD CR model including junctions motivates careful examination of such soft endpoint junctions. Also given that CR aims to minimise string lengths, one would naturally expect many junction systems to contain soft endpoints and soft gluon kinks near the junction. Thus in order for us to make firmer predictions, particularly but not only for low-$p_\perp$ heavy baryons, we here aim to construct a more physically robust model. The first step is to reexamine the determination of the average JRF.

\section{Junction Motion}
\label{sec:Theory}

\noindent The most intuitive way to construct the JRF is to consider balancing the force exerted by the three junction strings on each other. Given the constant force from strings and a simple three-parton junction configuration, the natural frame to consider is where the opening angle between the 3-momenta of each pair of partons is 120\textdegree~\cite{Sjostrand:2002ip}. In the following we will refer to this as the ``Mercedes frame'', an example of which can be seen in fig.~\ref{fig:juncFrag}. For a set of massless four-vectors, a boost to the Mercedes frame always exists (we will elaborate on this point in sec.~\ref{sec:120JRF}), however the same is not true for massive four-vectors. In the limit the Mercedes frame no longer exists, by considering the classical action of the string, the junction effectively becomes bound to the massive parton and follows its motion. This scenario, which we call ``pearl-on-a-string'', will be described in detail in sec.~\ref{sec:pearl}. 

In the following, we begin by describing the Mercedes frame and then proceed to consider soft-leg cases, examining both oscillatory motion of an endpoint around a junction and the pearl-on-a-string scenario. In these discussions, we will begin by remaining in the context of a simple three-parton junction configuration without the inclusion of gluon kinks. Then, we introduce a single gluon kink on a junction leg in sec.~\ref{sec:gluonKinks} and provide a theoretical description of the effect on junction motion. The practical handling of junction configurations generalised to multiple gluon-kinks is explored in sec.~\ref{sec:implementation}. 

\subsection{Mercedes frame}
\label{sec:120JRF}

The boost to the Mercedes frame can be determined using the method presented in~\cite{Sjostrand:2002ip}, which is summarised for context below (note this procedure is unchanged in the updated implementation). 
Let us label the parton four-vectors $p_i$ in the initial (arbitrary) frame of reference, with $p'_i$ being the four-vectors in the Mercedes frame. 
Given the Lorentz invariant four-products, $a_{ij}=p_i p_j$, and the fixed angle between the three-momenta in the Mercedes frame, $\theta_{ij} = 120^\circ = 2\pi/3$, the Lorentz invariant four-products can be defined in terms of the Mercedes-frame energy and momentum, 

\begin{equation}
	a_{ij} = p_ip_j = p'_i p'_j = E'_iE'_j - | \vec{p\,}'_i| |\vec{p\,}'_j| \cos{ \frac{2\pi}{3}} = E'_iE'_j + \frac{1}{2} | \vec{p\,}'_i| |\vec{p\,}'_j|.
	\label{aij}
\end{equation}

By rewriting the energies in terms of the mass and momenta, one can see that the only degrees of freedom in eq.~\eqref{aij} are $|\vec{p\,}'_i|$ and $|\vec{p\,}'_j|$. Thus by rearranging eq.~\eqref{aij} we can introduce a function, $f_{ij}$, such that solutions for $|\vec{p\,}'_i|$ and $|\vec{p\,}'_j|$ are found when $f_{ij}=0$. Note that the first two arguments of this function are monotonically increasing.
\begin{equation}
	f_{ij} = f(| \vec{p\,}'_i|,| \vec{p\,}'_j|;m_i,m_j,a_{ij}) = \sqrt{| \vec{p\,}'_i|^2 + m_i^2}\sqrt{ | \vec{p\,}'_j|^2 + m_j^2 } + \frac{1}{2} | \vec{p\,}'_i| |\vec{p\,}'_j| - a_{ij}
	\label{fij}
\end{equation}

By requiring $f_{12} = 0$ and $f_{13} = 0$ whilst allowing $| \vec{p\,}'_1|$ to vary freely in a kinematically allowed region, we can uniquely solve for the other momenta as a function of $|\vec{p\,}'_1|$,

\begin{equation}
	| \vec{p\,}'_j|(\vec{p\,}'_1) = \frac{2E'_1\sqrt{ 4a_{1j}^2 - m_j^2 (4E^{'2}_1 - | \vec{p\,}'_1|^2) } - 2| \vec{p\,}'_1| a_{ij} }{4E^{'2}_1 - | \vec{p\,}'_1|^2}.
	\label{pj}
\end{equation}

Using eq.~\eqref{pj}, we can rewrite $f_{23}(|\vec{p\,}'_2|, |\vec{p\,}'_3|)$ as a function of $| \vec{p\,}'_1|$. As eq.~\eqref{pj} is a decreasing function with $| \vec{p\,}'_1|$, the function $f_{23}$ is also a monotonically decreasing function with $| \vec{p\,}'_1|$. Thus we can use $f_{23}(|\vec{p\,}'_1|)=0$ to solve for $| \vec{p\,}'_1|$ and therefore solve for all $|\vec{p\,}'_i|$. In the case where all three junction legs are massless, there is always a physical solution which is given by 
\begin{equation}
E'_i = | \vec{p\,}'_i| = \sqrt{2a_{ij}a_{ik} / 3a_{jk}}~.
\end{equation}
For general masses, however, there is no simple analytical solution and instead iterative root-finding procedures are used. But for some three-parton configurations involving at least one massive parton, a physical solution to the above procedure may not exist at all, implying that a boost to the Mercedes frame does not exist. That is, for certain parts of massive three-body phase space, there is no Lorentz frame in which the opening angles between all three partons are 120 degrees. These cases constitute one class of ``soft-leg" cases, and will be treated separately (see below in sec~\ref{sec:pearl}). Note that in \Py 8.310 and prior, there is no special consideration for such scenarios, and the root-finding procedure would generally terminate in an error.

When the Mercedes frame exists and having determined the energies and momenta of each parton in that frame according to the above procedure, one can easily construct a boost to this frame. We use the centre-of-mass frame as a stepping stone as it is a well-defined frame and the momenta in this frame lie in a plane, which necessarily will be the same plane the Mercedes frame sits on. The boost to the centre-of-mass frame is given by $\vec{\beta\,}^{CM} = -\sum\vec{p}_i / \sum E_i$, with centre-of-mass momenta $p_i^{CM}$. Thus for boost $\vec{\beta\,}'$ from the centre-of-mass frame to the Mercedes frame, $\vec{\beta\,}'$ must obey $\gamma'E_i^{CM}+\gamma'\vec{\beta\,}'\vec{p\,}_i^{CM} = E'_i$. Dividing this equation by $E_i^{CM}$, and subtracting the equation for $i=j$ from $i=1$, we get

\begin{equation}
	\gamma'\vec{\beta\,}' \biggl( \frac{\vec{p\,}_1^{CM}}{E_1^{CM}} - \frac{\vec{p\,}_j^{CM}}{E_j^{CM}} \biggr) = \frac{E'_1}{E_1^{CM}} - \frac{E'_j}{E_j^{CM}}.
	\label{JRFboost}
\end{equation}
From here we can parameterize the boost to the Mercedes frame as a linear sum of the vector differences in eq.~\eqref{JRFboost} for $j=2$ and $j=3$. Then combining $\beta^{CM}$ and $\beta'$ gives the overall boost to the Mercedes frame from the original frame.

\subsection{Soft-leg Junction Motion}
\label{sec:soft}

We define a junction leg as ``soft'' if it does not have sufficient energy for a string break to occur between the endpoint and the  junction. In the following, we will consider junction motion in the absence of such a break, and will assume there is only a single soft leg. We consider both massless and massive limits for the soft leg, while we shall assume the other two legs to have massless endpoints and be much more energetic in comparison.

\subsubsection{Massless Soft-leg Case}
\label{sec:softMassless}

Consider the case of a single soft massless endpoint in the Mercedes frame. In the absence of string breaks, it will travel outwards from the junction until it has lost all of its momentum to the junction string. After that, it will change its direction of motion and begin moving back towards the junction. Eventually, it will ``hit'' the junction, at which time the junction will no longer remain at rest and instead begins moving in the direction of the soft endpoint. This process is repeated when the parton turns around again. Overall this behaviour results in some oscillation of the soft leg around the junction, with the junction itself moving in a sort of start-stop motion, at rest when the soft parton is on one side of it and moving when it is on the other. 

\begin{figure}[tp]
    \centering
    \includegraphics[width = \textwidth]{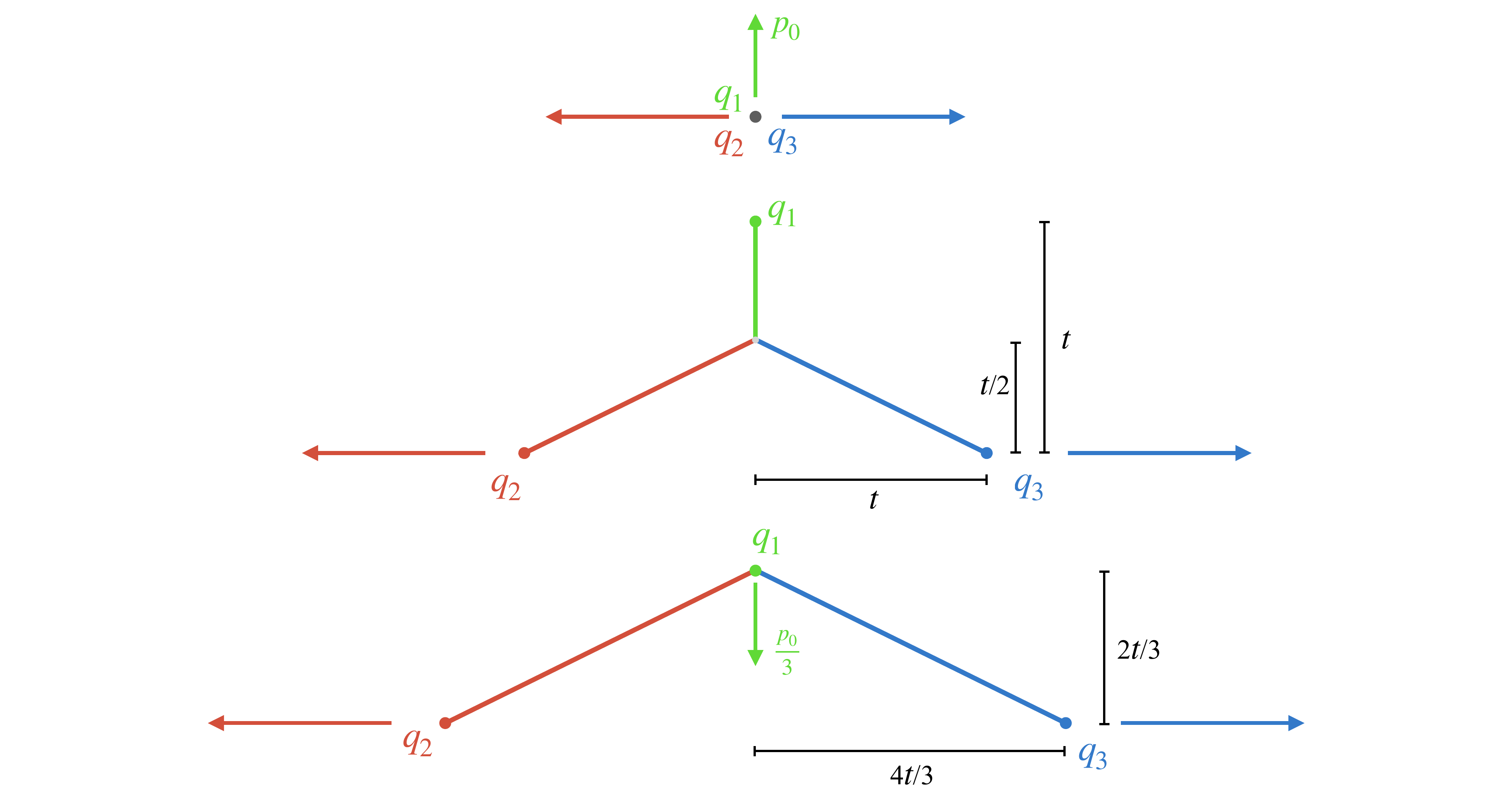}
    \includegraphics[width = \textwidth]{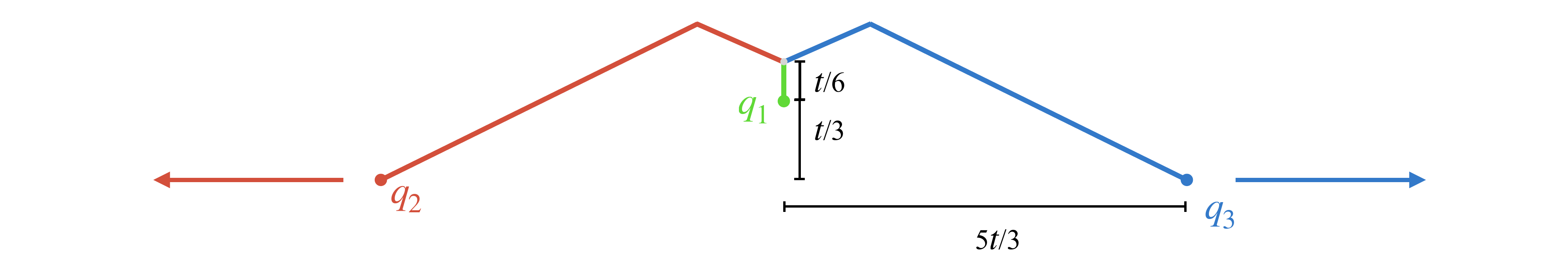}
    \includegraphics[width = \textwidth]{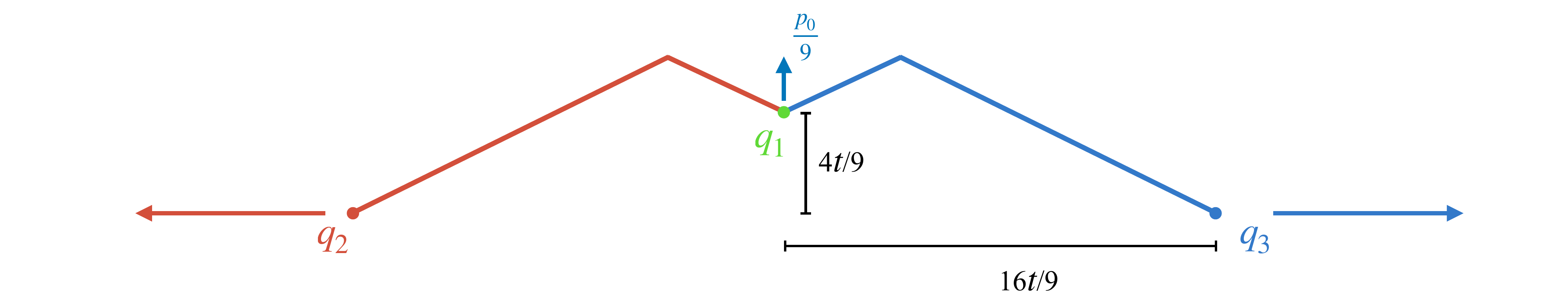}
    \includegraphics[width = \textwidth]{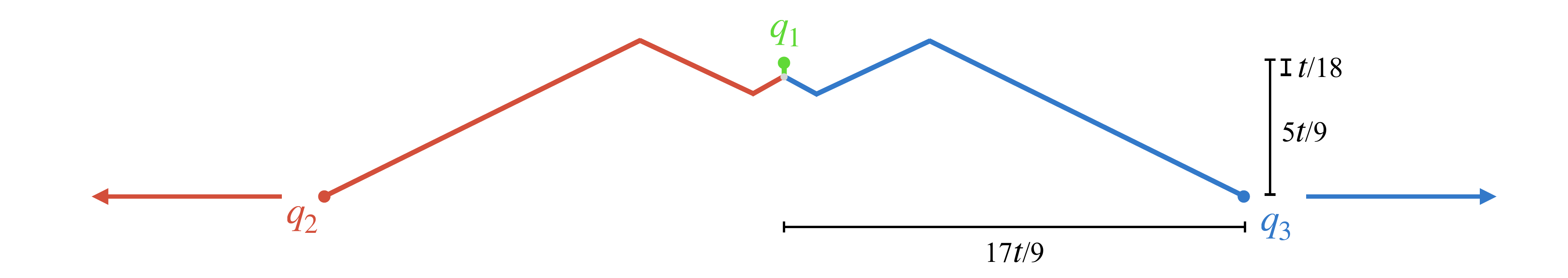}
    \caption{Given a simple three-parton junction configuration in the Ariadne frame with respect to $q_1$, here we show the oscillatory behaviour of parton $q_1$ around the junction at different times. This sketch is drawn assuming no string breaks occur and all massless partons. Here the initial momentum of $q_1$ is $p_0$, and the distance it traverses till it loses its momentum in the first iteration is given by $t = p_0/\kappa$. The initial momenta of $q_2$ and $q_3$ are assumed to be much greater than $p_0$. Note that the arrows here are only intended to illustrate the momentum directions, not their magnitudes.}
    \label{fig:softOsc}
\end{figure}

To study this oscillatory motion in detail, it is instructive to examine the behaviour in the so-called Ariadne frame with respect to the soft junction leg. As depicted in the first frame of fig.~\ref{fig:softOsc}, the Ariadne frame is defined such that the 3-momenta of the two more energetic legs are back-to-back (here we allign this with the $z$-axis) and the soft leg is at 90\textdegree\space to the other two legs (along the labelled $x$-axis in this case). This is a convenient frame to map the junction motion as the $x$-direction contribution is solely due to the soft endpoint. 
Given that the initial momenta of each parton are massless for now, the Mercedes frame exists and defines the JRF at early times. From a Mercedes frame supposing massless partons, a boost of $\beta=1/2$ is required to get to the Ariadne frame, implying a junction velocity of $1/2$ in this frame. Using the initial junction velocity of 1/2 and parton velocity of 1, the oscillatory motion can be fairly simply mapped out as in fig.~\ref{fig:softOsc}.

In fig.~\ref{fig:softOsc}, the initial 3-momentum of the soft endpoint $q_1$ is aligned with the positive $x$ direction, $\vec{p}_0 = p_0(1,0,0)$. The equation of motion for a string endpoint is $\mathrm{d}p/\mathrm{d}t=-\kappa$~\cite{Andersson:1998tv} (oriented along the string). Thus, the time it takes for $q_1$ to lose all of its momentum, and hence the distance it will traverse before it must turn around, is linearly proportional to its momentum. We label this time as $t = p_0/\kappa$. 

The second image in fig.~\ref{fig:softOsc} shows the configuration at time $t$, at which $q_1$ has lost its momentum and will begin to move in the negative $x$-direction. This change has not propagated to the junction yet, however, which will continue to move in the positive $x$-direction with velocity $\beta = 1/2$.
With these new velocities, one can calculate that after a further time $t/3$ the junction will catch up to parton $q_1$ as seen in the third image of fig.~\ref{fig:softOsc}. 
By this time $q_1$ will have gained a momentum $-\vec{p}_0/3$ from the string. 

As $q_1$ passes through the junction, the net force exerted on the junction from junction strings will change, and the junction will now begin to move in the negative $x$-direction with velocity $\beta = -1/2$.
This forms a zig-zag shape on the string around the junction as seen in fourth image of fig~\ref{fig:softOsc}.
One can see that the shape of strings close to the junction after the zig-zag has formed is the same shape as at initial junction string configuration, however scaled down by a factor of $1/3$. 

From here, the same calculations can be used to map out the next oscillation as shown in the bottom three frames of fig~\ref{fig:softOsc}, whereby given momentum $-\vec{p}_0/3$ at the start of the oscillation, $q_1$ will now lose its momentum after a further time $t/9$ and will next hit the junction with momentum $\vec{p}_0/9$. This oscillatory pattern then continues, scaled down by a factor of $1/3$ per oscillation. 

As this oscillatory motion scales down by $1/3$ each time and the initial assumption was that $p_0$ corresponded to a ``soft'' leg, the scale of these oscillations remains below that of hadronic sizes. 
Therefore the details of the zig-zag string shape from this oscillatory behaviour are presumably not significant and should not require detailed tracing for an accurate hadronization picture. 
Thus, in this work we consider it sufficient to find an average junction rest frame. Details on how the averaging treatment is carried out are given in sec.~\ref{sec:avgJRF}. 
One should also note that though these diagrams show a spatial representation of the strings for schematic simplicity and visualisation purposes, in practice we handle everything from JRF calculations to fragmentation in momentum space. 

To construct the average, we will need the junction velocity during each oscillation. This can be determined  without an explicit boost to the Ariadne frame and also generalised for massive endpoints. 
Consider a soft parton with momentum $p'_0$ in the Mercedes frame of a junction system (when such a frame exists). Given the equation of motion $\mathrm{d}p/\mathrm{d}t=-\kappa$, at time $p'_0/\kappa$ the endpoint will lose all its momentum and will change its direction of motion back towards the junction, however in this frame the junction itself remains at rest for now. This Mercedes frame will remain the JRF up until time $2p'_0/\kappa$, when the soft endpoint hits the at-rest junction, with 3-momentum $-\vec{p\,}_0'$ (irrespective of mass). This momentum, $-\vec{p\,}'_0$ will then dictate the direction of the pull on the junction contributed by the string piece connected to the soft parton. 
To calculate the pull from the other two junction legs we consider the momenta of these legs at time $2 p'_0/\kappa$, by which time the partons will have lost $2p'_0$ momentum in their respective directions of motion. Given these updated momenta for each junction leg, one can construct a new Mercedes frame (assuming it exists), which then defines the JRF for times after $2 p'_0/\kappa$ up till the next oscillation begins. For massive oscillations, this will rapidly approach a pearl-on-a-string scenario, which we discuss below. 

\subsubsection{Pearl-on-a-String}
\label{sec:pearl}

The next case to consider is a massive endpoint that is sufficiently soft that a boost to the Mercedes frame does not exist. In such scenarios, the solution to $f_{23}(|\vec{p\,}'_1|) = 0$ would return a negative-valued $| \vec{p\,}'_1|$ which of course is unphysical. Without resorting to testing for negative solutions of $f_{23}(|\vec{p\,}'_1|) = 0$, such configurations can be identified by looking at the rest frame of each massive parton. Whilst in the rest frame of a given massive endpoint, if the opening-angle between the other two legs is greater than 120\textdegree, a boost to a Mercedes frame does not exist. One can easily convince oneself of this by considering the boost required to reduce the angle of the other two legs to 120\textdegree, i.e. $f_{23}(|\vec{p\,}'_1|) = 0$, and the resulting momenta of the massive endpoint. Though not immediately obvious, this property is unique to at most one parton in any given 3-parton configuration. In principle, as we assume massless gluons these cases should only occur with endpoint partons. In such cases, we expect the junction to get ``stuck’’ to the massive endpoint, and consequently one can map the junction motion by considering the massive parton motion.
When the junction becomes bound to the massive quark in this way, we call that quark a ``pearl-on-a-string''. 

\begin{figure}[tp]
\centering
    \includegraphics[totalheight=4.8cm]{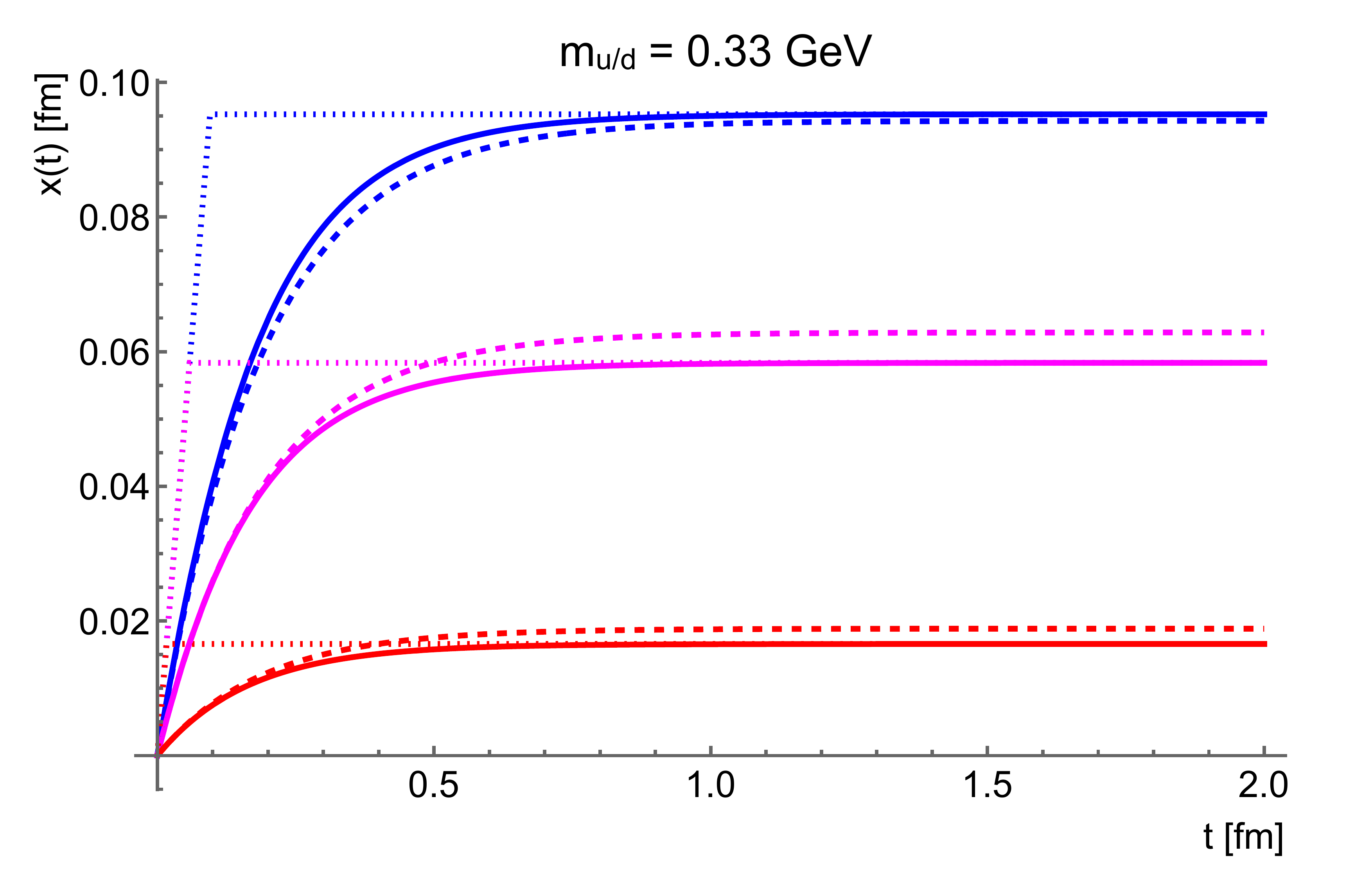}
    \includegraphics[totalheight=4.8cm]{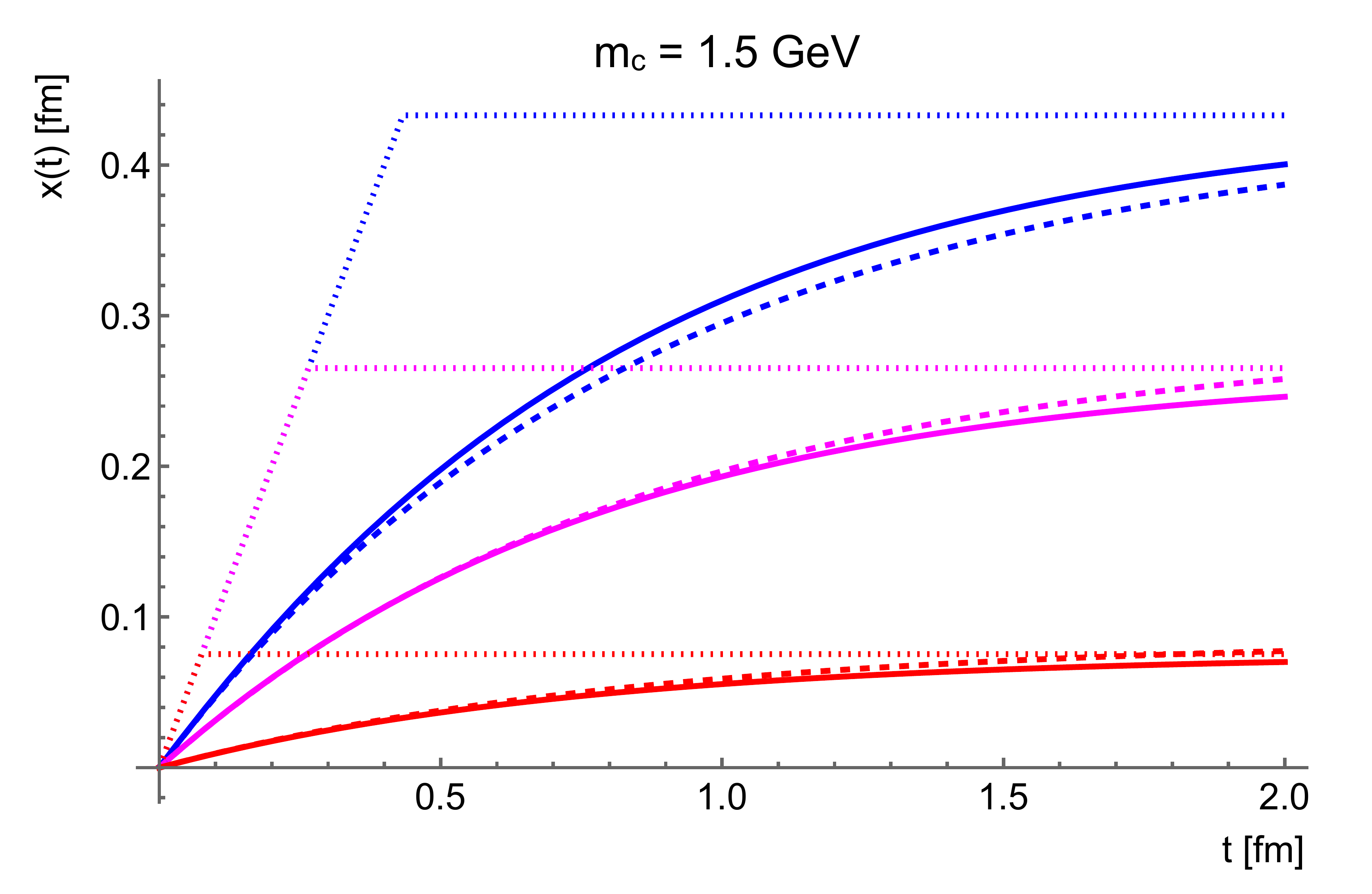}
    \includegraphics[totalheight=4.8cm]{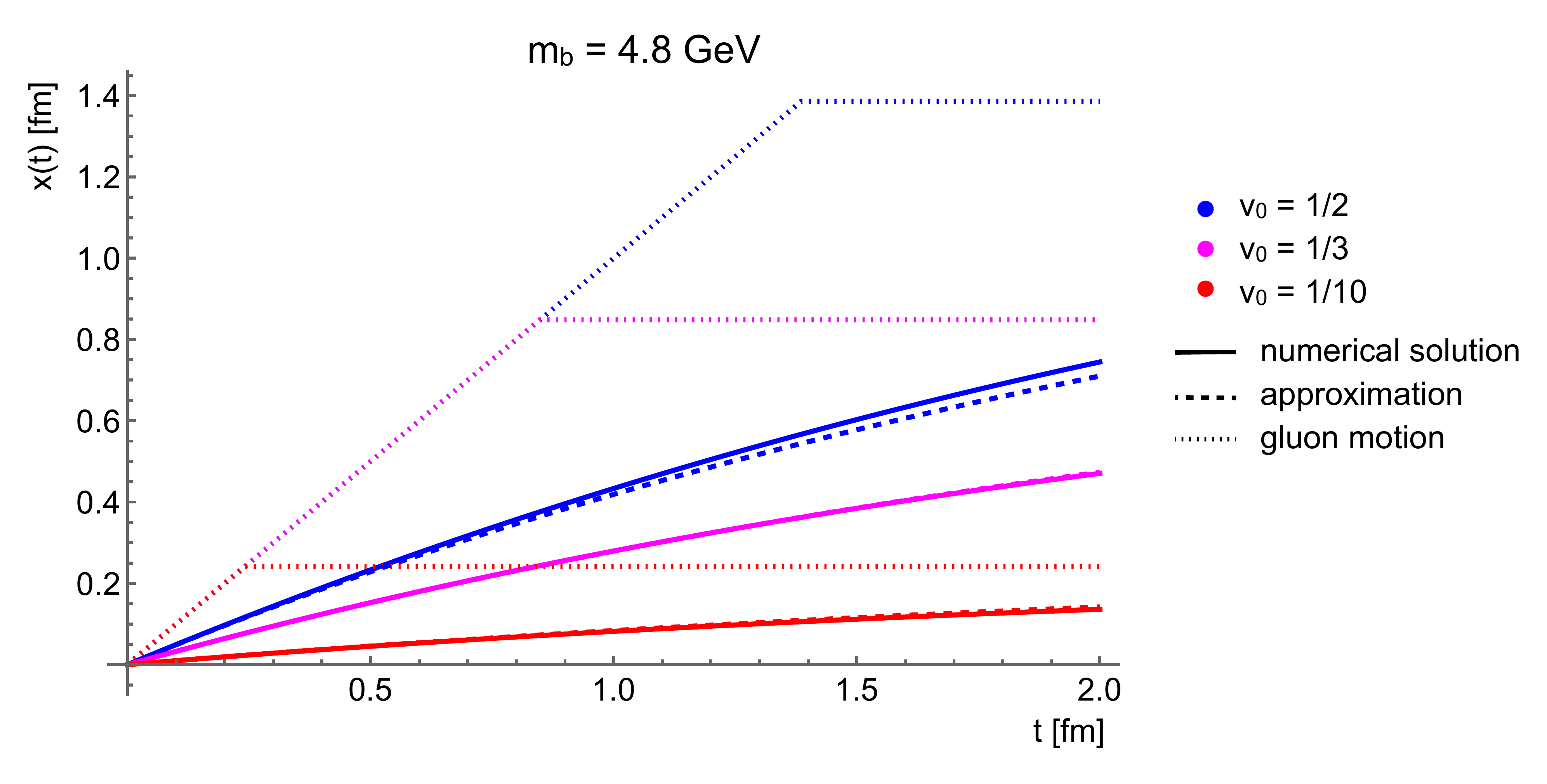}
    \caption{The pearl motion for an up/down, charm and bottom quark with constituent masses, considering initial pearl velocities of $\frac{1}{2}$, $\frac{1}{3}$ and $\frac{1}{10}$. Plots show the numerical solution to eq.~\eqref{eqn:vHeavy} (solid line), the simple approximation of the pearl motion given eq.~\eqref{eqn:approx} (dashed line), and the motion of a gluon-kink on a $q\bar{q}$ string (dotted line).}
    \label{fig:pearlOnAString}
\end{figure}

\begin{figure}[tp]
    \centering
    \includegraphics[width = 1.02\textwidth]{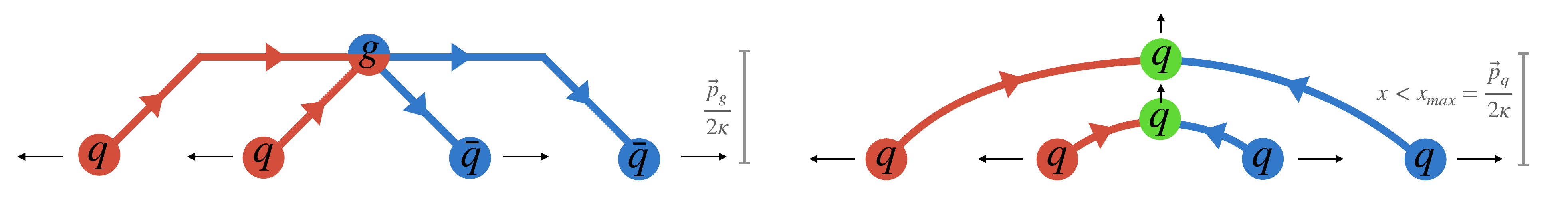}\hspace*{-5mm}
    \caption{Left: a gluon kink on a $q\bar{q}$ string, forming a flat region after the gluon momentum has been used up.  Right: schematic of the junction moving whilst ``stuck'' to a massive pearl, producing the shape of the pearl-on-a-string scenario. Note this is by no means a to-scale diagram but merely to show the general concept of the junction motion and the curvature of the string. 
    Each sketch shows the string configuration at two different times, with black arrows indicating direction of motion. Both sketches are shown in the Ariadne frame, the left-hand one with respect to the gluon, the right-hand one with respect to the pearl.}
    \label{fig:pearl}
\end{figure}

For the purpose of CR, simply using the initial rest frame of the pearl quark is sufficient as at early times this will be the JRF, which has now been explicitly implemented in \Py 8.311 onwards. For fragmentation purposes we wish to map out the junction motion. 
To do so, let us consider the simplest case; a junction with a single massive endpoint and two hard massless endpoints such that we have a pearl-on-a-string configuration. As with mapping the oscillatory motion of a massless endpoint, here it is again instructive to work with the equations of motion for this configuration in the Ariadne frame with respect to the soft massive parton. 
Given the string tension $\kappa$, the initial magnitude of the massive parton 3-momentum $p_0$, and the position $x(t)$ and velocity $v(t)$ of the massive parton, we can write the massive-parton momentum as a function of time, 
\begin{equation}
	p(t) = p_0 - 2\kappa x(t) = \frac{mv(t)}{\sqrt{1-v(t)^2}}~.
	\label{pHeavy}
\end{equation}
Note the factor of 2 here as the junction/pearl are connected to two other string segments. Using the equations of motion of relativistic strings, eq.~\eqref{pHeavy} is derived by considering the energy gained by the string $\mathrm{d}E_s$, over time $\mathrm{d}t$, given the massive parton moves some distance $\mathrm{d}x$. Detailed derivations can be found in Appendix~\ref{app:derivation}. This gives a maximal distance of $x_{max} = p_0/2\kappa$, at which we expect the velocity to approach zero. Rearranging eq.~\eqref{pHeavy} for the massive parton velocity, we get the following:

\begin{equation} 
	\frac{\mathrm{d}x}{\mathrm{d}t} = \frac{1}{\sqrt{1 + \frac{m^2}{(p_0 - 2\kappa x(t))^2}}}~.
	\label{eqn:vHeavy}
\end{equation}
This differential equation is non-trivial to solve as the time dependence is not straightforward, thus it is not very useful in this form for a practical implementation. Instead of an exact solution, an approximation of the time-dependent velocity can be used in its place. By numerically solving eq.~\eqref{eqn:vHeavy}, the general shape of the solution can be approximated by a simple exponential,

\begin{equation}
    \frac{\mathrm{d}x}{\mathrm{d}t} = v_0 \exp \left( \frac{-1.75t}{m} \right)~,
    \label{eqn:approx}
\end{equation}
with $m$ being the mass of the pearl, $v_0$ the initial velocity in the Ariadne frame, and using the assumption that $\kappa\sim 1$ GeV/fm. A comparison of the solution $x(t)$ between the numerical solution and approximate functional form is given in fig.~\ref{fig:pearlOnAString}, showing the motion of a pearl quark over a time-scale of hadronization ($\sim$2 GeV) for several different $p_0$ and $m$ values. Given this non-uniform motion of the junction bound to a pearl with a decelerating velocity, the resulting string shape is curved and appears as shown in the right image of fig.~\ref{fig:pearl}.

While this may seem like an oversimplification, we do  believe it would be overkill at this point to develop a more complicated functional form, also just given the uncertainties, e.g., on the precise value of the string tension to use, and the fact that the real physics is that of a curved string with non-trivial time dependence which is not well understood. Rather than attempting to fragment the actual curved string, we look to reuse techniques for fragmenting dipole strings again in this context. The key behaviours we wish to model in this pearl motion is the position (in momentum space) of the junction baryon and some sense of the pearl $p_\perp$ that is propagated along the strings to create the curve. The simple form of eq.~\eqref{eqn:approx} should be sufficient for our purposes.

For the case of light quarks with sufficiently large $p_0$, one can see in fig.~\ref{fig:pearlOnAString} that the motion somewhat mimics that of a gluon kink on a $q\bar{q}$ string. 
In the Lund string model, gluons form transverse excitations (or kinks) on strings. These initially form a sharp corner on the string. Once they have lost their momentum to the string, a flat so-called ``region’’ forms on the string, spanned between two kinks travelling in opposite directions. 

A massless gluon with initial momentum $|\vec{p}_g|$ and initial velocity of 1  will lose its momentum after a time $|\vec{p}_g|/2\kappa$. 
Once it loses all its momentum to the string, the flat region forms and the effective gluon velocity becomes zero (there is no longer a force acting on the string in the direction of the original gluon). The general shape of such a gluon-kink can be seen in the left image of fig.~\ref{fig:pearl}. For a detailed description of string regions and gluon kinks see \cite{Sjostrand:1984ic, Sjostrand:1984iu}. Given this flat region formed by soft gluons, one naturally can make a comparison to the curve in the pearl-on-a-string case; the pearl-on-a-string essentially acts as a massive gluon. 

Hence for light-quark cases with sufficiently large $p_0$, we may approximate the massive pearl as a massless gluon, in order to mimic the curved string and the $p_\perp$ propagated to the string from the pearl quark. 
As such we can recycle the standard $q-g-\bar{q}$ fragmentation procedure. In the following we will explain the procedure for the case of a junction with quark endpoints; the same obviously holds for an antijunction system with ($q\leftrightarrow\bar{q}$).

In the gluon-pearl approximation, the massive quark and junction are replaced by a massless gluon, $g_q$, with energy and momentum of $p_0$ in the Ariadne frame. The excess energy from the mass of the pearl is ``stored'' at the junction itself. 
From here the $q-g_q-q$ string can be mapped onto a $q-g-\bar{q}$ string by taking $q\rightarrow\bar{q}$ for one of the junction legs, and then standard dipole fragmentation procedures can be used. To simplify the treatment, we initially force fragmentation from only the $\bar{q}$ end of the $q-g_q-\bar{q}$ string. For each hadron produced from this $\bar{q}$ string end, the conjugate is taken ($q\leftrightarrow\bar{q}$) which maps it back to the initial $q-g_q-q$ system. This procedure is continued until a string break steps over the junction, at which point the pearl quark and the excess energy from the pearl quark mass is gained, forming the junction baryon. As with standard junction fragmentation, the energy of the produced hadrons is used as a measure of when the junction is reached. 
Once the junction baryon has been made, all that remains is a typical $q-\bar{q}$ string for which standard fragmentation is carried out including the left-right randomization of string ends. 

Notably this gluon-pearl approximation only holds for light-quark cases with sufficiently large initial velocity $v_0$, such that the change in velocity $\Delta v$, is approximately 1/2 over some reasonable hadronization time $\sim 2$ GeV. It is evident in fig.~\ref{fig:pearlOnAString} that the approximation no longer holds for heavier flavours or quarks with sufficiently small initial velocity. 
Let us consider the limit where $p_0$ approaches zero and the limit of an infinitely heavy quark. In such cases the pearl does not move and no $p_\perp$ is propagated along the string. As such, we would expect the standard junction hadronization method of fragmenting each leg towards the at rest junction to work well. For large mass quarks or light quarks with low non-zero $v_0$, although the junction motion is non-uniform, the change in velocity is small. Thus in these cases one can use a perturbed JRF, which simply takes the average velocity according to the approximate functional form in eq.~\eqref{eqn:approx} over some hadronization time-scale. From here one can use the standard junction fragmentation framework given this perturbed JRF. 

There is of course ambiguity in which treatment to use in cases with intermediate $p_0$ for light quarks and for charm quarks. Hence to create a smooth transition between the two treatments, a probabilistic choice can be made based on $\Delta v$ over hadronization times, with velocity changes of zero and $1/2$ corresponding to the standard junction fragmentation and gluon-pearl approximations respectively. 

\subsection{Gluon Kinks on Junction Motion}
\label{sec:gluonKinks}

As with a simple $q\bar{q}$ string, each junction leg can also have any number of intermediate gluons (kinks) between the junction and the endpoint. 
In this section we will consider the effects of intermediate gluons on junction motion. First we will look at a simplified model of only considering the pull from the gluon extending out from the junction. This simplification is what is implemented in \Py 8.311 as outlined in sec.~\ref{sec:implementation}. 
Then we will allow the gluon kinks to propagate back towards the junction, and the effect on the junction motion as the kink passes through the junction. This is proposed simply as a theoretical model and not yet implemented into \Py as a general description of multiple gluon kinks has not yet been solved. However we do not expect the corrections due to gluon kinks propagating through the junction to be large.

\begin{figure}[t]
    \centering
    \includegraphics[width = 0.43\textwidth]{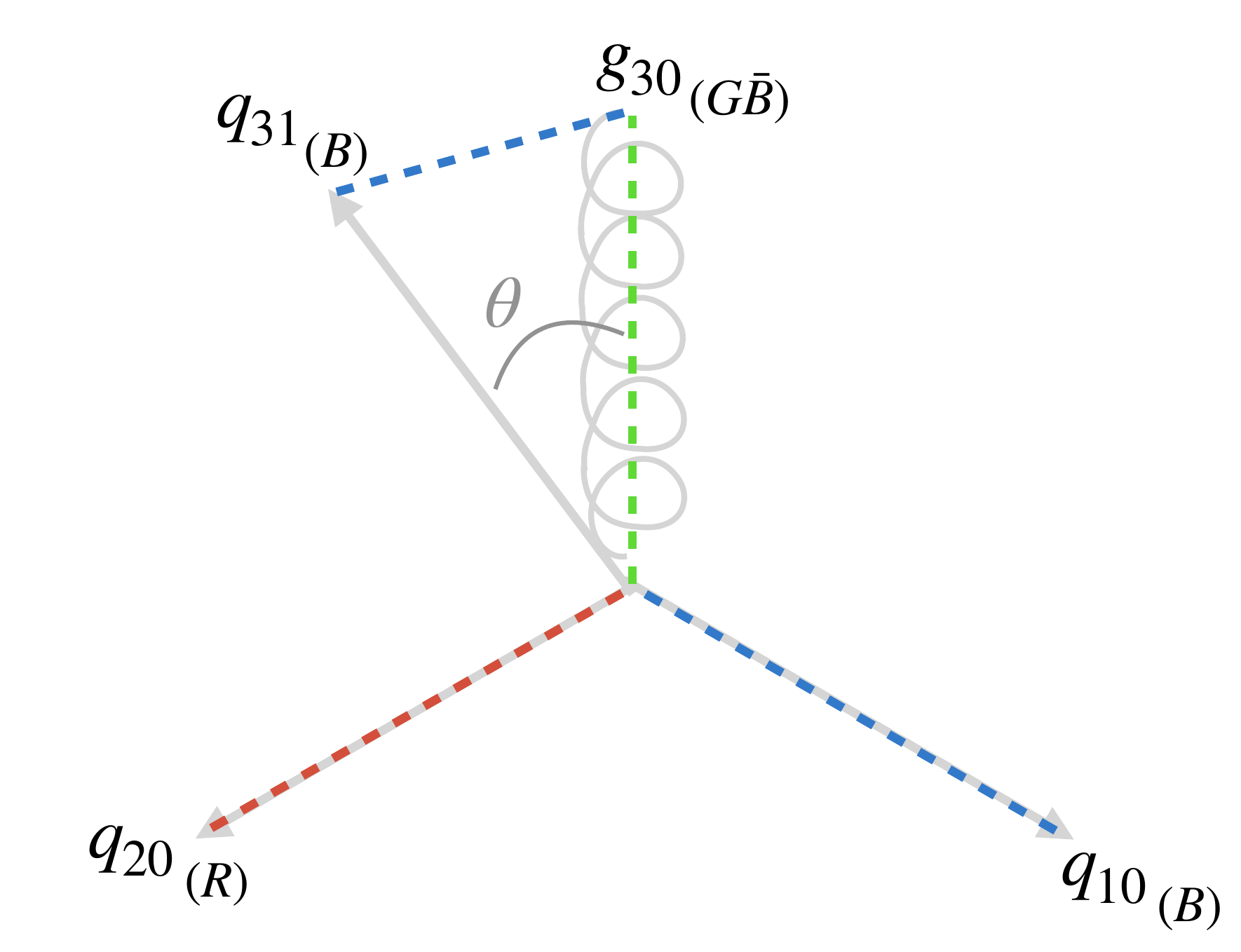}
    \caption{The example setup considered throughout sec.~\ref{sec:gluonKinks}, with a single gluon kink junction leg 3 and all massless parton. This shows the string configuration at early times, depicted in the JRF of the first parton on each leg, with the coloured dashed lines representing string segments.}
    \label{fig:gluonSetup}
\end{figure}
We first consider the simplest example; a junction system with a single gluon kink on one leg, and all  partons massless. To examine the junction motion of such a configuration, we consider the pull on the junction at different times. As information is propagated along the string at the speed of light, the initial junction velocity will be dictated by the first parton on each junction leg, and the pull from partons further out along the leg would only contribute to the junction motion at later times. 

The string configuration of such a junction system with a single gluon kink in its initial JRF is illustrated in fig.~\ref{fig:gluonSetup}, where this early-time junction motion is dictated by the partons labelled $q_{10}$, $q_{20}$, and $g_{30}$. 
This frame is expected to remain the JRF until the first of these partons loses its momentum, which is determined by the string equation $dp/dt = -\kappa$. This equation of motion assumes a connection to a single string with tension $\kappa$, however gluons are colour octets and therefore connected to two other colour charges via string pieces, thus they lose momentum twice as quickly compared to endpoints. The case of an endpoint being the softest parton results in endpoint oscillations as described for massless partons in sec.~\ref{sec:softMassless}. Here let us instead consider the alternative scenario where the gluon is the first to lose its momentum, i.e., $|\vec{p}_{g_{30}}|/2 < |\vec{p}_{q_{10}}|$ and $|\vec{p}_{g_{30}}|/2 < |\vec{p}_{q_{20}}|$. 
The gluon will lose its momentum at time $|\vec{p}_{g_{30}}|/2\kappa$, at which time $q_{10}$ and $q_{20}$ will have lost momentum $|\vec{p}_{g_{30}}|/2$ in their respective directions of motion. 

Once the gluon has lost its momentum, the next parton on that leg, $q_{31}$, will determine the pull on the junction, along with the reduced momenta of the other two legs. Given the momentum of $q_{31}$ and the reduced momenta of $q_{10}$ and $q_{20}$, one can construct a new JRF valid for times after $|\vec{p}_{g_{30}}|/2\kappa$. 

This procedure can be easily generalised to consider multiple gluons on each junction leg by iteratively stepping outwards on each junction leg, and this is the model we use for the implementation in \Py which we elaborate on in sec.~\ref{sec:implementation}. 

Though this modelling of gluon kinks is adequate, similarly to expecting endpoint oscillations one would expect the gluon kink to propagate back towards the junction once it has lost its momentum, the behaviour of which is ignored in the above description. Below we consider, for theoretical reference without a full-fledged implementation, the effect on junction motion when mapping out these gluon kinks in full detail. 

\begin{figure}[t]
\centering  
\includegraphics[width = 0.95\textwidth]{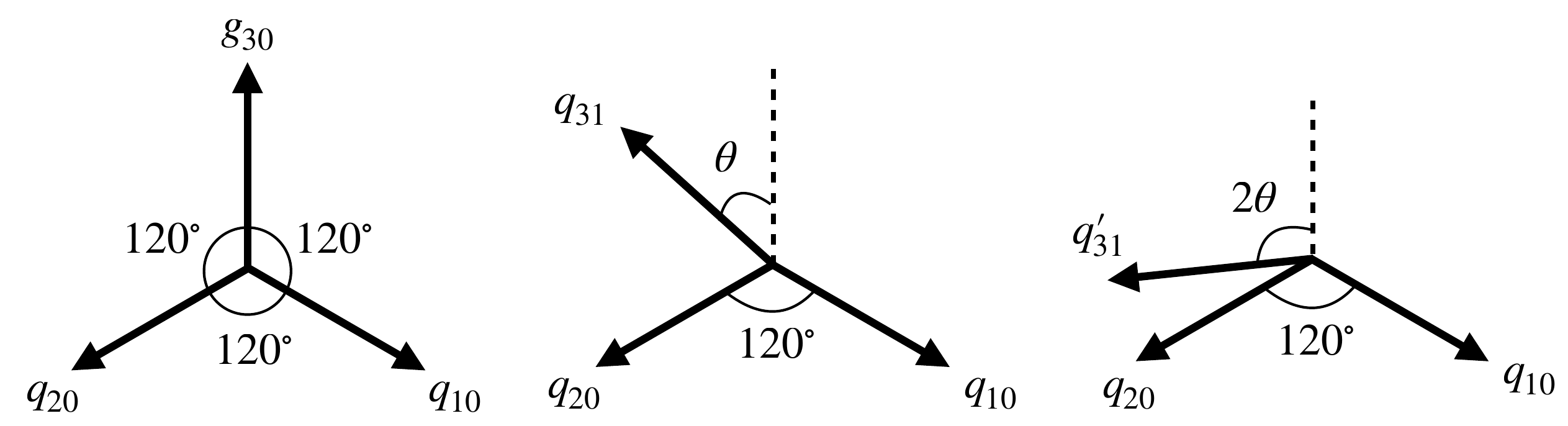}   
    \caption{For the junction configurations portrayed in fig.~\ref{fig:gluonSetup}, each image here shows which partons  dictate the junction motion and the direction of the pull from each such parton over a given time interval. From left to right, the time intervals depicted are 0 to $|p_g|/2\kappa$, $|p_g|/2\kappa$ to $|p_g|/\kappa$, and after $|p_g|/\kappa$. Note these momenta directions are all shown relative to the initial JRF, and such arrows simply represent the direction of the pull on the junction and not the magnitude.}
    \label{fig:gJRF}
\end{figure}

In the same scenario depicted in fig.~\ref{fig:gluonSetup}, at time $|\vec{p}_{g_{30}}|/2$ the gluon $g_{30}$ will have lost its momentum, after which the gluon kink begins to propagate back towards the junction. 
From here it will take another $|\vec{p}_{g_{30}}|/2$ for the gluon kink to return and ``hit'' the junction. This means that only after time $|\vec{p}_{g_{30}}|/\kappa$ will the junction feel the effects of the kink propagating through the junction. 
Thus we can split up the junction motion into three relevant time intervals relative to the initial JRF; between 0 to $|\vec{p}_{g_{30}}|/2\kappa$, $|\vec{p}_{g_{30}}|/2\kappa$ to $|\vec{p}_{g_{30}}|/\kappa$, and after $|\vec{p}_{g_{30}}|/\kappa$. 

Fig.~\ref{fig:gJRF} depicts each of these time intervals and the respective partons which dictate the pull on the junction. As described prior, between times 0 to $|\vec{p}_{g_{30}}|/2\kappa$ the JRF is defined by $q_{10}$, $q_{20}$, and $g_{30}$ (left image of fig.~\ref{fig:gJRF}), and at times $|\vec{p}_{g_{30}}|/2\kappa$ to $|\vec{p}_{g_{30}}|/\kappa$ the JRF is defined by $q_{10}$, $q_{20}$, and $q_{31}$ (middle image of fig.~\ref{fig:gJRF}). 

\begin{figure}[t]
    \centering
    \includegraphics[width = 0.56\textwidth]{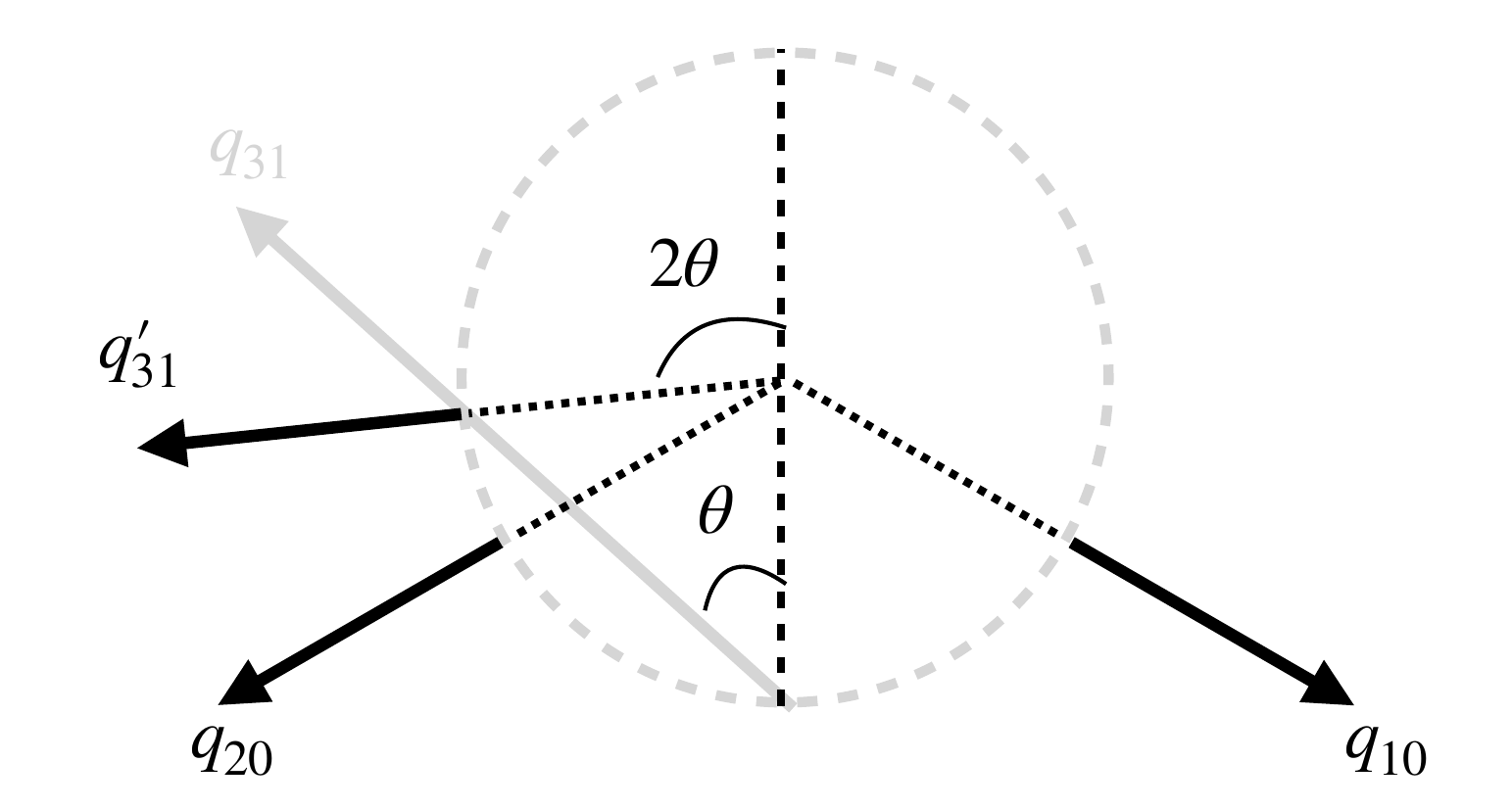}
    \caption{Given the configuration set up in fig.~\ref{fig:gluonSetup}, here shows a sketch demonstrating the calculation of the pull leg 3 exhibits on the junction at times after $|\vec{p}_g|/\kappa$.}
    \label{fig:kinkProp}
\end{figure}

After time $|\vec{p}_{g_{30}}|/\kappa$, the gluon kink ``hits'' the junction and propagates through to the other side with momentum $-\vec{p}_{g_{30}}/2$, whilst the pull from parton $q_{31}$ remains the same. 
Thus in order to determine the combined effect on the junction due to pulls from $q_{31}$ pull and the gluon kink, 
one simply translates the momentum vector of $q_{31}$ along the direction of the propagating kink. 
This translation of the $q_{31}$ momentum is shown by the grey arrow in fig.~\ref{fig:kinkProp}. 

As all partons in this example are assumed to be massless, the endpoints and the kink move at the speed of light, and thus will expand out along the unit circle. 
This means that the intersection of this kink-transported $q_{31}$ vector with the unit circle now can be used to define the overall direction of pull on the junction from leg 3. 
This effective pull is labelled by $q'_{31}$ in fig.~\ref{fig:kinkProp}, which extends outwards from the centre of the unit circle to the circle intersection with the kink-transported $q_{31}$. 

Given an angle of $\theta$ between momenta of $g_{30}$ and $q_{31}$ in the initial JRF, the angle between the initial $g_{30}$ direction and $q'_{31}$ will be $2\theta$. Hence at times after $|\vec{p}_{g_{30}}|/\kappa$, the JRF is determined by momenta $q_{10}$, $q_{20}$, and $q'_{31}$ as shown in the right image of fig.~\ref{fig:gJRF}. Note here we have also assumed the endpoints are sufficiently hard such that there are no endpoint oscillation effects. 

Though the above provides a prescription for the treatment of a single soft gluon on one junction leg, the calculation of these new pull vectors and their associated times becomes increasingly complex when including variables such as multiple gluons, massive endpoints, and/or soft endpoints. Hence we have not developed this model to the point of being able to implement it into the JRF-finding system as of \Py 8.311. One would also expect that the other half of the gluon momenta that propagates out towards the endpoint of the junction leg would eventually return towards the junction. However due to hadronization timescales, the junction motion at late times are not expected to contribute significantly if at all, and such detailed modelling may well be overkill. 

\section{Implementation}
\label{sec:implementation}

In the preceding section we explored how to map the junction motion out over time, including oscillation effects, pearl-on-a-string scenarios, and junctions with gluon kinks. Here we explore how to apply these ideas to a practical hadronization model. 

For pearl-on-a-string cases (when a junction is effectively bound to a soft/slow endpoint), the implementation follows the procedure described in sec.~\ref{sec:pearl}. The approximation of using a gluon to represent the soft/slow endpoint turns out to only be practically useful in a somewhat limited set of circumstances; we only use it in cases when gluon-kinks do not cause changes in the junction motion over a time scale characteristic of hadronization (see below).
This means that only the first parton on each junction leg dictates the junction motion, whether it be because the non-pearl junction legs are sufficiently hard or are endpoints. 
We have not generalised the gluon approximation to junction topologies with soft gluon kinks that cause changes in the junction motion as in such cases there is no longer a straightforward method to approximate the massive pearl as a massless gluon.
The gluon approximation is also only considered if the pearl has at least its constituent mass in order to protect against incorrectly assigned quark masses. 

The scale that determines the softest kink for which the gluon-approximation may be used is set by the parameter \texttt{StringFragmentation:pNormJunction}, which we label $p_{norm}$ below. This should reflect an average (inverse) time for string breaks to occur (i.e. $\langle\Gamma\rangle = \langle\kappa\tau\rangle \sim 2$ GeV).

When standard junction fragmentation is used instead (the majority of cases), the JRF is needed to create the fictitious legs; this is especially important to predict the junction baryon kinematics. However, to deal with junction systems with gluon-kinks where the junction motion changes over time, instead of mapping out the changing junction motion in detail (and e.g.\ dividing it into different string worldsheet regions), we here simply use a single average JRF. This notion of an average JRF was also used in the previous implementation~\cite{Sjostrand:2002ip}, however here we provide a new method to determine the average junction motion in a way that is more stable and reliable when handling gluon kinks and soft/slow endpoints. In the following subsections we outline the new iterative procedure to calculate the average JRF, and explain the prescription used for fragmenting such  systems. 

\subsection{Time dependence of Junction Motion}
\label{sec:iterativeProcedure}

We first briefly outline the general idea of the procedure implemented in \Py 8.310 and prior (with full details in Ref.~\cite{Sjostrand:2002ip}) --- the ``old'' procedure --- before turning to the new one.

In the old JRF finding procedure, weighted sums of the four-momenta of the partons on each junction leg are made, which are called ``pull vectors’’. The JRF is defined as the frame in which these pull vectors form a Mercedes configuration. The weightings used in the construction of these pull vectors depend on parton energies and are thus frame-dependent. The determination of the JRF is therefore an iterative procedure. For a given set of pull vectors, one boosts to the Mercedes frame defined by those pull vectors; one then updates the pull vectors. These updated vectors are not necessarily in a Mercedes configuration; one then moves to the Mercedes frame defined by the updated vectors, and iterate until, ideally, the procedure converges. Convergence is not guaranteed however, especially if/when large-mass pull vectors result due to the summation of four-momenta. Indeed, the iterative procedure fails in around 10\% of minimum-bias events at LHC energies, in which case the procedure reverts to the centre-of-mass frame of the junction system as a fall-back frame in which to perform the  fragmentation instead. Moreover, the procedure assumes the JRF must be of Mercedes type, and allows root-finding for the Mercedes frame to return an unphysical answer in the would-be pearl-on-a-string cases (when the Mercedes frame does not exist). This prescription was largely formulated with high-energy legs in mind, and does work well in that context however falls short with many soft gluon kinks and soft endpoints. 

The procedure we propose here has been formulated to be valid for arbitrary combinations of gluon kinks and endpoint masses. It does not presuppose the existence of a Mercedes frame nor does it rely on convergence of an iteration. Instead, it finds a time-ordered sequence of well-defined JRFs each of which is valid in a given time window, and then makes a time-weighted average over these successive JRFs to find an overall average JRF. To determine the motion of the junction, the procedure steps sequentially through the partons on each junction leg. As per the simple description of gluon kinks in sec.~\ref{sec:gluonKinks}, at early times the partons immediately nearest to the junction will dominate the pull on the junction. After the momentum of the nearest parton on a leg has been depleted, the next parton on that junction leg will take over in dictating the junction motion, and so on. To carry out this procedure, we need to keep track of several pieces of information per iteration; the junction velocity, the time interval the velocity is relevant for, and the four-vectors that will dictate the next iterative JRF.

To simplify the language used below, we freely use momentum magnitude as a measure of time. This is justified as the time it takes each parton to lose its momentum will be proportional to the 3-momentum magnitude, $|\vec{p\,}|$, of the parton according to string equation of motion $dp/dt = -\kappa$, with an additional factor of 2 for gluons. We also simplify notation by distinguishing between endpoint parton momenta (inclusive of (anti)quarks and (anti)diquarks) and gluon momenta with subscripts $q$ and $g$ respectively.

Junction equations of motion and the calculation of  junction velocities were discussed in sec.~\ref{sec:Theory}. For a given set of four-vectors, one finds either the Mercedes frame should it exist, or one uses the pearl-on-a-string notion and find the average velocity using the approximation in eq.~\eqref{eqn:approx}. We expect each JRF to remain the rest frame until either a gluon depletes its momentum, or an endpoint parton oscillates and returns to hit the junction. For a given configuration, the time until the next change in junction velocity is given by whichever has the smallest momentum among the three partons that are currently adjacent to the junction. This ``currently smallest momentum'' is labeled $p_{\rm small}$. 

For a Mercedes-frame topology, $p_{\rm small}$ is defined as the smallest of any $|\vec{p}_g|/2$ or $2|\vec{p}_q|$ in the given three-parton configuration. The factor of $1/2$ on $|\vec{p}_g|$ is to account for a gluon being connected to two string pieces, hence it loses energy twice as fast and half of its momentum propagates outward, away from the junction, while the factor of 2 on $|\vec{p}_q|$ helps to account for endpoint oscillations as will be described below. It also accounts for the fact that after an endpoint has lost all of its energy, it takes the same amount of time again for that information to propagate back to hit the junction. For pearl-on-a-string cases, we define $p_{\rm small}$ as the smallest momentum of the two non-pearl junction legs in the Ariadne frame. As this is with respect to the Ariadne frame and not the JRF, this time is multiplied by a $\gamma-$factor to transform from the Ariadne frame to the perturbed JRF. As these $p_{\rm small}$ times (in both the Mercedes and pearl-on-a-string cases) are measured in each iterative JRF, an additional $\gamma$-factor is used to translate these associated times back to the laboratory frame. 

Once the junction velocity (in the lab frame) and the associated time is calculated for a given set of four-vectors, the momenta are then updated for the next time interval. For pearls, the momentum at time $p_{\rm small}$ is simply determined by the velocity at time $p_{\rm small}$ according to eq.~\eqref{eqn:approx}. 
Otherwise using the momentum-loss relation $\mathrm{d}p/\mathrm{d}t = -\kappa$ again, each parton will lose momentum at a constant rate irrespective of the parton mass. This means after a time $p_{\rm small}$, the partons will now have updated 3-momenta of $\vec{p}\,'_i = \hat{\vec{p}}_i (|\vec{p}_i| - p_{\rm small})$, with their energies scaled accordingly to preserve their mass. If the parton that defines $p_{\rm small}$ is an endpoint, this momentum-updating mechanism naturally incorporates oscillations about the junction as it will update the 3-momentum to be $-\vec{p}_{q}$. If the parton that defines $p_{\rm small}$ is a gluon, one simply updates the pull from that junction leg by stepping to the next parton on said leg.

\begin{figure}[t]
    \centering
    \includegraphics[width = 0.92\textwidth]{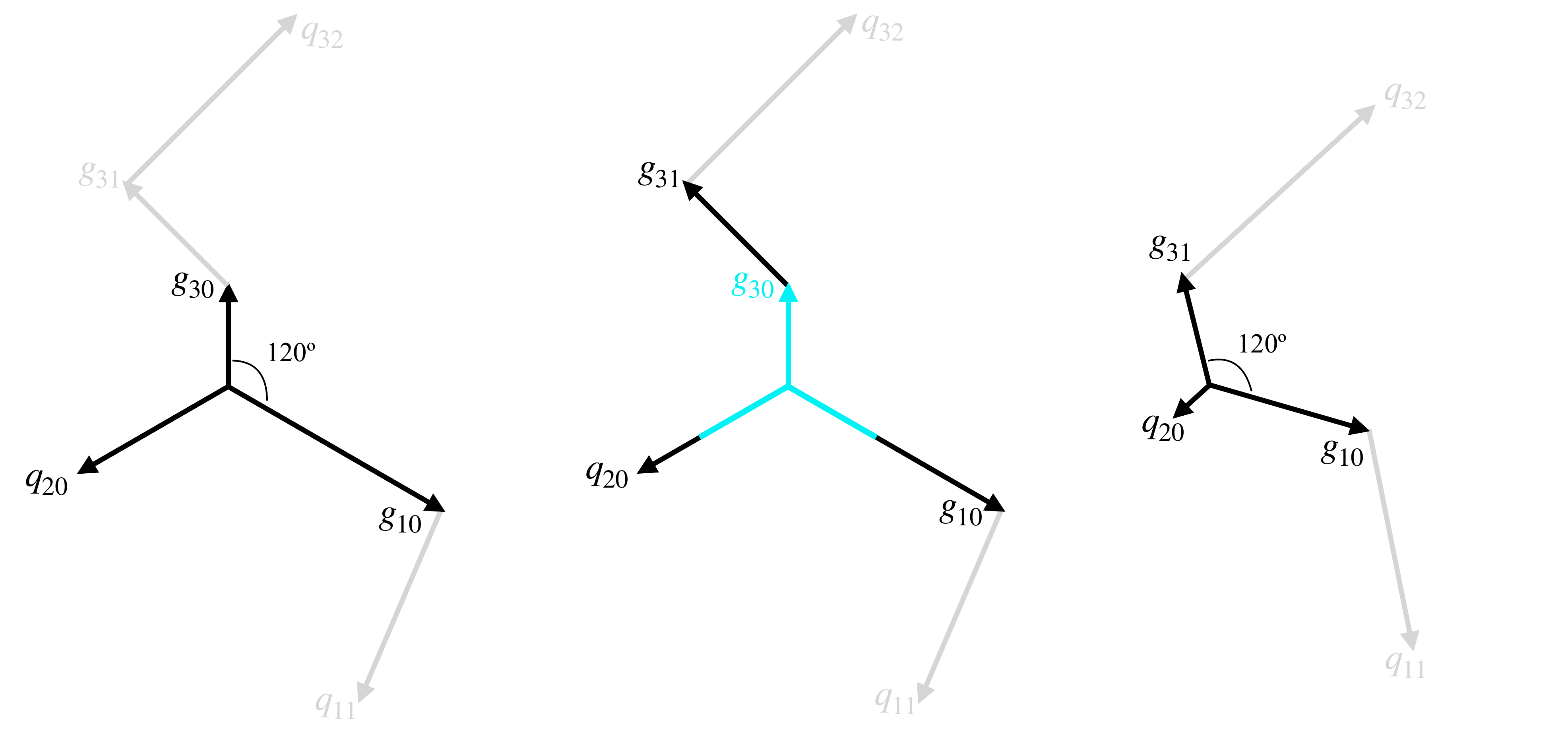}
    \caption{Illustration of the JRF averaging procedure, by stepping through individual JRF frames sequentially. The three-momenta of partons along each leg are lined up, stepping outwards along the junction legs. The black arrows highlight the momenta that dictate junction motion. The left image shows the first step in the sequential procedure: finding the initial JRF, dictated by the first parton on each leg. The middle image illustrates the combination of steps 2 and 3 of the sequential procedure procedure. In this example $g_{30}$ defines $p_{\rm small}$. After that amount has been lost from each leg, highlighted in cyan, the pull for leg 3 is now determined by the next parton on that leg, $g_{31}$, and the momenta remaining on the other legs are updated. The right image depicts the start of the next iteration, showing the next JRF given the updated momenta for each junction leg. Note that angles and vector lengths here are not to scale  and are only intended to give a visual representation of the sequential procedure.}
    \label{JRFiter}
\end{figure}

Putting the above together, we can now construct a full sequential procedure for finding the junction motion over time. The steps of this process are as follows, with the initial 3-parton configuration being the first parton on each junction leg. A schematic example of a corresponding sequence is given in fig.~\ref{JRFiter}.

\begin{enumerate}

\item Check whether we have a pearl-on-a-string configuration or a Mercedes type JRF. Compute and store the velocity $v_i$ with respect to the initial frame of reference, and boost to this frame.

\item Find $p_{{\rm small}_i}$ for the given 3-parton configuration.

\item Update the four-vectors that dictate the junction motion, by stepping to the next parton on the leg associated with $p_{{\rm small}_i}$ if possible, else update momenta according to the time $p_{{\rm small}_i}$.

\item Boost the system back to the initial frame and repeat the process till either the sum of all $p_{\rm small}$ exceeds a threshold value of $p_{\rm max}$ (definition proceeds in the next section), or till two endpoints have been reached. 

\end{enumerate}

\subsection{Handling of Exceptional Topologies}
Though the above procedure is generally stable, there are still a few topologies we have made exceptions for: massive gluons and collinear massless partons.

Though the default gluon mass in \Py is zero, non-zero mass gluons can be encountered at times either from user input or the clustering of two or more nearby gluons into a single small-mass gluon. Should the Mercedes frame not exist for a configuration due to a massive gluon being soft, these are not treated as pearl-on-a-string cases. Instead the contribution from the soft gluon to the JRF is considered negligible and we step to the next parton on that leg.

The other special case considered is collinear massless partons, which is mostly expected to occur with massless gluons. In such cases a boost to a Mercedes frame will never be possible, nor does it make sense to think about a pearl-on-a-string. Perfectly collinear partons are not expected to be encountered often in \Py event generation, however numerical precision of nearly collinear partons can lead to issues in finding a Mercedes frame given the root-finding procedure used. Additionally we wish to ensure stability if given unphysical user-inputted parton configurations. 
For a pair of exactly collinear partons in a three-parton configuration, the junction motion becomes ill defined however one would expect the junction to be highly boosted in the direction of the collinear pair. Hence the following fail-safe is implemented not so much to describe the exact physics of the scenario, but to provide some good approximation to reflect the boost in the collinear pair direction and to give stability to the procedure. 

To handle these configurations, we form a four-vector in the direction of motion of the collinear pair and assign it a diquark mass, and then use this diquark-type velocity as the junction velocity. To construct this, in the centre-of-mass frame of the 3-parton configuration, the summed momenta of the collinear partons defines the direction and energy of this diquark-mass four-vector. The three-momentum magnitude is then fixed by the constituent mass of diquark $ud_0$ which is the lightest diquark according to the particle data table in \Py. From here the velocity of this four-vector defines the JRF, and the three-momentum magnitude defines $p_{\rm small}$. The momenta used in the iterative JRF are then updated to the next parton on all junction legs. 

If any of the partons were already endpoints and hence one cannot step out along the leg further, we simply stop the iterative procedure here. 
Should the collinear pair have insufficient energy to form a diquark mass and both partons are gluons, we simply step to the next parton on each of these legs and continue with the next iteration. 
If there is insufficient energy and one of the collinear partons is an endpoint, no junction velocity is recorded for this iteration and the iterative procedure is stopped here. 
In the rare case we encounter this scenario and there were no previous iterative JRFs found, i.e. the soft collinear partons are the first partons on their respective junction legs and at least one is an endpoint, then we resort to a fail-safe of defining the JRF as the centre-of-mass frame. 
Indeed junctions should not be forming directly between collinear partons, however nonetheless we have ensured to protect against such occurrences and to ensure procedural stability.

\subsection{Average Junction Rest Frame}
\label{sec:avgJRF}

Using the set of sequentially calculated velocities $\vec{v}_i$, according to the above procedure, an average JRF, and hence the associated average junction velocity, $\vec{v}_{\rm jun}$, can be determined. In the following, the first iteration of the sequential procedure is marked by $i=1$. Since fragmentation will gradually happen, we expect that the pull on the junction at early times will be more important than those at late times to determine its motion over the time scales relevant for hadronization. We introduce a time-dependent exponential decay to weight each $\vec{v}_i$. As explained above, the length of time each JRF is relevant for is given by $p_{\rm small}$. Importantly this time measure is defined in the JRF, and thus must be translated to the initial frame of reference by a Lorentz factor, $\gamma_i = 1/\sqrt{1-|\vec{v}_i|^2}$. This allows the time interval for junction motion $\vec{v}_i$ to be defined from times $p_{i-1}$ to $p_i$, where $p_i$ is a sum of times defined by $p_i =\sum_{j=0}^{i}\gamma_j p_{{\rm small}_j}$ with $p_0=0$. Using this, the calculation of the average junction velocity is given by, 

\begin{equation}
	\vec{v}_{\rm jun} = \frac{\sum_{i=1}^{i_{\rm max}} \vec{v}_{i} (e^{-p_{i-1}/p'_{\rm norm}} - e^{-p_{i}/p'_{norm}}) }{1 - e^{ -p_{i_{\rm max}} /p'_{\rm norm} }}.
	\label{exp_weight}~,
\end{equation}
where the parameter $p’_{\rm norm}$ acts as a reference scale characteristic of the time scale of the hadronization process. It is defined via the parameter $p_{\rm norm}$ (set by \texttt{StringFragmentation:\newline pNormJunction}), which is assigned the default value of 2~GeV. If we want $p_{\rm norm}$ to define a (proper) time in the JRF, we need to redefine this normalisation parameter for the initial frame of reference in which we are calculating the average junction velocity, which we here we call $p_{\rm norm}'$. To do so, we consider the sum of times $p_{{\rm small}_i}$ up till they add to $p_{\rm norm}$, then incorporate the changing junction motion by a Lorentz factor. The summation of these $\gamma-$scaled $p_{\rm small}$ values then defines $p'_{\rm norm}$, 

\begin{equation}
	p'_{\rm norm} = \sum_{i=1}^{N} \gamma_i p_{{\rm small}_i} + \gamma_{N+1} \left( p_{\rm norm} - \sum_{i=1}^{N} p_{{\rm small}_i} \right). 
	\label{pNorm}
\end{equation}

The second term in eq.~\eqref{pNorm} ensures the sum of $p_{\rm small}$ add to $p_{\rm norm}$ exactly, where $N$ is defined such that $\sum_{i=1}^{N+1} p_{{\rm small}_{i}} > p_{\rm norm}$ and $\sum_{i=1}^{N} p_{{\rm small}_{i}} \leq p_{\rm norm}$. In the case $p_{{\rm small}_1} > p_{\rm norm}$, we simply have $p'_{\rm norm} = \gamma_1 p_{\rm norm}$. We also use the parameter $p_{\rm norm}$ to control the value of $p_{\rm max}$ which dictates the stopping point in the sequential JRF averaging procedure. We choose $p_{\rm max} = 5p_{\rm norm}$ so that we keep calculating junction velocities beyond the average hadronization time whilst having these JRFs heavily suppressed by the exponential decay in eq.~\eqref{exp_weight}.

Once the average JRF has been constructed, the fictitious endpoints are formed by summing the momenta on each junction leg and reflecting this summed momenta on the other side of the junction, after which standard fragmentation can then be carried out, as described in \cite{Sjostrand:2002ip,Bierlich:2022pfr}.

\section{Results}
\label{sec:results}

We begin this section by examining the theoretical expectations from the updated modelling of junctions. We then inspect the effects in hadron-collision events by comparing to both experimental data and to the previous junction modelling. In the following, the treatment described in this paper (and implemented in \Py 8.311) is labelled ``new’’ while the treatment in \Py 8.310 is labelled ``old’’~ \cite{Sjostrand:2002ip,Christiansen:2015yqa}.

\subsection{Theoretical Implications}
\label{sec:theoryResults}
We have made three key alterations to the junction modelling implementation in \Py 8.311; the $\lambda$-measure used in the QCD CR algorithm, the junction rest-frame finding procedure, and the use of a gluon approximation for fragmentation of (a subset of) junction systems with soft light-flavour legs.

\begin{figure}[t]
    \centering
    \includegraphics[width = 0.475\textwidth]{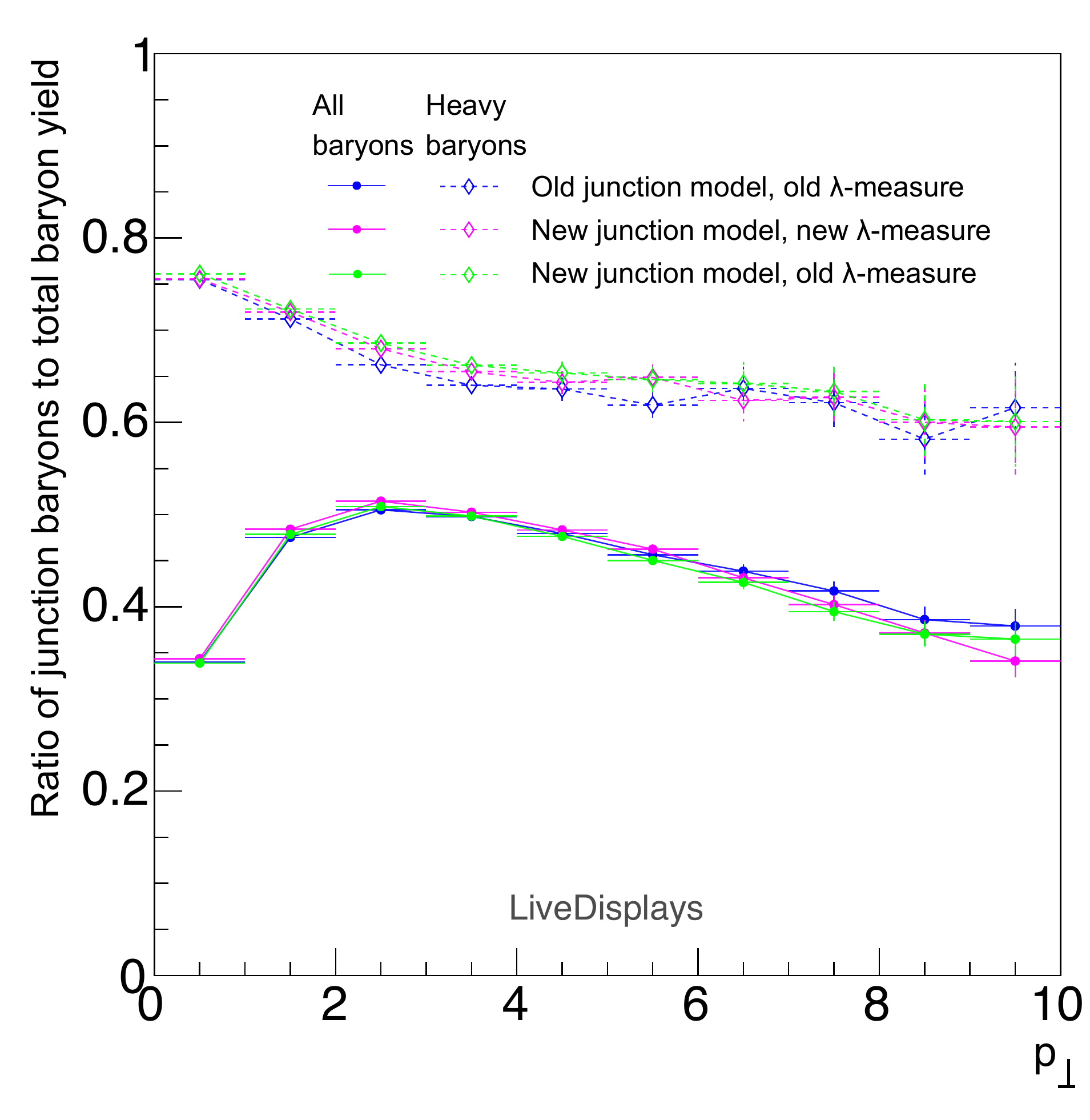} ~~
    \includegraphics[width = 0.475\textwidth]{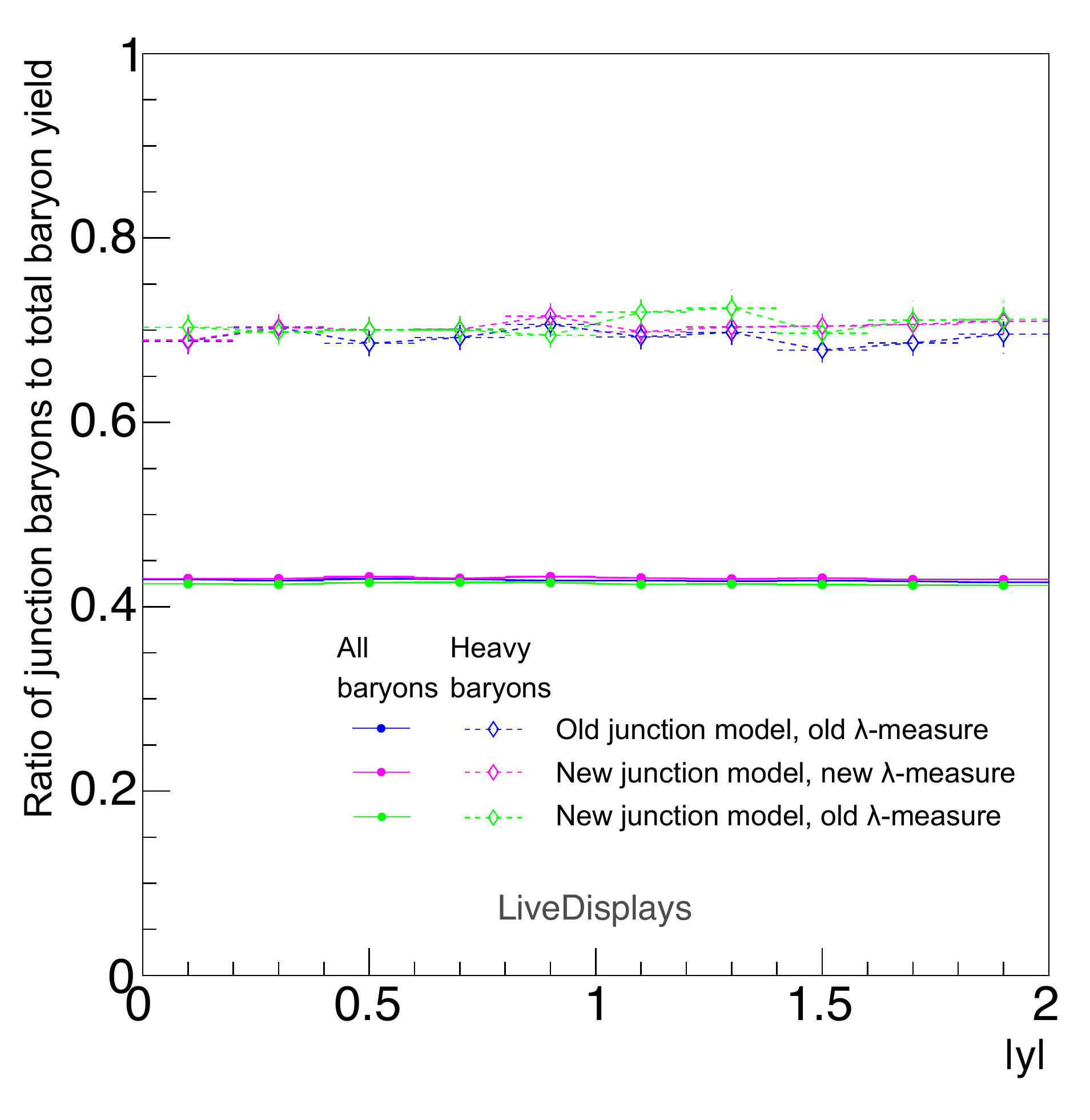}
    \caption{Junction-baryon to total baryon yield ratios, considering both ratios of all flavours of baryons (solid lines) and heavy-flavour baryons only (dashed lines). We compare both the old $\lambda$-measure for both the new and old junction modelling, to the updated $\lambda$-measure in conjunction with the new junction model. Simulations were for NSD events at $\sqrt{s}=7$ TeV, looking at hadrons produced directly from hadronization (i.e., with status codes between 80 and 90).}
    \label{fig:junctionRatios}
\end{figure}

\textbf{String-length measure changes:} We first verify that the change to the new $\lambda$-measure,  eqs.~\eqref{eqn:lambdaNew} and \eqref{eqn:lambdaJunc}, does not cause unexpected large effects in comparison to the previous form eq.~\eqref{eqn:lambdaOld}. In fig.~\ref{fig:junctionRatios} we show the ratio of all junction baryons to the total baryon yield (solid lines), as well as specifically the heavy-junction baryon to total heavy-baryon yield (dashed lines), as  a function of baryon $p_\perp$ (left) and rapidity (right).
This demonstrates that there are no major changes at the level of total yields. It also highlights the importance of rigorous junction-motion modelling as up to 50\% of light-flavour baryons and up to 75\% of heavy-flavour ones are coming from junctions. We return to this below.

We note that we do expect small differences, e.g., due  to the parameter $j_0$ in eq.~\eqref{eqn:lambdaJunc}\footnote{Set by \texttt{ColourReconnection:junctionCorrection}, with default value 1.2.}, which modifies the probability of junction reconnections. Although the same parameter is used in both the new and old $\lambda$-measures, the impact of the parameter is not identical between them.

For completeness, although our study is focused on junction topologies we note that changes to the string-length measure will also affect dipole connections, which may in turn affect, e.g., heavy-flavour meson production and potentially the frequency of usage of the ministring fragmentation \cite{Norrbin:2000zc} procedure. To facilitate investigations of differences, both forms of the $\lambda$-measure are available in \Py 8.311, controlled by the parameter \texttt{ColourReconnection:lambdaForm}. For the remainder of the paper, when using the new junction modelling in \Py 8.311, we use the new string-length measure (\texttt{ColourReconnection:lambdaForm} = 0 as of \Py 8.311).

\textbf{Junction Rest-Frame Determination:} The changes to the JRF-finding procedure comprise the new junction velocity-averaging method as well as taking into account the pearl-on-a-string motion for soft legs. The former is relevant for junction topologies with one or more gluons along the legs, and is difficult to illustrate in a simple/clean way. We therefore defer discussion of these cases to sec.~\ref{sec:experiment} on experimental comparisons. Here, we focus on the extent to which the JRF changes for pearl-on-a-string cases affects total hadron multiplicities as well as the $p_\perp$ spectra of the junction baryon and of the hadrons from the other junction legs.

\begin{figure}[t]
    \centering
    \includegraphics[width = 0.9\textwidth]{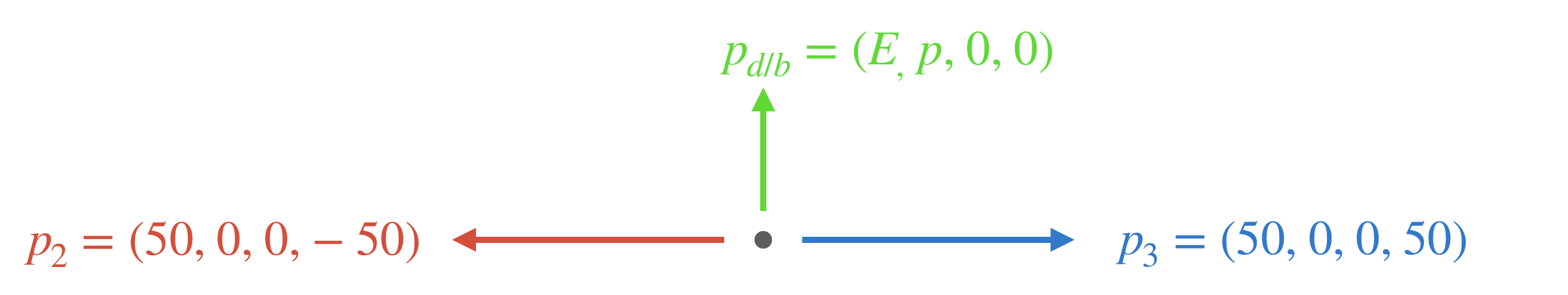}
    \caption{A simple setup for a junction system with two massless junction legs aligned along the z-axis and defined by $p_2$ and $p_3$, and a single masive quark defined by $p_{d/b}$ with momentum along the positive x-axis. In this setup, the massive quark is either a down or bottom quark with constituent masses 0.33 GeV and 4.8 GeV respectively, and down quarks are used for the massless quarks defined by $p_2$ and $p_3$. This configuration is constructed in the Ariadne frame with respect to the massive quark, with $p_\perp$ measured relative to the z-axis.}
    \label{fig:juncSetup}
\end{figure}
Consider the simple generic test case illustrated in fig.~\ref{fig:juncSetup}: a three-parton junction configuration with two energetic massless quarks and a single massive one, whose properties we can vary.  
The momenta of the partons are set up in the Ariadne frame of the single massive quark. That is, the two massless quarks are aligned to be back-to-back along the $z$-axis, and the momentum of the massive quark is aligned with the positive $x$-axis. The energies of the massless quarks are arbitrarily both set to 50 GeV, mainly just to ensure that the massive quark is always the softest leg and that there are no soft effects associated with the other two legs. 

This particular configuration is chosen for convenience; given that the two energetic legs are massless and back-to-back, a boost by $v=1/2$ will reduce their opening angle to 120\textdegree\space, as required in the Mercedes frame. Thus, pearl-on-a-string cases are easily identifiable and correspond to massive-quark velocities of less than 1/2.

To further simplify these tests and isolate junction-baryon production more cleanly, we also turn off diquark production from standard string breaks (\texttt{StringFlav:probQQtoQ} = 0) and the popcorn mechanism which allows for meson production from diquark endpoints (\texttt{StringFlav:popcornRate} = 0). 

We consider two limiting cases for the massive quark; a light down quark with a constituent mass of 0.33 GeV and a heavy bottom quark with a constituent mass of 4.8 GeV.

First, let us discuss the expectations from the old   JRF-finding procedure. When the Mercedes frame does not exist (i.e., for massive-quark velocities less than 1/2 in the Ariadne frame, a.k.a.\ pearl-on-a-string cases), the old algorithm accidentally converges on the rest frame of the pearl quark. Hence, for such cases the junction velocity (in the Ariadne frame) will simply be the quark velocity. For Ariadne-frame quark velocities greater than 1/2, the Mercedes frame exists and the old algorithm takes this as the JRF; oscillations of soft legs around the junction are not mapped out and therefore the junction velocity (in the Ariadne frame) will simply be 1/2. 

The new modelling seeks to take into consideration the deceleration of the pearl quark due to the connection to two other string pieces, determining an average JRF which, for pearl-on-a-string scenarios, should have a lower velocity (in the Ariadne frame) than that of the massive quark. For cases where the Mercedes frame does exist, the junction velocity may also still be lower in the new treatment, if the endpoint energy is low enough that oscillations about the junction become relevant (taken into account in the new treatment, ignored in the old one). 

The old and new treatments will coincide when the massive-quark leg velocity $v > 1/2$ and it is hard enough  that no oscillations need be considered. 
The scale of ``hardness’’ here when talking about oscillations is relative to $p_{\rm norm}$ (\texttt{StringFragmentation:pNormJunction}), where the junction motion is most heavily weighted for times within $p_{\rm norm}$. For the two limiting cases we consider here, a $b$ quark with mass 4.8 GeV and a light quark with mass 0.33 GeV, the no-oscillation limit (for the default value of $p_{\rm norm} = 2\,\mathrm{GeV}$) corresponds to quark velocities above 0.64 and 0.98, respectively. 

\begin{figure}[t]
    \centering
    \includegraphics[width = 0.45\textwidth]{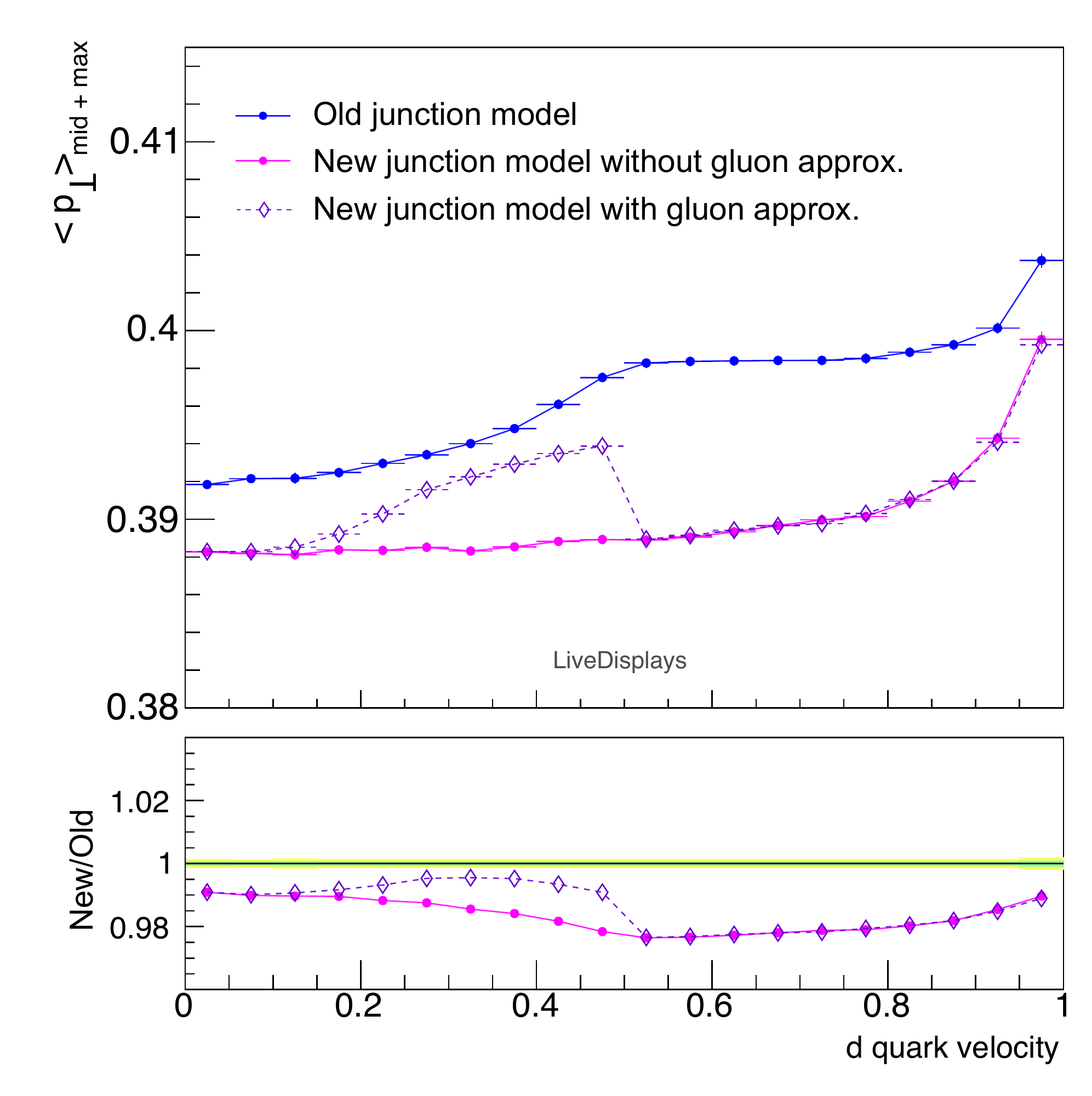}~
    \includegraphics[width = 0.45\textwidth]{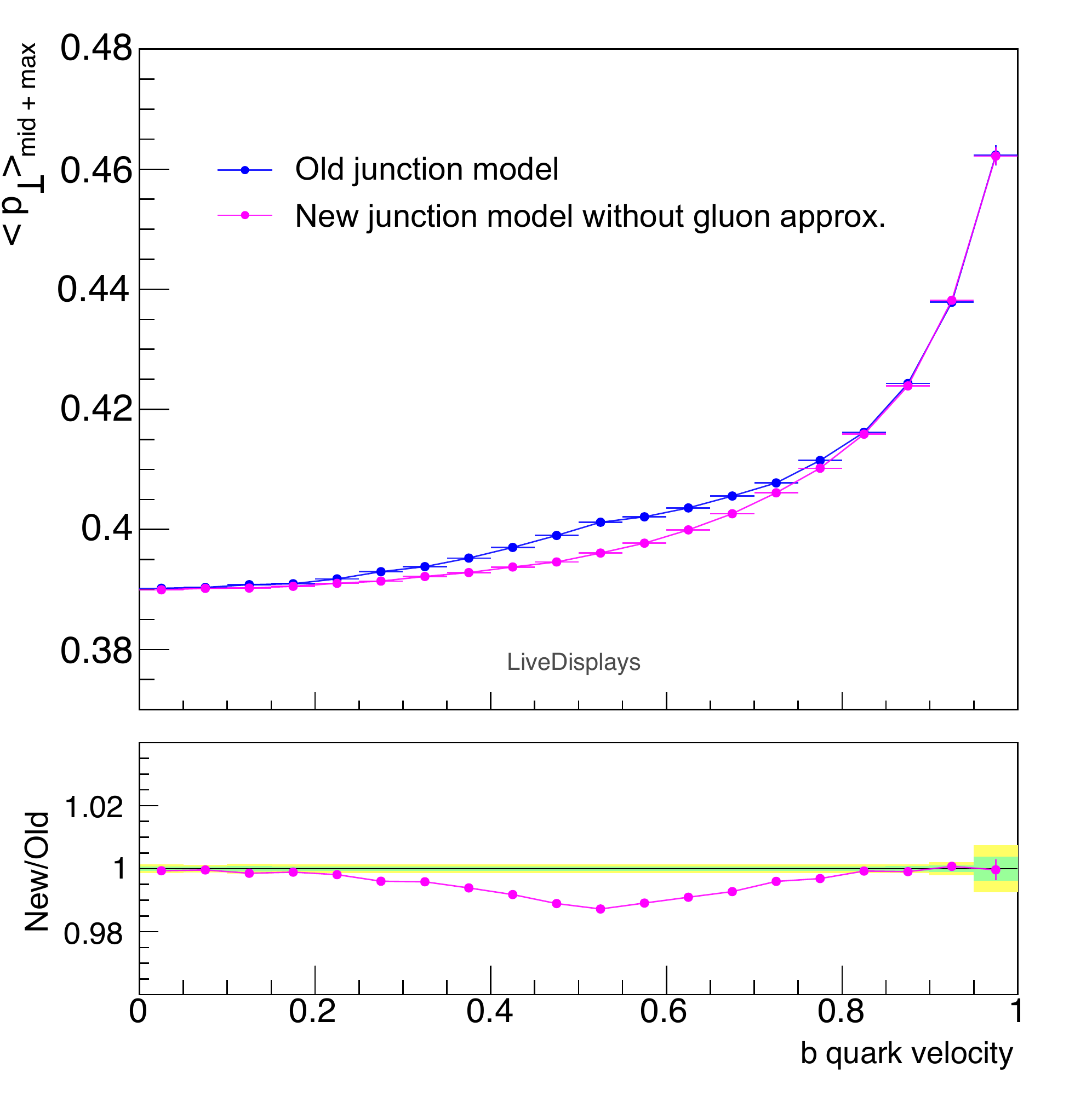}
    \includegraphics[width = 0.45\textwidth]{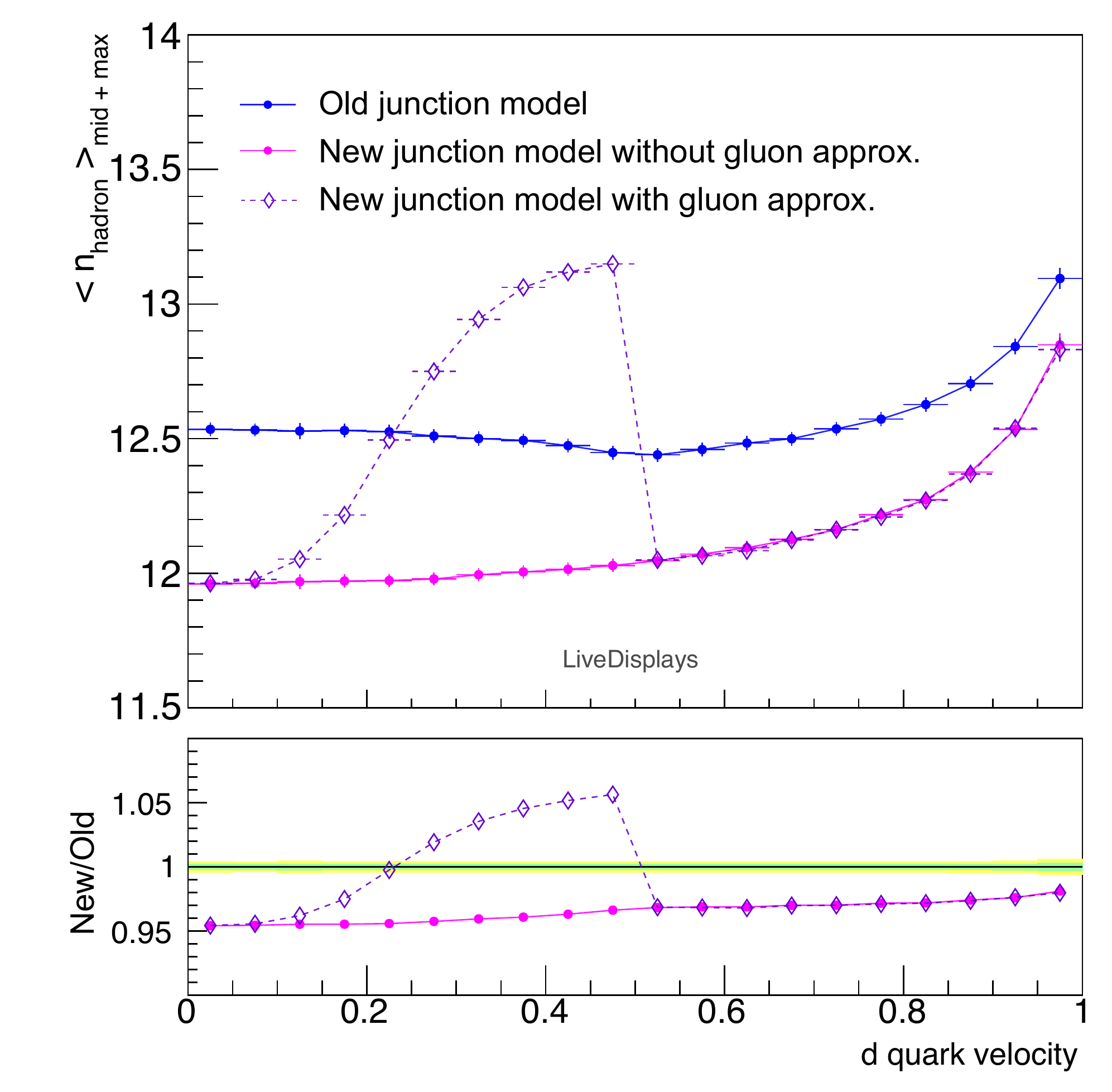}~
    \includegraphics[width = 0.45\textwidth]{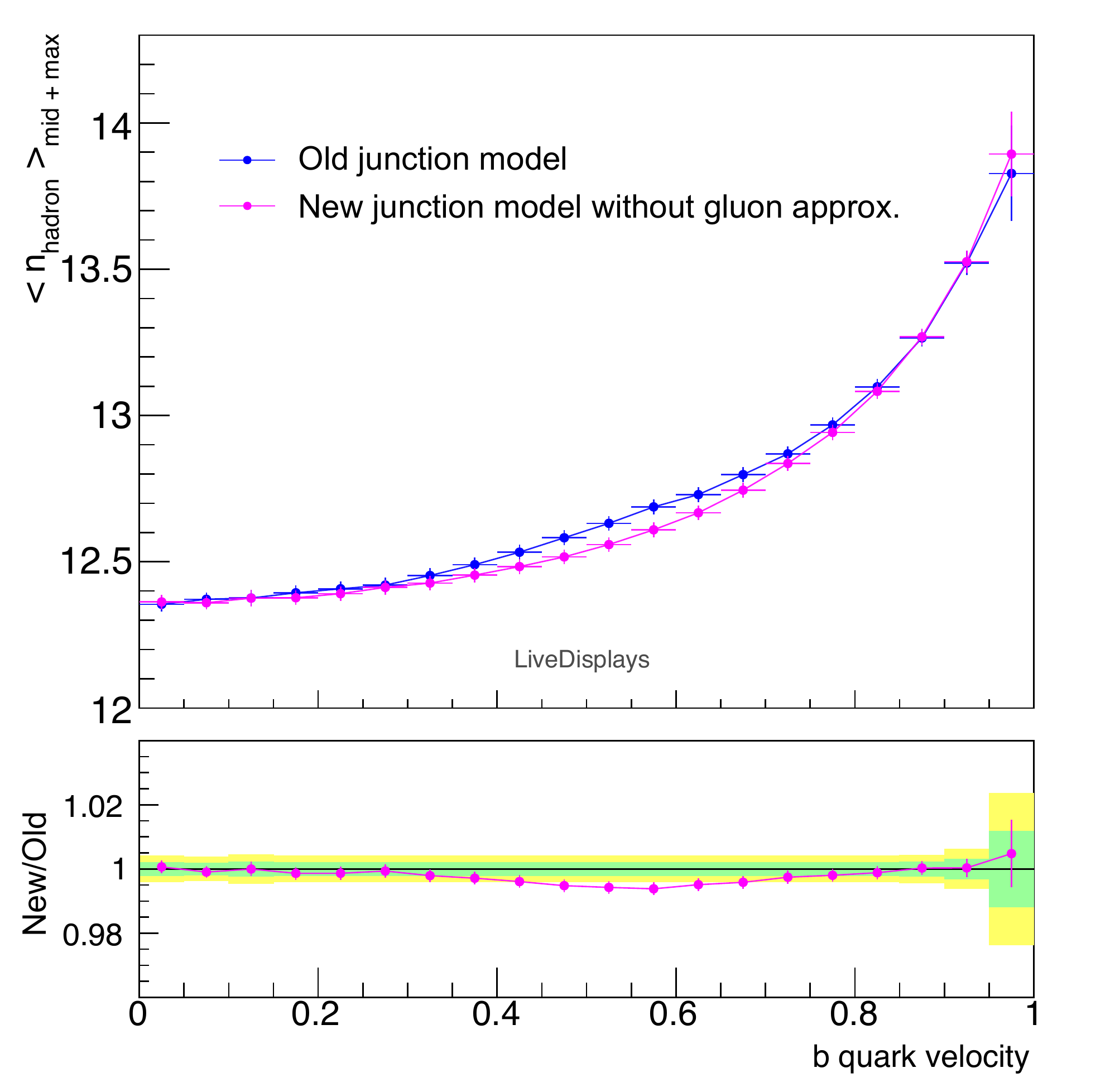}
    \caption{The average transverse momentum (top row) and multiplicity (bottom row) of hadrons produced from the fragmentation of the two junction legs defined by the massless quarks given the Ariadne frame setup described in fig.~\ref{fig:juncSetup}. These distributions are given as a function of the massive quark velocity (in the Ariadne frame), where the massive quark is a down quark in the left column of plots, and a bottom quark in the right column.}
    \label{fig:testJuncStrings}
\end{figure}

\begin{figure}[t]
    \centering
    \includegraphics[width = 0.45\textwidth]{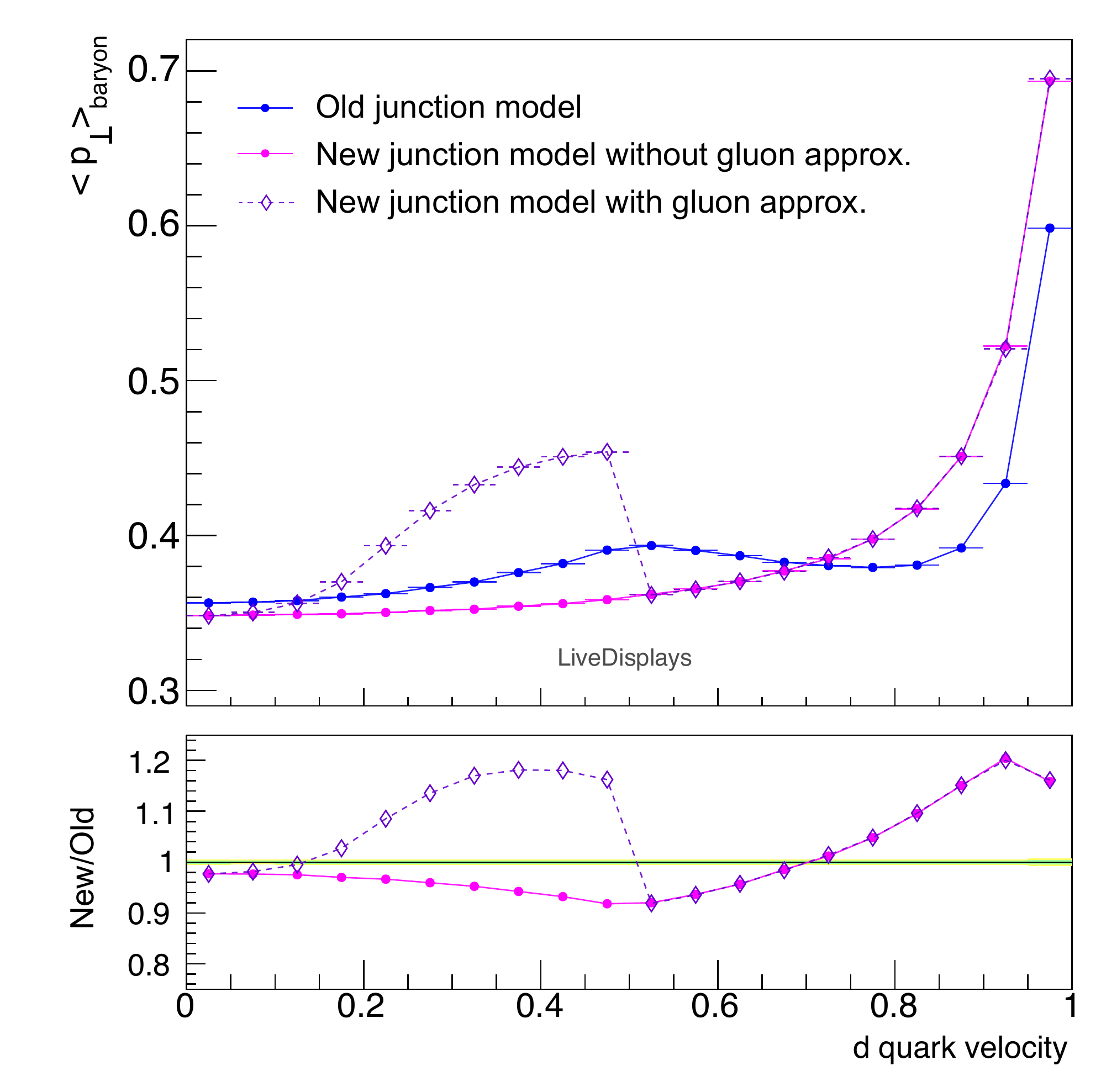}~
    \includegraphics[width = 0.45\textwidth]{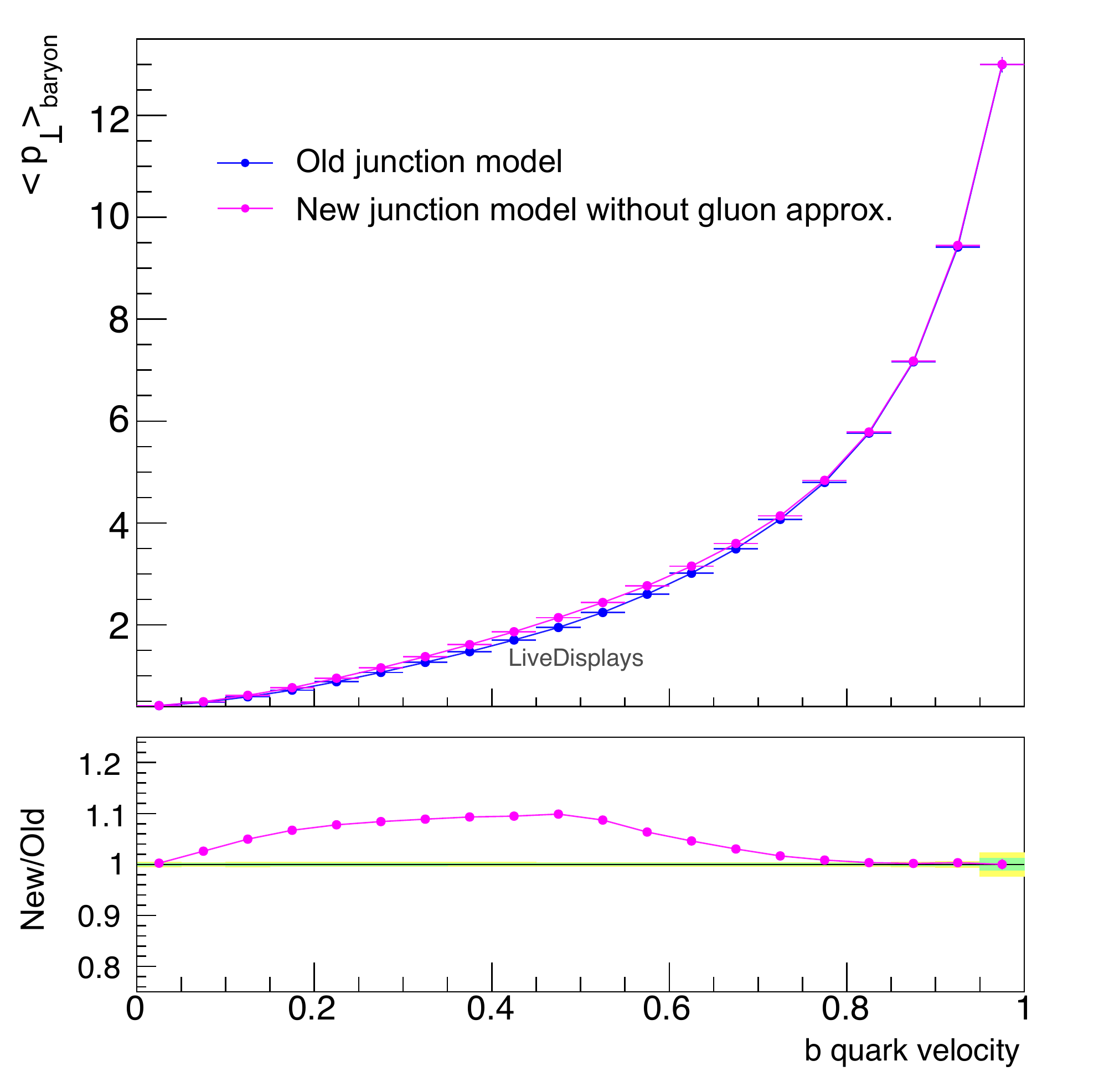}
    \includegraphics[width = 0.45\textwidth]{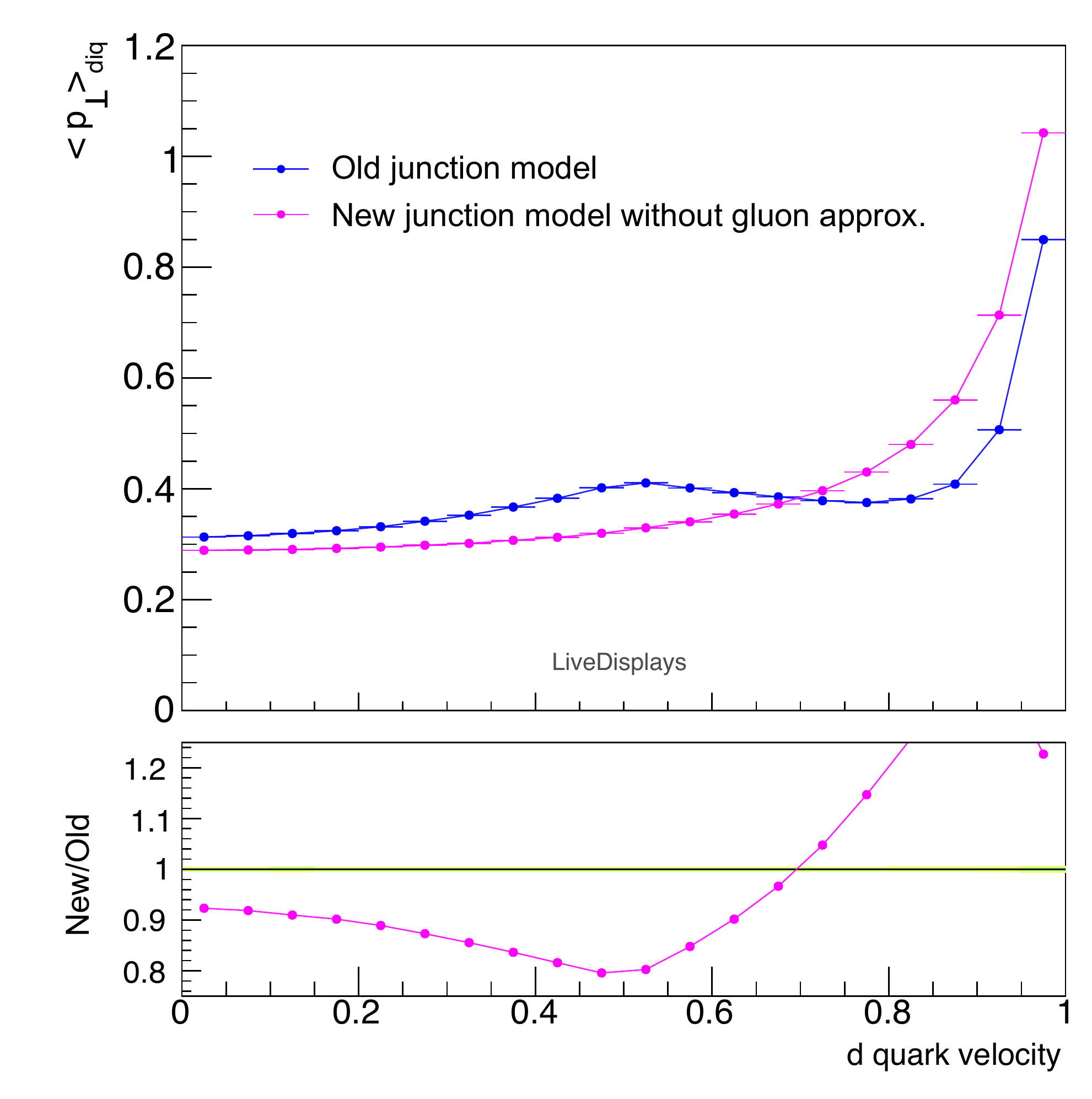}~
    \includegraphics[width = 0.45\textwidth]{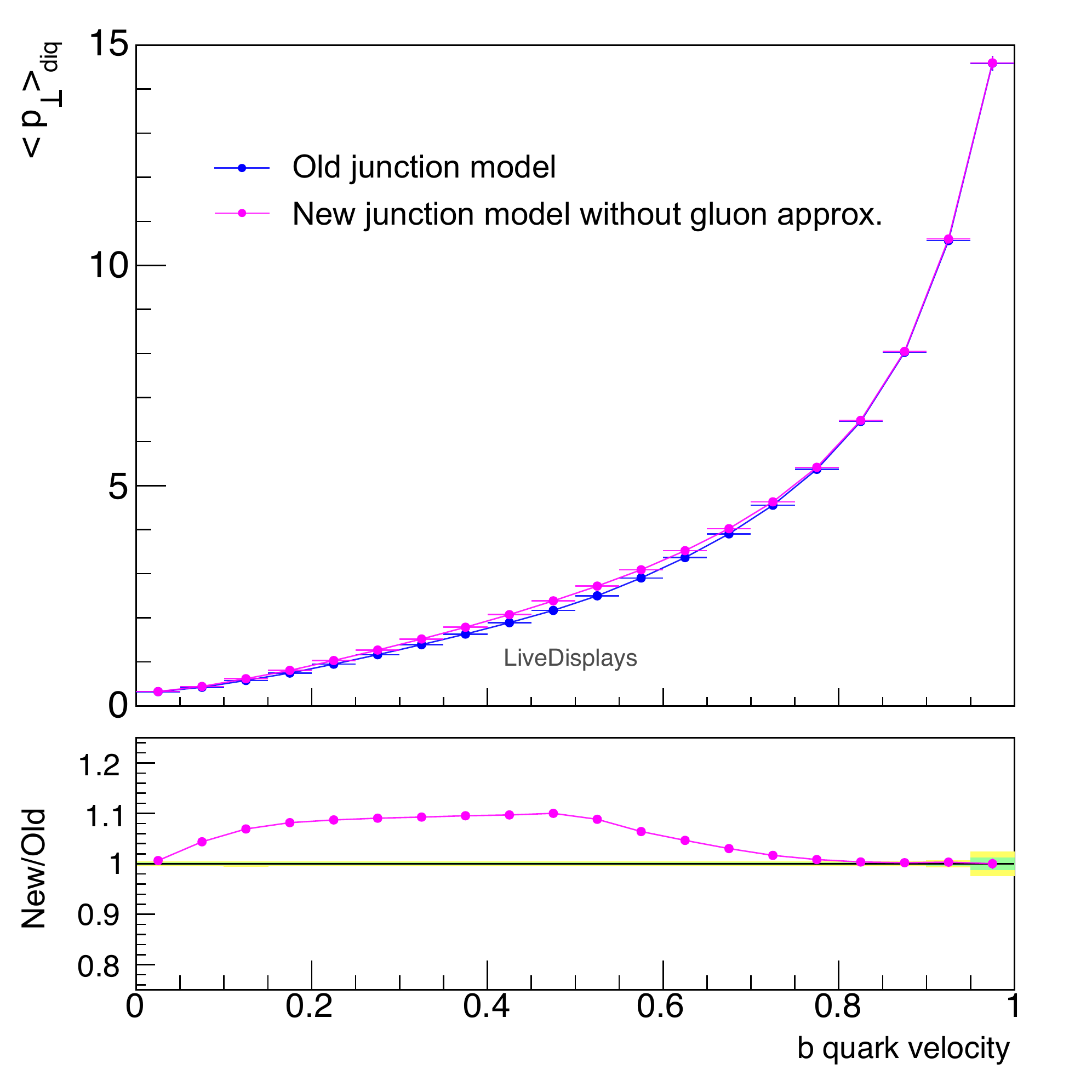}
    \caption{The average $p_\perp$ distributions for the junction baryon (top row) and junction diquark (bottom row), given the setup in fig.~\ref{fig:juncSetup}, such that the massive quark is contained within the junction baryon. The left and right columns are for massive quark flavour of down and bottom respectively.}
    \label{fig:testBaryon}
\end{figure}

\begin{figure}[t]
    \centering
    \includegraphics[width = 0.45\textwidth]{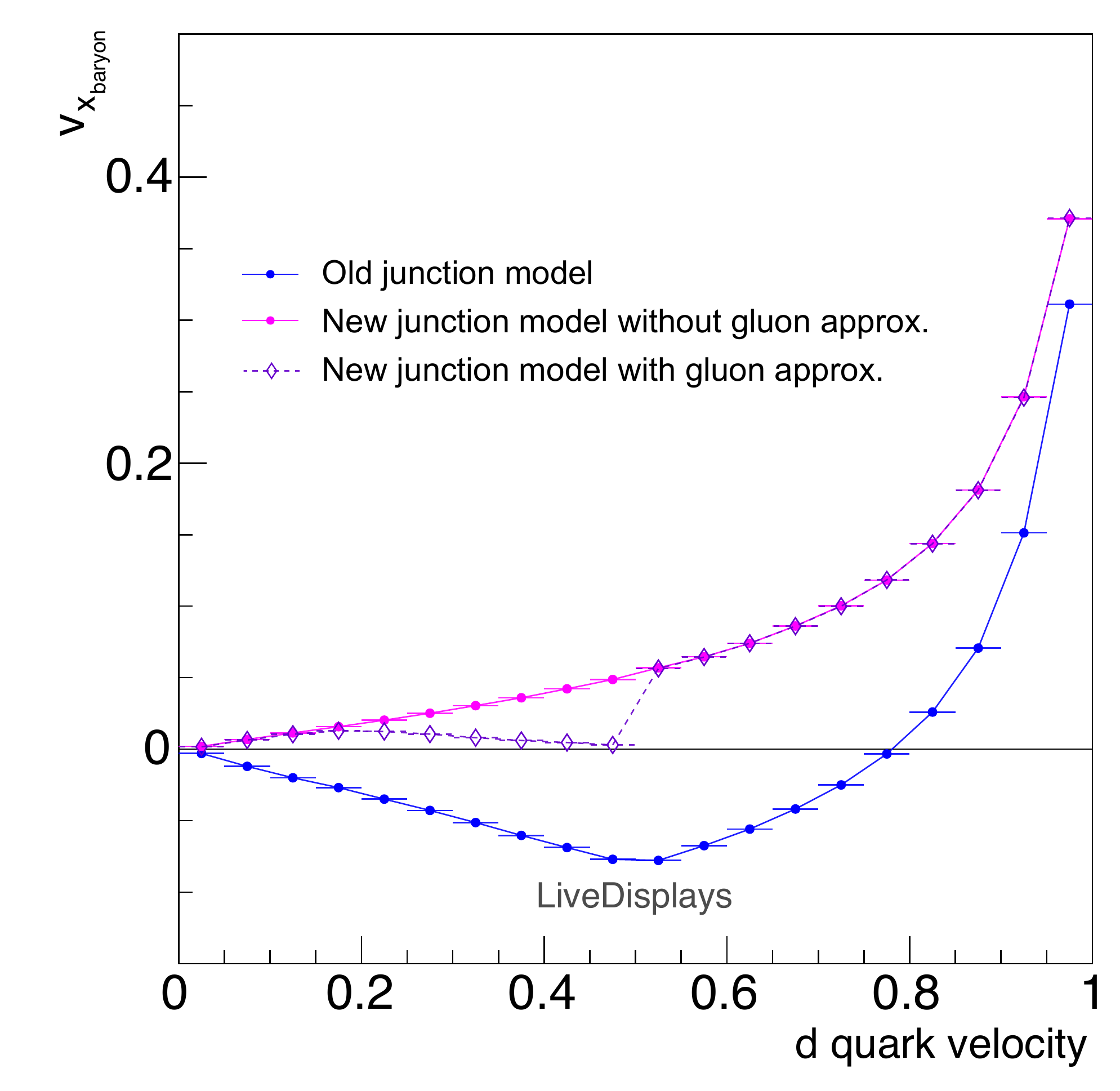}~
    \includegraphics[width = 0.45\textwidth]{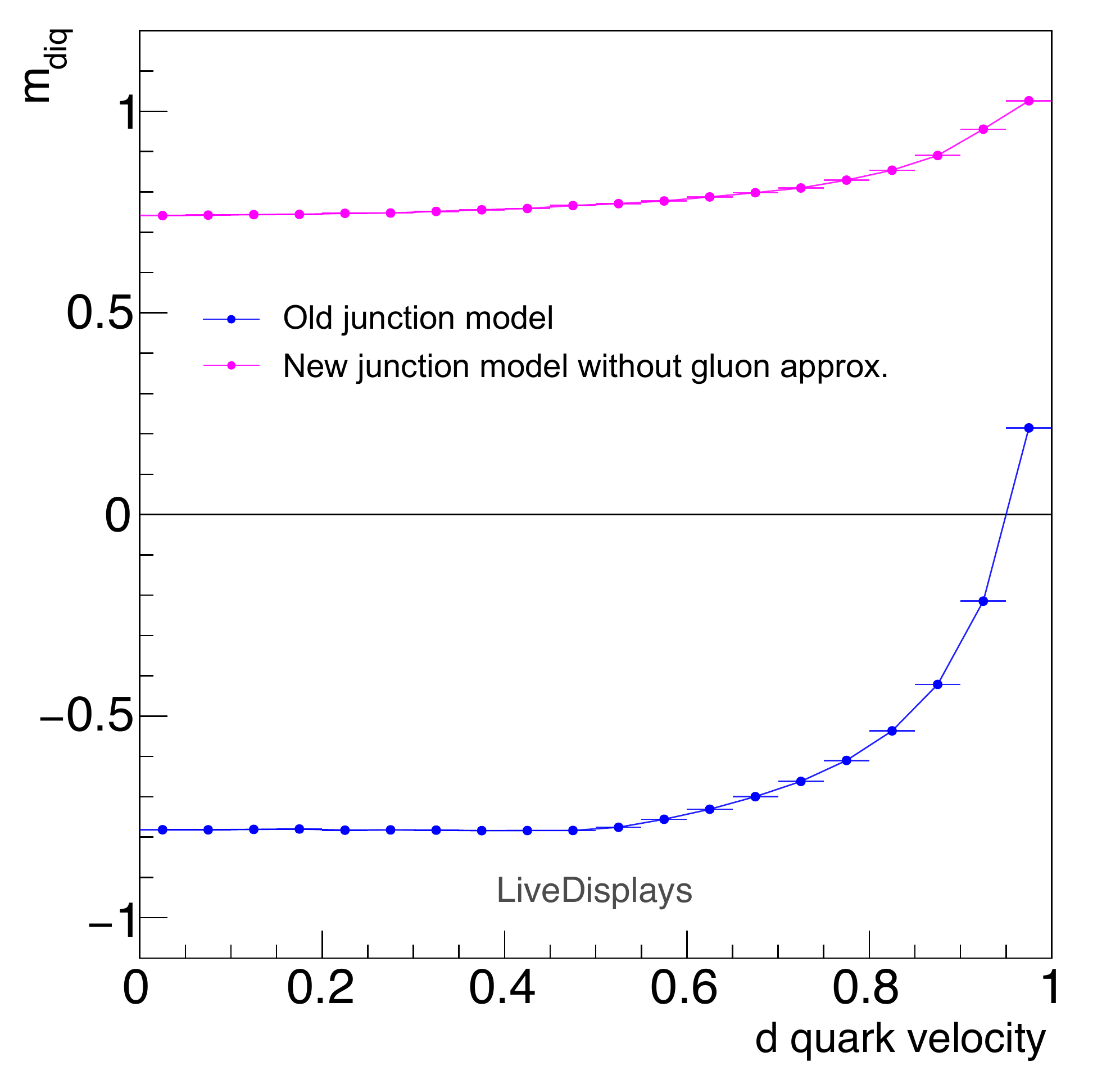}
    \caption{The distributions here are for the setup given in fig.~\ref{fig:juncSetup} with a down quark of mass 0.33 GeV for the massive quark. As a function of the massive quark velocity the left panel shows the average $v_x$ of the junction baryon produced, and the right panel shows the mass of the junction diquark produced by the standard junction fragmentation procedure. The junction baryons and diquarks considered here are only those which contain the massive down quark in the initial setup. }
    \label{fig:testDiq}
\end{figure}

Hence, for the light-quark cases in figs.~\ref{fig:testJuncStrings}, \ref{fig:testBaryon} and \ref{fig:testDiq}, which are all distributions as a function of the Ariadne-frame massive-quark velocity, the convergence is not quite visible as it happens at the right edge of the plots. 
We also note that all distributions in figs.~\ref{fig:testJuncStrings}, \ref{fig:testBaryon} and \ref{fig:testDiq} are for the hadron spectra given the massive quark leg does not fragment, or in other words the massive-quark endpoint becomes contained within the junction baryon itself.

We will begin by making a comparison of the JRF finding procedures given the standard junction fragmentation method (with the use of fictitious legs), then proceed to compare to the gluon-approximation approach. 
For this test scenario as per fig.~\ref{fig:juncSetup}, first let us examine the effect on the fragmentation of the two more energetic junction legs defined by $p_2$ and $p_3$.

Since the boost between the Ariadne frame and the JRF is lower in the new treatment than in the old (in which the  massive-quark rest frame was used for the JRF), the magnitudes of $p_2$ and $p_3$ will be less in the new treatment. Thus when constructing the fictitious endpoints used to approximate each leg as a dipole, the string lengths would be a bit smaller in the new average JRF compared to if the rest frame of the massive quark were used. Hence one would expect the new JRF-finding procedure to result in an overall smaller number of hadrons produced from the two harder junction legs. This result is evident in the bottom row of plots in fig.~\ref{fig:testJuncStrings}, which shows the average hadron multiplicities from junction legs defined by $p_2$ and $p_3$ as in fig.~\ref{fig:juncSetup}. 

Not only would the lower junction velocity affect the hadron multiplicities, it should also impact the $p_\perp$ distributions of the hadrons. The construction of the dipole using the fictitious endpoint does not just determine the string length but also dictates the string axis which defines both the longitudinal direction of string breaks (according to the symmetric fragmentation function in eq.~\eqref{symFrag}), and the axis relative to which $p_\perp$ kicks from string breaks are determined (according to the Schwinger mechanism in \eqref{eqn:schwinger}). 
The lower average junction velocity in the new treatment will result in the fictitious dipoles of the massless quark legs being less boosted, hence should result in a softer average  $p_\perp$ distribution for the produced hadrons. This effect can be seen in the top row of plots in fig.~\ref{fig:testJuncStrings}. 

Most interesting is of course the effect on the junction baryon itself. Predicting this in the updated treatment is less straightforward however. Instinctively one would expect a lower junction velocity to correspond to a lower junction baryon velocity. However this effect is skewed by the old treatment overpredicting the $p_\perp$ distribution of the more energetic junction legs. When carrying out the standard junction fragmentation procedure, after the two softest legs are fragmented, the momentum of the junction diquark (see fig.~\ref{fig:juncFrag} for reference) is determined using energy-momentum conservation according to eq.~\eqref{eqn:pDiq}. As the $p_\perp$ of the produced hadrons is overpredicted when using the massive-quark rest frame, this is compensated for by a softer junction diquark. Hence when looking at the $p_\perp$ distributions of the junction baryon when the massive quark is a bottom quark (top right panel of fig.~\ref{fig:testBaryon}), 
the new junction modelling actually results in a slightly harder baryon despite having a lower junction velocity. 

Interestingly, for the case of the massive quark being a light-flavoured quark (with a constituent mass), we observe the opposite behaviour. 
However though the $p_\perp$ of the junction baryon is greater as seen in fig.~\ref{fig:testBaryon}, the direction of motion of the baryon is oppositely oriented to the junction velocity itself. As evident in the left-hand panel of fig~\ref{fig:testDiq}, the old modelling results in the junction baryon traveling in the negative $x$-direction, despite the JRF per the old modelling being the pearl-quark rest frame, which in this example is moving in the positive $x$-direction. Naturally, we would not expect the junction baryon to move in the opposite direction to the junction itself; this unwanted behaviour in the modelling is a consequence of the incorrect JRF being used. Using the light-quark rest frame as the JRF overestimates the junction velocity, thus when fragmenting the second-softest junction leg in this frame, the hadrons produced take too much positive $p_x$, which then is compensated for when constructing the junction diquark, and then the junction baryon. 

Another contributor to this discrepancy is the negative-mass junction diquarks in the old treatment, cf.\ the right-hand panel of fig.~\ref{fig:testDiq}. This issue arises from only using the energy as a measure of when to stop fragmenting the two soft legs in towards the junction, and is avoided by the additional constraint added in the new procedure that checks both the remaining energy of each junction leg and the mass of the potential junction diquark. We note that the positivity of the diquark masses in the new treatment also noticeably reduces the number of errors encountered during the fragmentation of the last string piece.

Overall we conclude that, for the standard junction fragmentation procedure, using an average JRF instead of simply the early-time JRF (as per the old modelling) results in lower hadron multiplicities and lower average $p_\perp$ distributions (excluding the junction baryon). The effect on the junction baryon itself is less straightforward but the treatment of the junction diquark in the updated modelling is more stable and results in fewer errors\footnote{Typically of the kind \texttt{``StringFragmentation::fragment: stuck in joining''}, which in \Py 8.311 is now encountered approximately a factor 15 less (for dijet events at LHC energies).}. 

\begin{figure}[t]
    \centering
    \includegraphics[width = 0.45\textwidth]{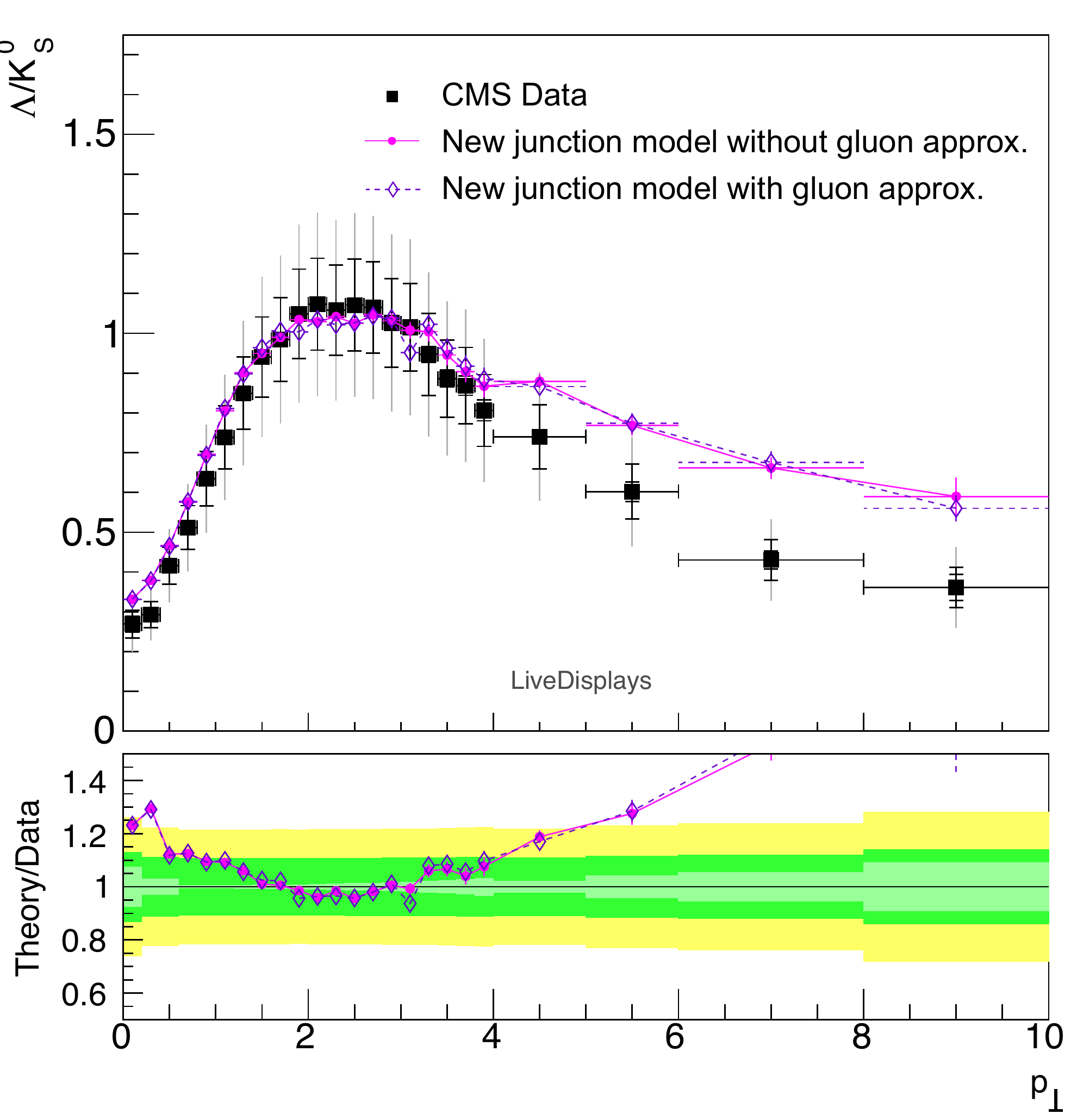}~
    \includegraphics[width = 0.45\textwidth]{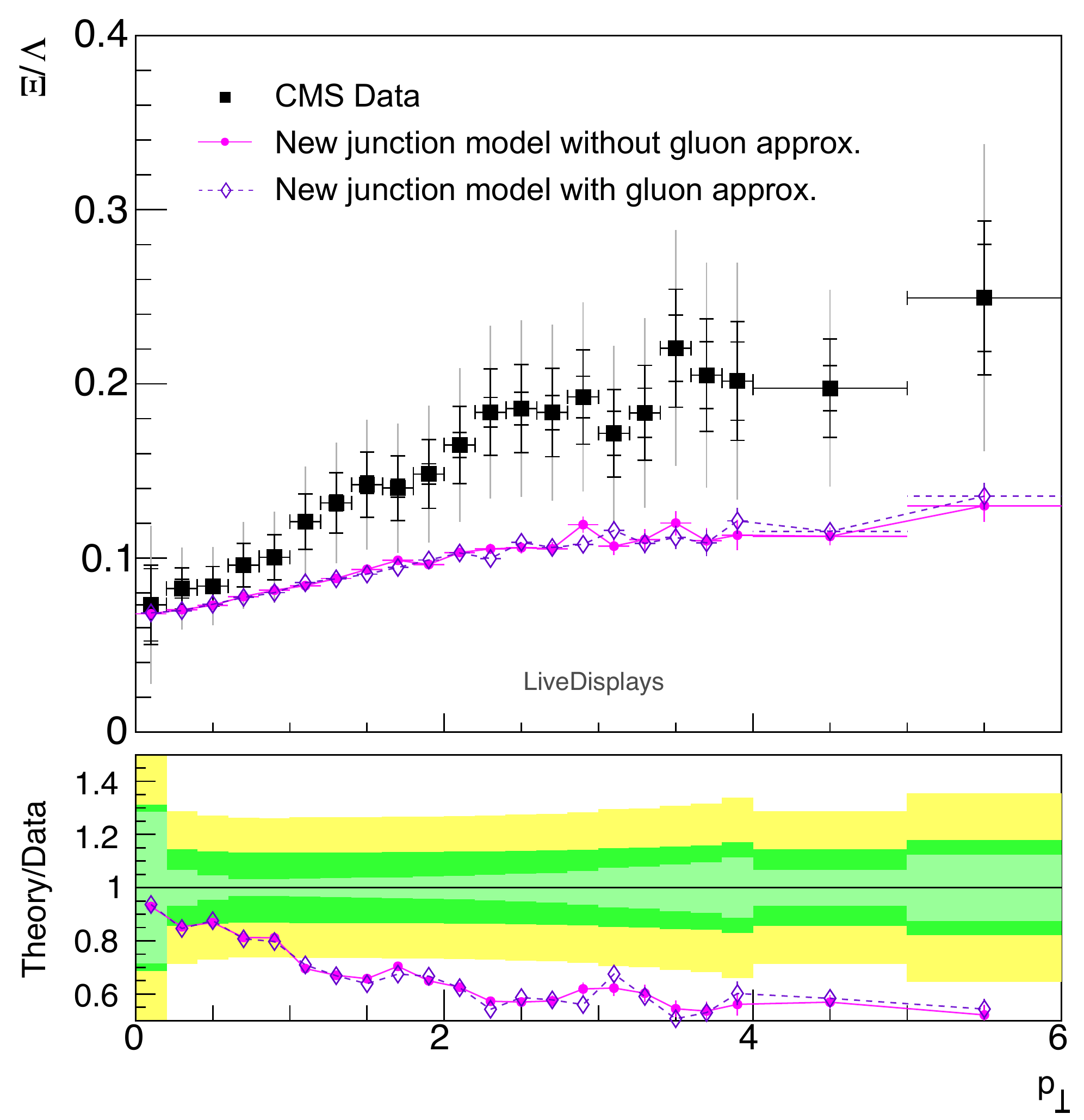}
    \caption{The $\Lambda/K^0_S$ ratio with respect to $p_\perp$ (left panel) and $|y|$ (right panel), with data from CMS \cite{CMS:2011jlm}. The models shown here are using \Py 8.311 with and without the gluon approximation allowed for pearl-on-a-string cases. All events are 7 TeV NSD events, where \Py simulations have a lifetime cut of $\tau_{max} = 10$ mm/c and no $p_\perp$ cuts on final state particles.}
    \label{fig:pearlRatios}
\end{figure}

\textbf{Gluon Approximation for Light Pearls:}
Next we consider the effect of using the gluon-approximation approach for fragmenting light-flavoured pearl-on-a-string cases. 
Note here we neglect the bottom-quark case as the gluon approximation only makes sense in light quark cases, cf.\ the arguments given in sec.~\ref{sec:pearl}. 
The aim of the gluon approximation is to capture the effects of the motion of the light pearl quark, which as seen in fig.~\ref{fig:pearlOnAString} traverses a similar distance over time \texttt{pNormJunction} as a massless gluon kink with the same 3-momentum. This should ideally mimic the $p_\perp$ from the pearl being propagated out along the two non-pearl junction legs,
which in turn should mimic the curve in the strings better than approximating each junction leg as half of a dipole string.

As the $p_\perp$ in this case would be explicitly propagated along the string by the gluon kink, one would expect a harder $p_\perp$ spectrum overall in comparison to the standard fragmentation approach, which is evident in the top left panel of fig.~\ref{fig:testJuncStrings}.
The $p_\perp$ spectrum given the gluon approximation still remains below that of the old modelling. 
Interestingly the hadron multiplicity is higher than the standard fragmentation approach and even becomes higher than the old modelling as seen in the bottom left panel of fig.~\ref{fig:testJuncStrings}. 
Given a higher $p_\perp$ and $n_{\text{had}}$, this means reduced momenta longitudinally. 
This is somewhat surprising and appears to be a consequence of the modelling of gluon kinks. 
Whether this achieves the desired result is somewhat difficult to determine as there is ambiguity in how well one can approximate a massive pearl as a massless gluon. We retain this case mostly out of theoretical interest and to allow to explore modelling ambiguities.

As the gluon approximation aims to distribute the $p_\perp$ of the pearl along the string, one may assume that the junction baryon itself would retain less $p_\perp$. However interestingly the junction baryon $p_\perp$ is predominantly harder in these cases compared to the standard fragmentation procedure. 
From looking at the $p_\perp$ of the junction baryon alone, one may initially assume the baryon becomes harder in the junction direction of motion (the $x$-direction, in our example). However by looking at the velocity in the $x$-direction of motion in the left panel of fig.~\ref{fig:testDiq}, one can see that it is actually reduced when using the gluon approximation. This means that the $\langle p_\perp \rangle$ from string breaks is greater, however less biased in the positive $x$-direction. 

One would reasonably expect the standard procedure to bias the junction direction of motion more in the positive $x$-direction, as the junction baryon necessarily retains the momentum of the massive quark. 
In contrast, in the gluon approximation, the $p_x$ of the pearl is explicitly propagated outwards along the hard-leg string pieces, resulting in a junction baryon with less positive $p_x$. 
It is therefore somewhat surprising that the $p_\perp$ of the junction baryon in the gluon-approximation case is larger, however this again is a result of the treatment of gluon-kinks and the $p_\perp$ from string breaks. 
Comparatively, the old junction model produced junction baryons predominantly moving in the negative $x$-direction, which is opposite to the junction motion. This negative average $v_x$ is another consequence of negative-mass junction diquarks in the old treatment.

Note that in the current implementation, the gluon approximation is only used in cases where the pearl forms at initial time zero, meaning that only cases with quark velocities of less that 1/2 in this Ariadne frame setup are considered for the gluon-approximation approach. 
Though the treatment could in principle be extended to pearls that form at later times which in turn would smooth out these distributions, this is not considered in the current implementation. 
This is partially as there is no clear approach to what would be a reasonable gluon momentum to use in these cases, 
but also these gluon approximation cases are not encountered often in real events which will be further elaborated on below.

The strict cutoff for the gluon approximation at $v=1/2$ results in the curves in fig.~\ref{fig:testJuncStrings} and fig.~\ref{fig:testBaryon} showing a sharp change at down-quark velocity of 1/2. 
One should also keep in mind that the approximation of the massive pearl quark as a massless gluon kink is not perfect. Though approximating the pearl as a gluon results in an overall similar displacement of the pearl as shown in fig.~\ref{fig:pearlOnAString}, the velocities of a pearl quark and a gluon kink are considerably different, with the pearl having a maximum initial velocity of 1/2 whereas a massless gluon will have a velocity of 1. Given this discrepancy between a realistic pearl quark and a gluon kink, it is somewhat difficult to tell whether the gluon approximation or simply using an average JRF is a more faithful approach in such cases.

Another factor to consider is that the gluon approximation is also only well defined in a simple three-parton configuration as described above. In the case of soft gluon kinks, there is no longer a distinct Ariadne frame to perform the gluon construction in, nor would the pearl motion necessarily be well mimicked by a gluon in such cases anyway. Given this constraint, when looking at a 
hadron collision event such as a $\sqrt{s} = 7$ TeV $pp$ collision,
only around 0.01\% of junction baryons are formed using the gluon approximation. We therefore expect this modelling to have minimal practical consequences. 

Additionally, any observable effects of this modelling would be further diluted by the portion of light-flavour baryons produced via standard string breaks via diquark-antidiquark pair creation. Thus though this model provides an in-principle elegant parallel between the pearl-on-a-string junction cases and gluon kinks, in practice it is of quite limited use as these cases rarely occur. This is evident in fig.~\ref{fig:pearlRatios} where there is no significant observable impact on the $p_\perp$ distribution of either the $\Lambda/K$ ratio or the $\Xi/\Lambda$ ratio with the inclusion of the gluon approximation. Hence, for the comparisons in sec.~\ref{sec:experiment}, we do not include the gluon approximation. 

\subsection{Experimental Comparisons}
\label{sec:experiment}

Here we examine the results of the revised junction fragmentation modelling in comparison to experimental data, with a particular focus on baryon-to-meson ratios. Note that in the following we do not attempt a full retuning but leave most parameters untouched relative to the ``CR Mode 2'' tune of the QCD CR model~\cite{Christiansen:2015yqa}. This allows for a direct comparison between the old and new junction modelling. We also will compare results to the Monash (2013) tune~\cite{Skands:2014pea}, which provides a baseline model without junctions. A full detailed tuning effort will be left for a future study. 

\begin{figure}[t]
    \begin{center}
    \includegraphics[width = 0.45\textwidth]{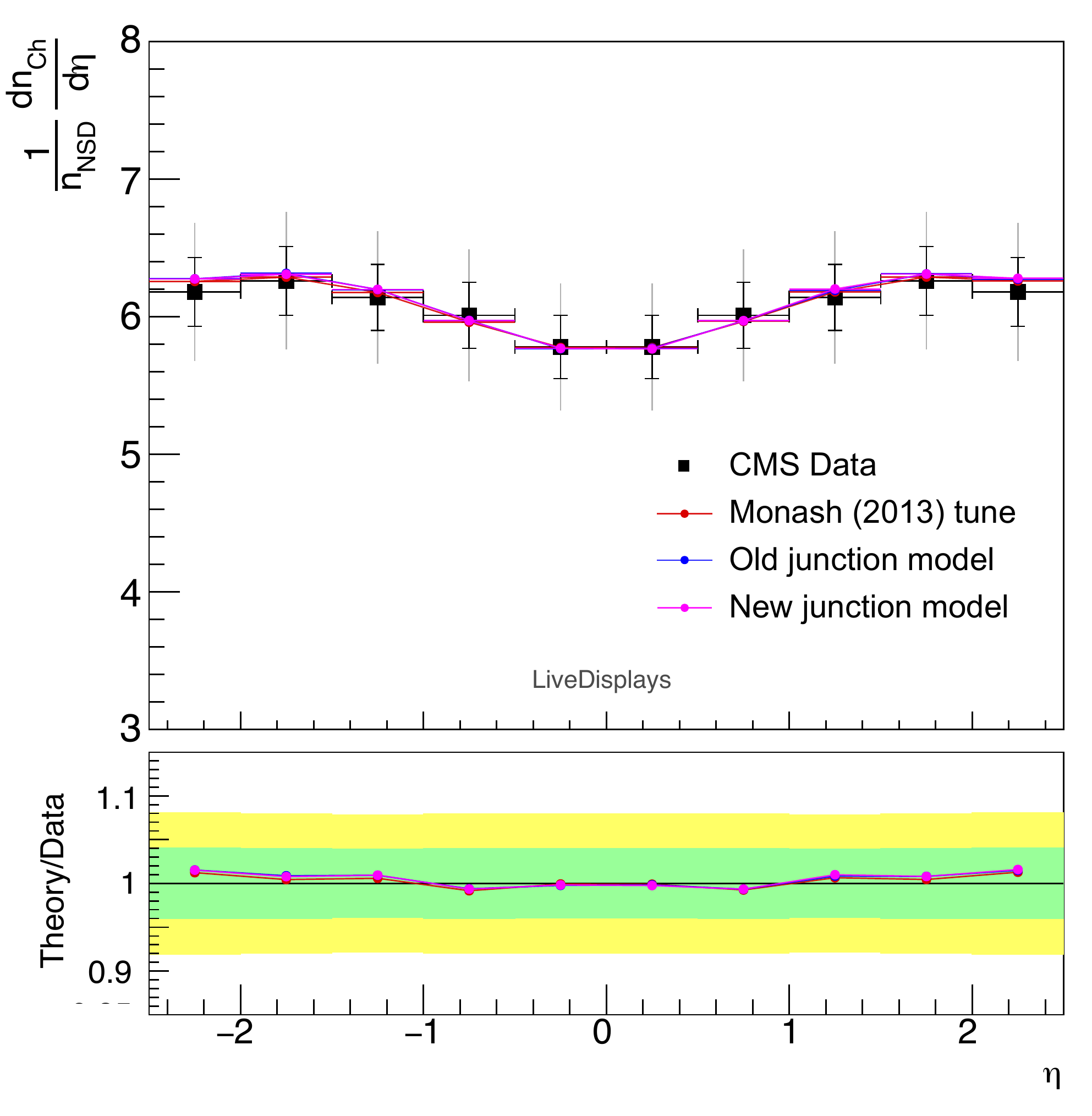}~
    \includegraphics[width = 0.45\textwidth]{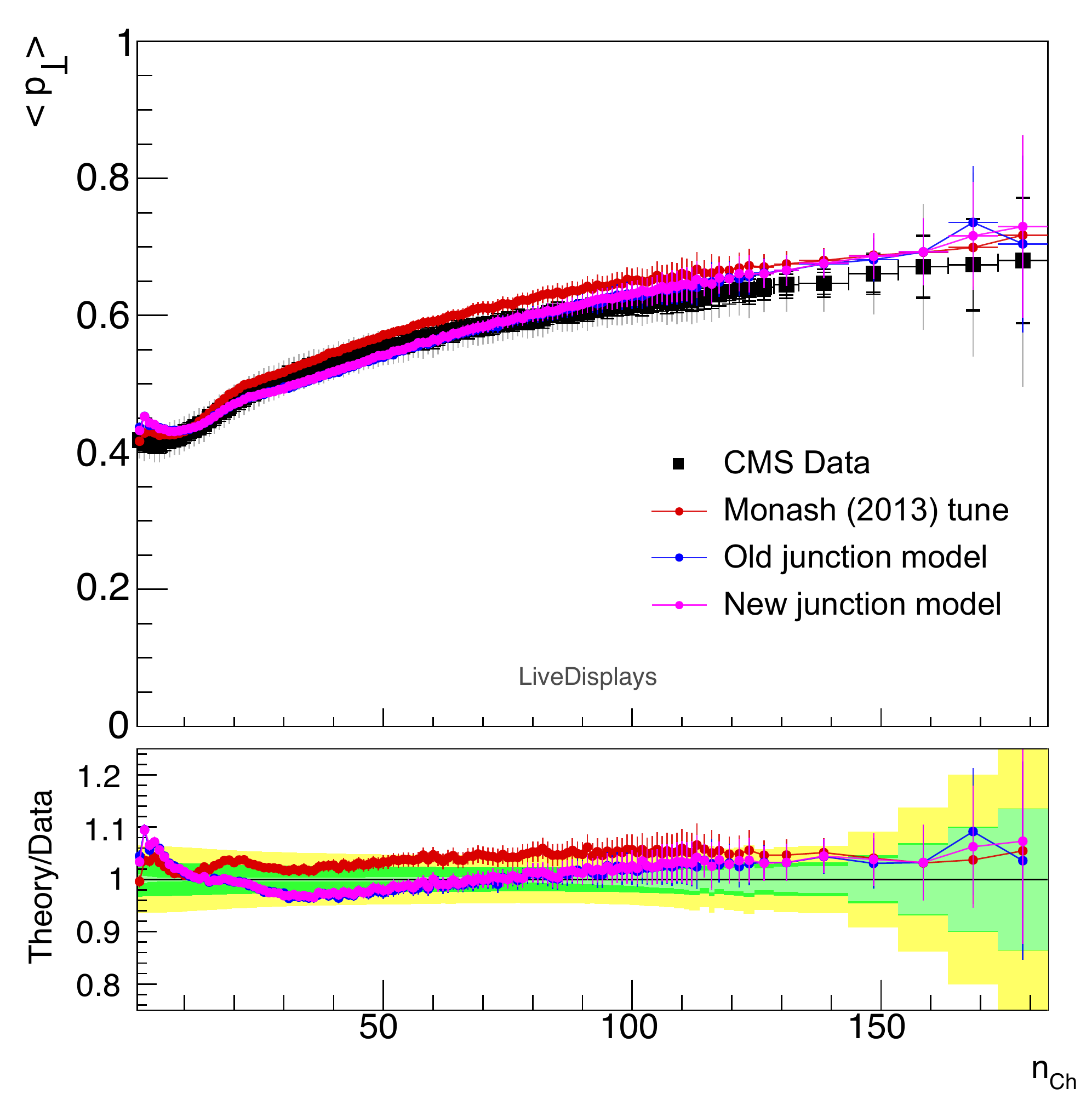}
    \end{center}
    \caption{The charged multiplicity average as a function of pseudorapidity on the left \cite{CMS:2010tjh}, and the right plot shows the average $p_\perp$ as a function of charged multiplicity \cite{CMS:2010qvf}. All events are 7 TeV NSD events from CMS, where \Py simulations have a lifetime cut of $\tau_{max} = 10$ mm/c and no $p_\perp$ cuts on final state particles. }
    \label{fig:chargeDist}
\end{figure}
\begin{figure}[t]
    \begin{center}
    \includegraphics[width = 0.45\textwidth]{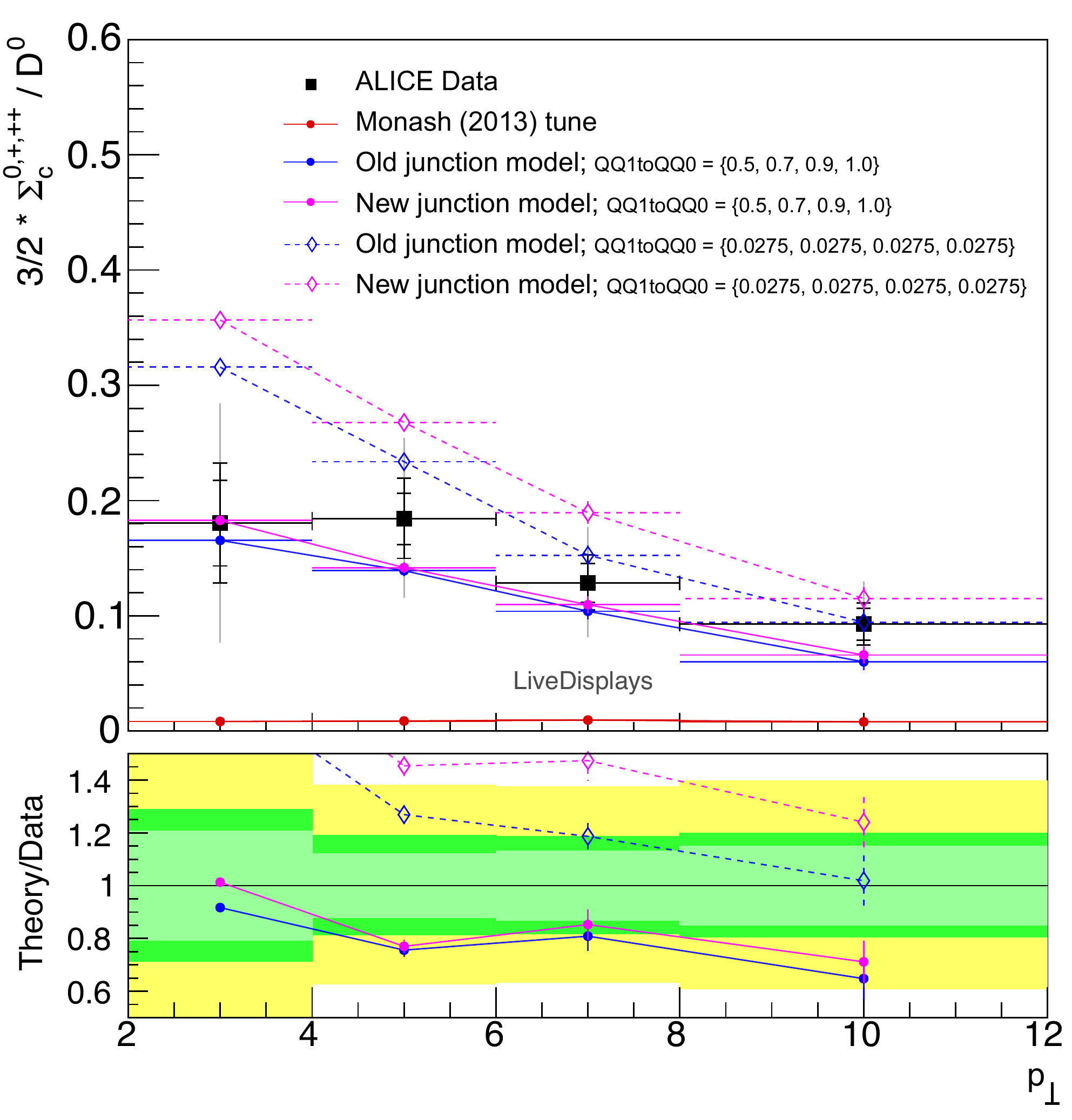}~
    \includegraphics[width = 0.45\textwidth]{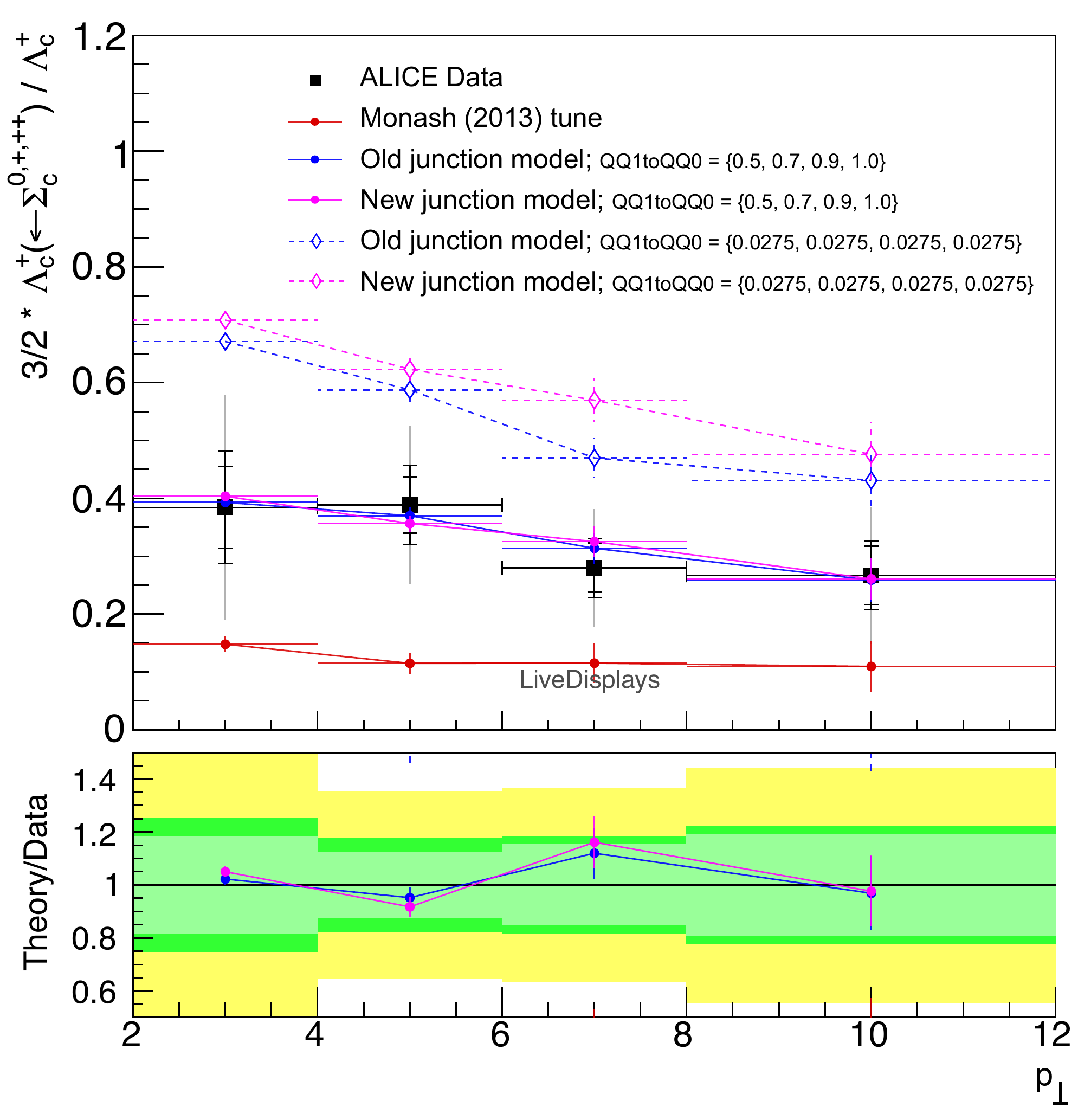}
    \end{center}
    \caption{$\Sigma^{0,+,++}_c/D^0$ (left panel) and $\Lambda^+_c(\leftarrow \Sigma^{0,+,++}_c)/\Lambda_c^+$ (right panel) distributions from ALICE \cite{ALICE:2021rzj} data, showing the Monash (2013) tune, and both the old (blue) and new (magenta) junction modelling. The new values of \texttt{StringFlav:probQQ1toQQ0join} = \{0.5, 0.7, 0.9, 1.0\} are shown with solid lines. The original CR Mode 2 values of \{0.0275, 0.0275, 0.0275, 0.0275\} are shown with dashed lines.}
    \label{fig:QQ1toQQ0}
\end{figure}

Only two of the CR Mode 2 baseline parameters were altered:
\begin{itemize}
    \item 
\texttt{MultiPartonInteractions:pT0Ref} was increased slightly, from 2.15 GeV to 2.25 GeV, to re-establish agreement with the $dn_{ch}/d\eta$ distribution, as shown in fig.~\ref{fig:chargeDist}.
\item \texttt{StringFlav:probQQ1toQQ0join}, which is really a vector of four parameters, control the spin of the junction diquark, via a suppression of spin-1 diquark formation relative to spin-0. This suppression factor is additional to the factor 3 enhancement of spin-1 from counting the number of states. By controlling spin-1 diquark production, one can directly modify probabilities of spin-3/2 baryon production from junctions. The parameter has four components that set the suppression factor given the heaviest quark flavour present is $u/d$, $s$, $c$, or $b$ respectively. In CR Mode 2, these parameters were all (arbitrarily) set to be equal to the corresponding suppression factor in diquark string breaks, \texttt{StringFlav:probQQ1toQQ0join} = \{0.0275, 0.0275, 0.0275, 0.0275\}, whereas the default values in \Py are (likewise arbitrarily) \{0.5, 0.7, 0.9, 1.0\}. The heavy suppression in the CR Mode 2 values was not particularly well motivated and not tuned to any particular set of data. 
Here, in view of new measurement results that have shown comparably large $\Sigma^{0,+,++}_c/\Lambda_c^+$ ratios, we choose to revert back to the default values instead of those provided by CR Mode 2. The $\Sigma^{0,+,++}_c/D^0$ and $\Lambda^+_c(\leftarrow \Sigma^{0,+,++}_c)/\Lambda_c^+$ distributions are shown in fig.~\ref{fig:QQ1toQQ0}. 
Though $\Sigma_c$ nor $\Lambda_c$ are spin-1/2 baryons, increasing \texttt{StringFlav:probQQ1toQQ0join} values enhances spin-3/2 baryon production, resulting in more excited $\Sigma_c$ and $\Lambda_c$ states. These spin-3/2 excited states decay favourably to $\Lambda_c$ rather than $\Sigma_c$ \cite{Workman:2022ynf}, and thus results in the reduction of the $\Sigma^{0,+,++}_c/D^0$ and $\Lambda^+_c(\leftarrow \Sigma^{0,+,++}_c)/\Lambda_c^+$ ratios hence describing the data more accurately. For this reason in the remainder of this paper, we use the default \Py values of \texttt{StringFlav:probQQ1toQQ0join} = \{0.5, 0.7, 0.9, 1.0\}. We note that these choices are obviously still to be regarded as preliminary and  subject to further smaller revisions in a more careful tuning study.
\end{itemize}

\begin{figure}[t]
    \begin{center}
    \includegraphics[width = 0.45\textwidth]{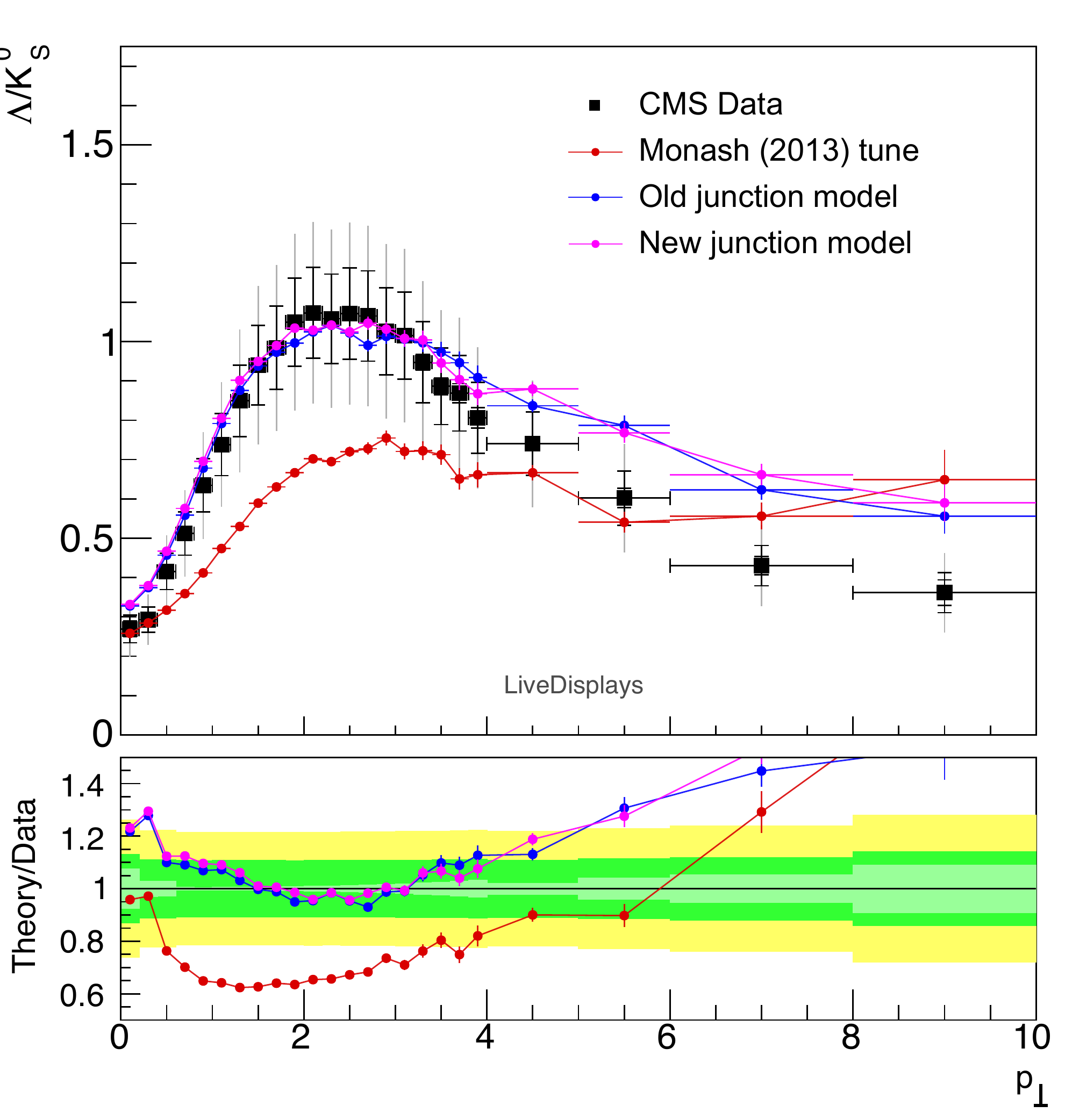}~
    \includegraphics[width = 0.45\textwidth]{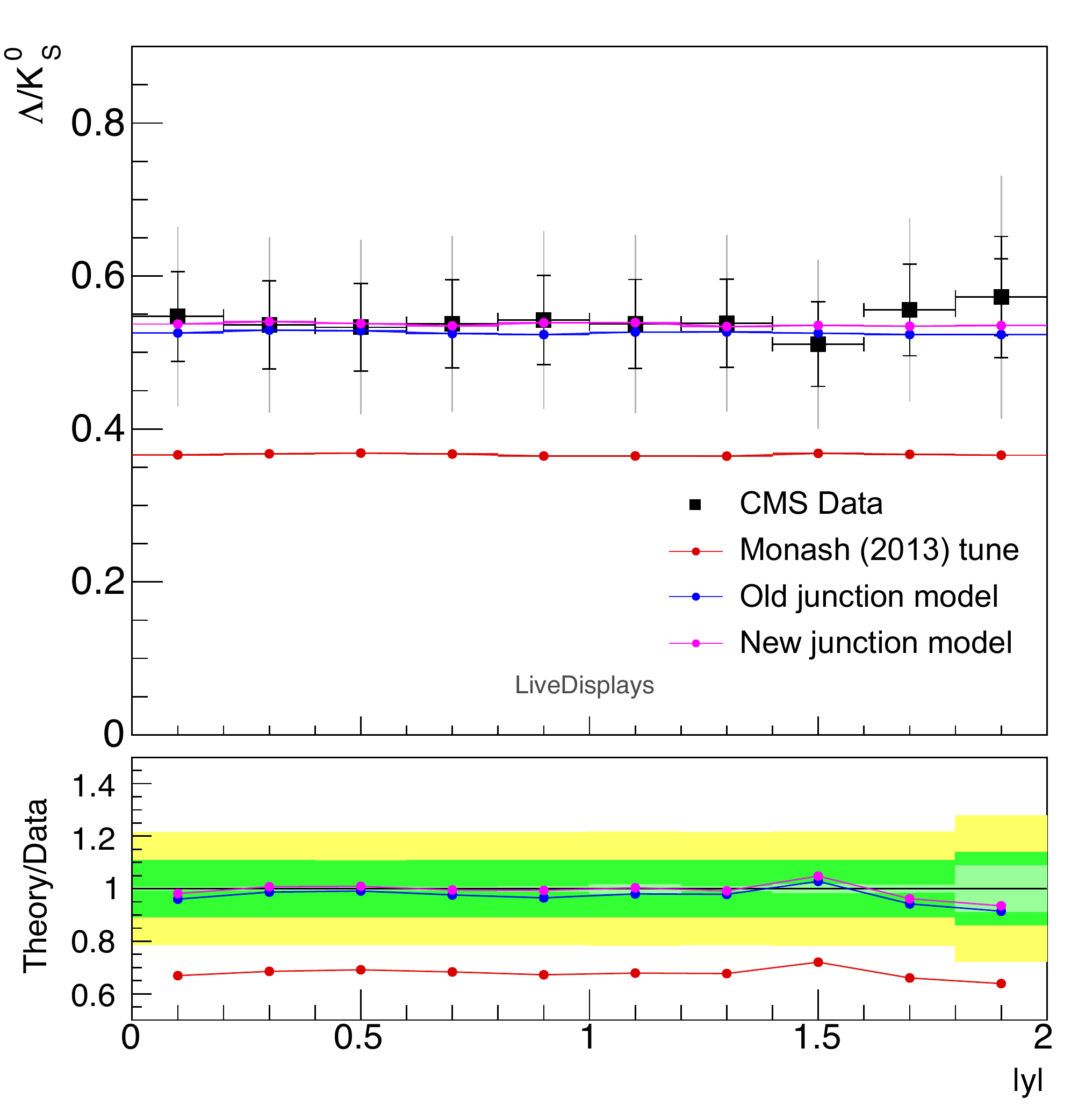}
    \includegraphics[width = 0.45\textwidth]{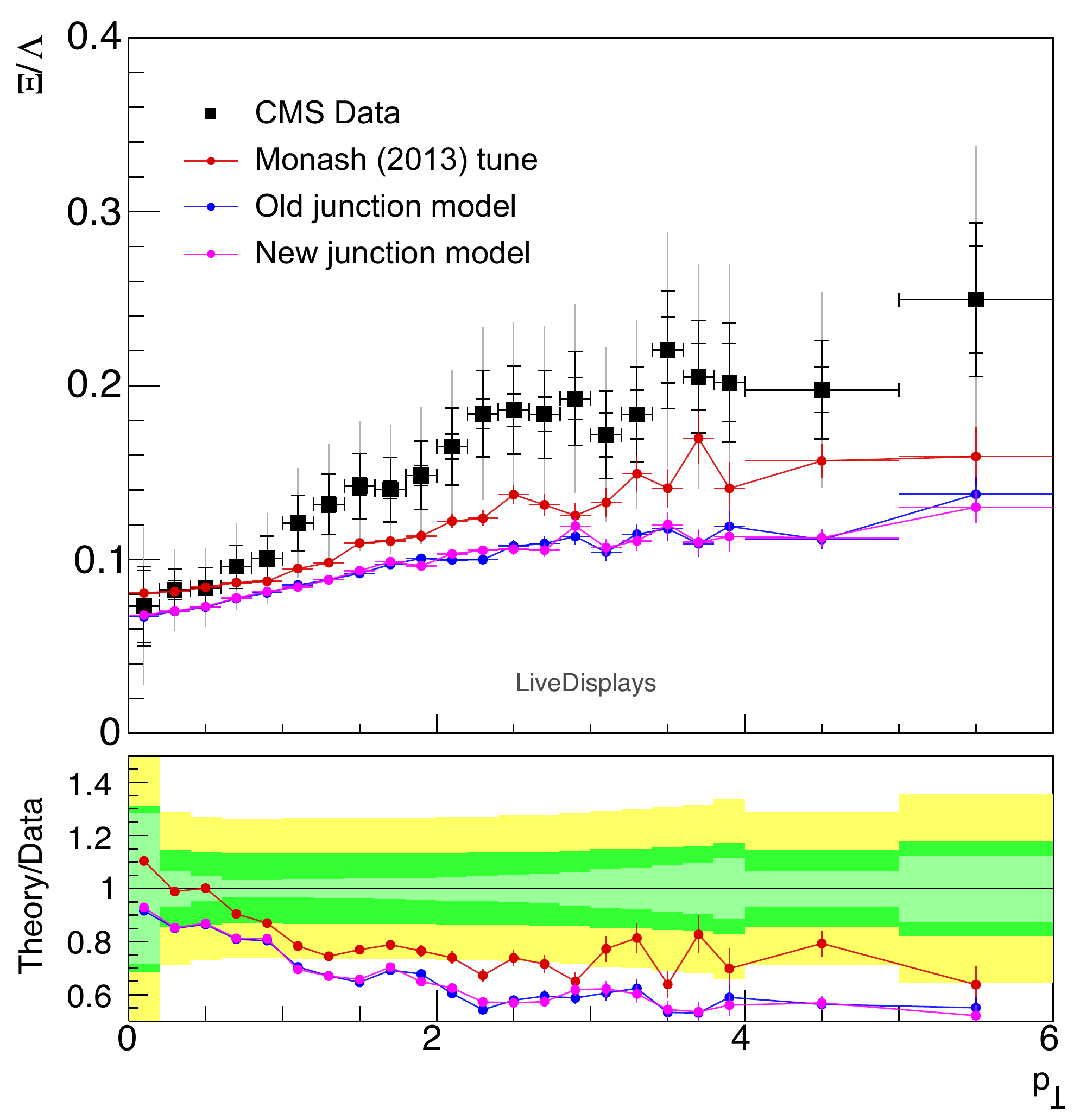}~
    \includegraphics[width = 0.45\textwidth]{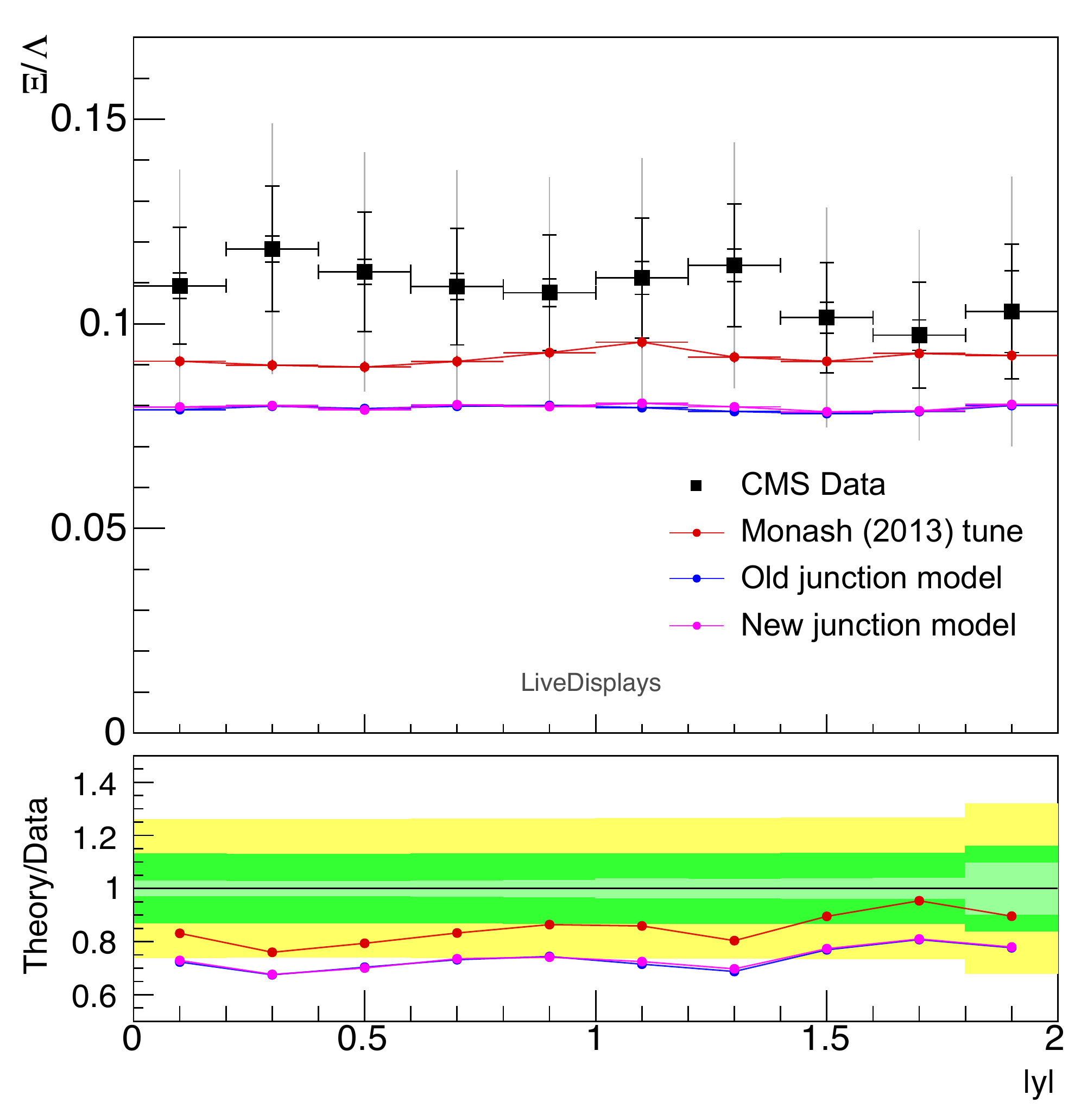}
    \end{center}
    \caption{The $\Lambda/K^0_S$ (top row of plots) and $\Xi/\Lambda$ (bottom row of plots) ratios. The left and right panels show these ratios as functions of $p_\perp$ and $|y|$ respectively. All events are 7 TeV NSD events from CMS \cite{CMS:2011jlm}, where \Py simulations have a lifetime cut of $\tau_{max} = 10$ mm/c and no $p_\perp$ cuts on final state particles.}
    \label{fig:CMSratios}
\end{figure}
We shall consider the impacts of the revised junction modelling on baryon-to-meson and baryon-to-baryon ratios, first in the light-flavour sector and then in the heavy-flavour sectors. In the light-flavour sector, we do not expect noticeable changes, as the revised modelling should only affect a small corner of the production phase space. Fig.~\ref{fig:CMSratios} depicts the ratio of light-flavour distributions, with the $\Lambda/K$ and $\Xi/\Lambda$ ratios with respect to transverse momentum and rapidity. As expected, the new modelling is very similar to the old, with both predicting roughly the same rapidity and $p_\perp$ distributions. The $\Lambda/K$ ratio remains somewhat overpredicted at high $p_\perp$, but since this is already the case for the Monash tune it presumably originates from diquark-antidiquark string breaks and not from junctions. Another noticeable feature of fig.~\ref{fig:CMSratios} is the underprediction of the $\Xi/\Lambda$ ratio, which is present in all models shown. The $\Xi/\Lambda$ ratio is a baryon-to-baryon ratio of a double-strange to single-strange baryon, thus this underprediction appears indicative of a need for strangeness enhancement which cannot be described by the inclusion of junctions alone. We plan to return to the question of strangeness enhancement in a separate study.

\begin{figure}[t]
    \begin{center}
    \includegraphics[width = 0.45\textwidth]{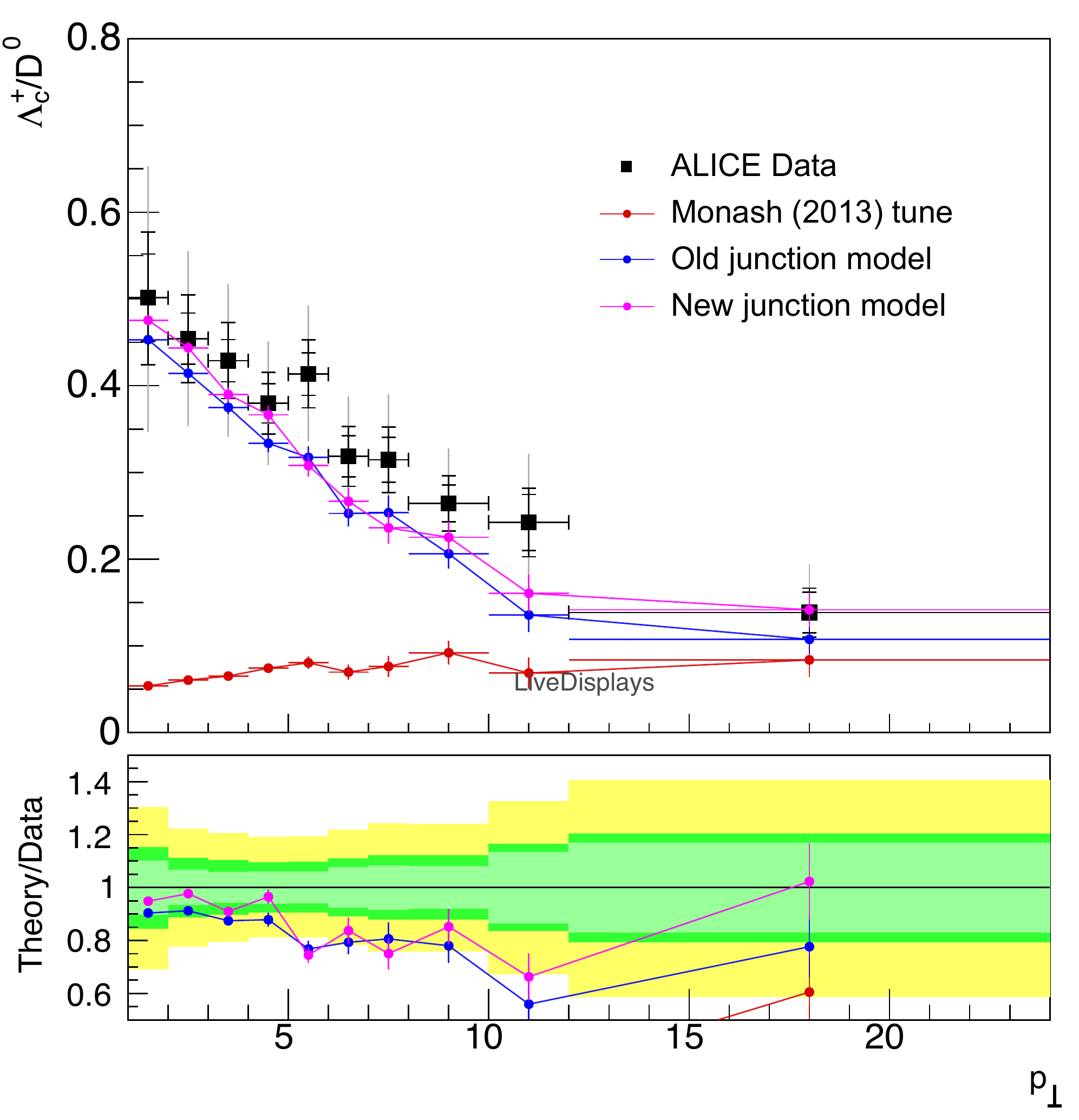}~
    \includegraphics[width = 0.45\textwidth]{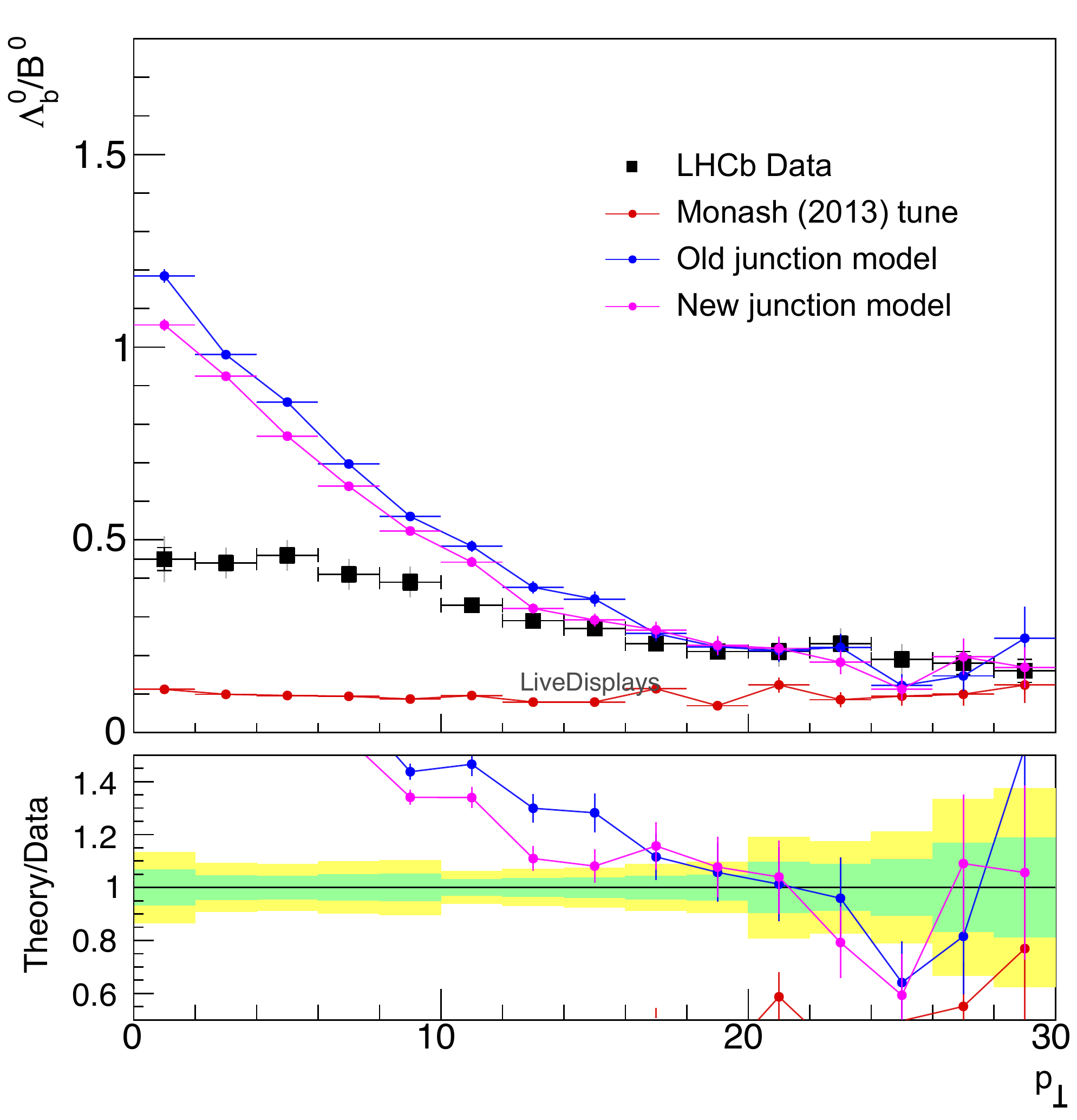}
    \includegraphics[width = 0.45\textwidth]{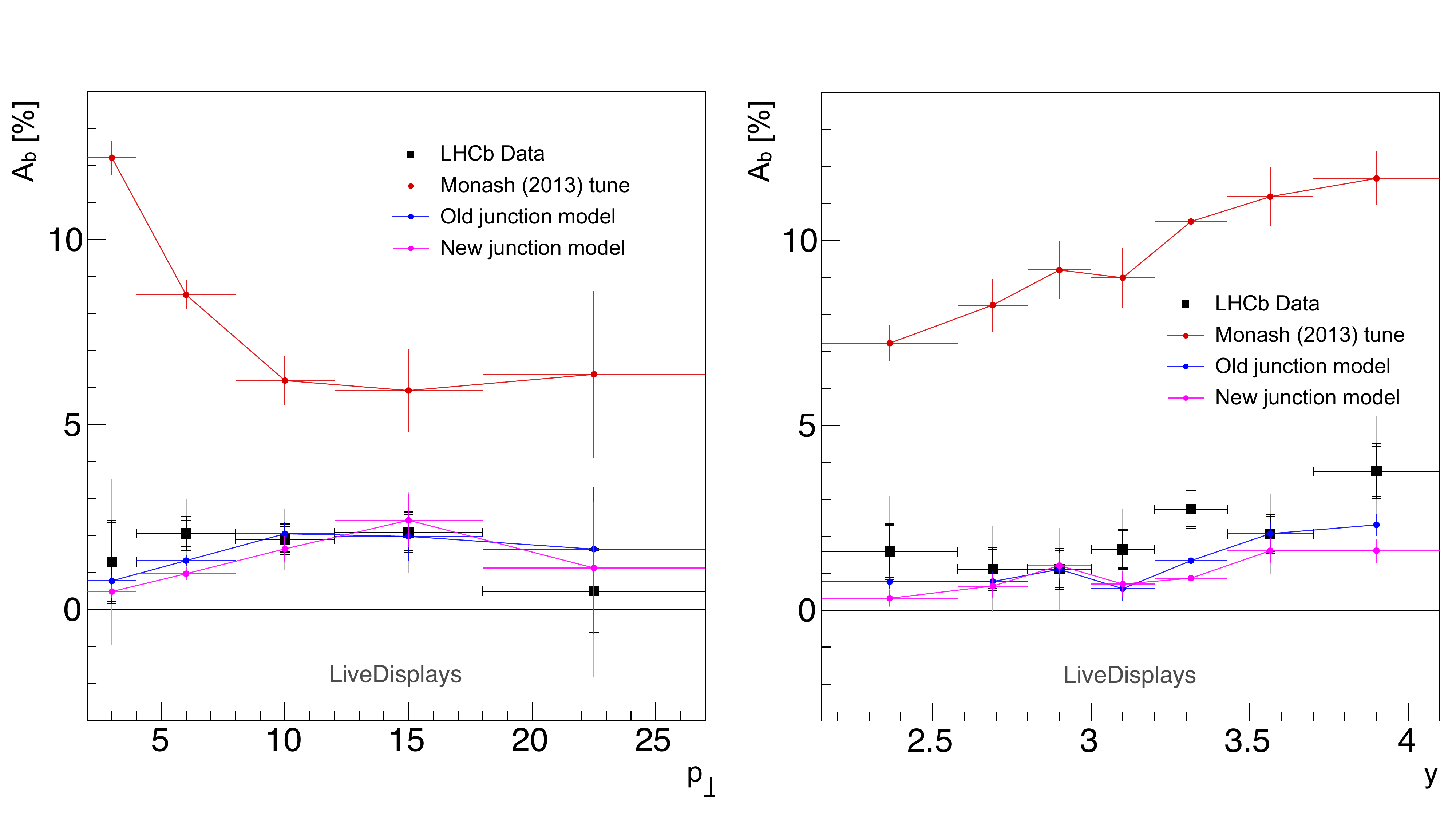}~
    \includegraphics[width = 0.45\textwidth]{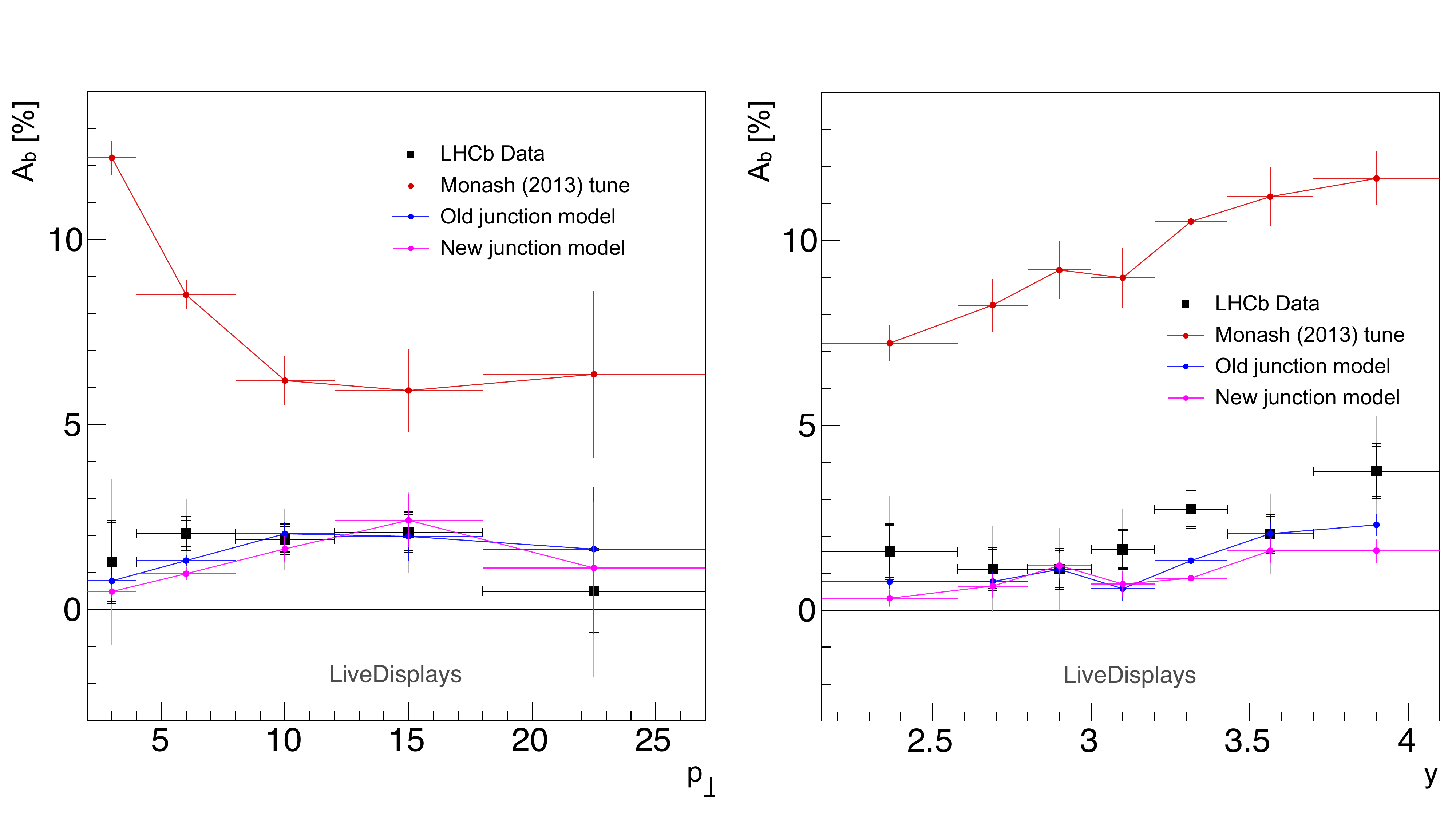}
    \end{center}
    \caption{The top row shows the $p_\perp$ distributions of baryon-to-meson ratios, with the left panel showing the prompt $\Lambda_c^+/D^0$ ratio from the ALICE collaboration \cite{ALICE:2021rzj} and the right panel shows the $\Lambda_b^0/B^0$ from the LHCb collaboration \cite{LHCb:2023wbo}. Both sets of data are for $\sqrt{s}=13$ TeV inelastic events, with rapidity ranges $|y|<0.5$ and $2<|y|<4.5$ respectively. The bottom row of plots shows the $\Lambda_b$ asymmetry~\cite{LHCb:2021xyh} for $\sqrt{s} = 7$ TeV events as a function of $p_\perp$ (left panel) and $y$ (right panel) in the rapidity range $2.15 < y < 4.10$ and transverse momentum range $2 < pT < 27$ GeV.}
    \label{fig:heavyDist}
\end{figure}

Turning now to heavy-flavour baryon-to-meson ratios, fig.~\ref{fig:heavyDist} shows the prompt $\Lambda_c^+/D^0$ and $\Lambda_b^0/B^0$ ratios as a function of $p_\perp$. When examining the success of the CR models, the $\Lambda_c^+/D^0$ ratio has been shown to 
fit the data fairly well, however in much previous literature the quality of the description of the $\Lambda_b^0/B^0$ ratio had not been studied. Given the new junction modelling, the $\Lambda_c^+/D^0$ is largely unchanged and exhibits similar behaviour to the old model, with perhaps a slightly higher relative enhancement at very low $p_\perp$. (We note that the old-model distributions here are slightly lower than those typically shown in other literature such as \cite{ALICE:2021rzj} due to the modified values of \texttt{StringFlav:probQQ1toQQ0join}.) However the $\Lambda_b^0/B^0$ ratio is overpredicted by a factor of two in both the old and new junction modelling. We regard this as the main issue that remains to be addressed and would like to follow up with a dedicated study of CR (and related) effects in $b$ vs.\ $c$ production. At this point, we restrict ourselves to the following remarks: though these ratios are for different rapidity ranges, the distributions according to \Py are largely unaffected by the the chosen rapidity range. Interestingly the data shows both ratios follow roughly the same $p_\perp$ dependence, starting at values of around 0.5 at low $p_\perp$, which neither model appears to be able to replicate. The new junction model goes in the right direction (but not far enough), resulting in a slightly lower $\Lambda_b^0/B^0$ ratio compared to the old junction model. 
This is not surprising as per the results in fig.~\ref{fig:testBaryon}, which show the new modelling results in slightly harder b-baryons from junctions, which in turn would result in slightly less soft $\Lambda_b$ baryons. 
From these distributions alone however it is unclear whether this discrepancy between the model and data is from the overprediction of $\Lambda_b$ production or underprediction of $B^0$ production.  It would be particularly insightful 
to compare to experimental results of other ratios for these heavy-flavour hadrons, such as a $\Lambda_b/\Lambda_c$ ratio or $B^0/D^0$ ratio, to help identify where the model predictions are falling short. 

We also validate that the new modelling well predicts the overall behaviour of the $\Lambda_b$ asymmetry~\cite{LHCb:2021xyh}, which the QCD-CR model in \Py has already been shown to predict quite well. This is evident in the bottom row of plots in fig.~\ref{fig:heavyDist} with respect to $p_\perp$ and $y$, and shows the dilution of the $\Lambda_b$ asymmetry particularly at low $p_\perp$ from junction baryons. If anything, the dilution effect seems to be somewhat overpredicted, consistent with what we see in the $\Lambda_b^0/B^0$ distribution.

\begin{figure}[t]
    \begin{center}
    \includegraphics[width = 0.45\textwidth]{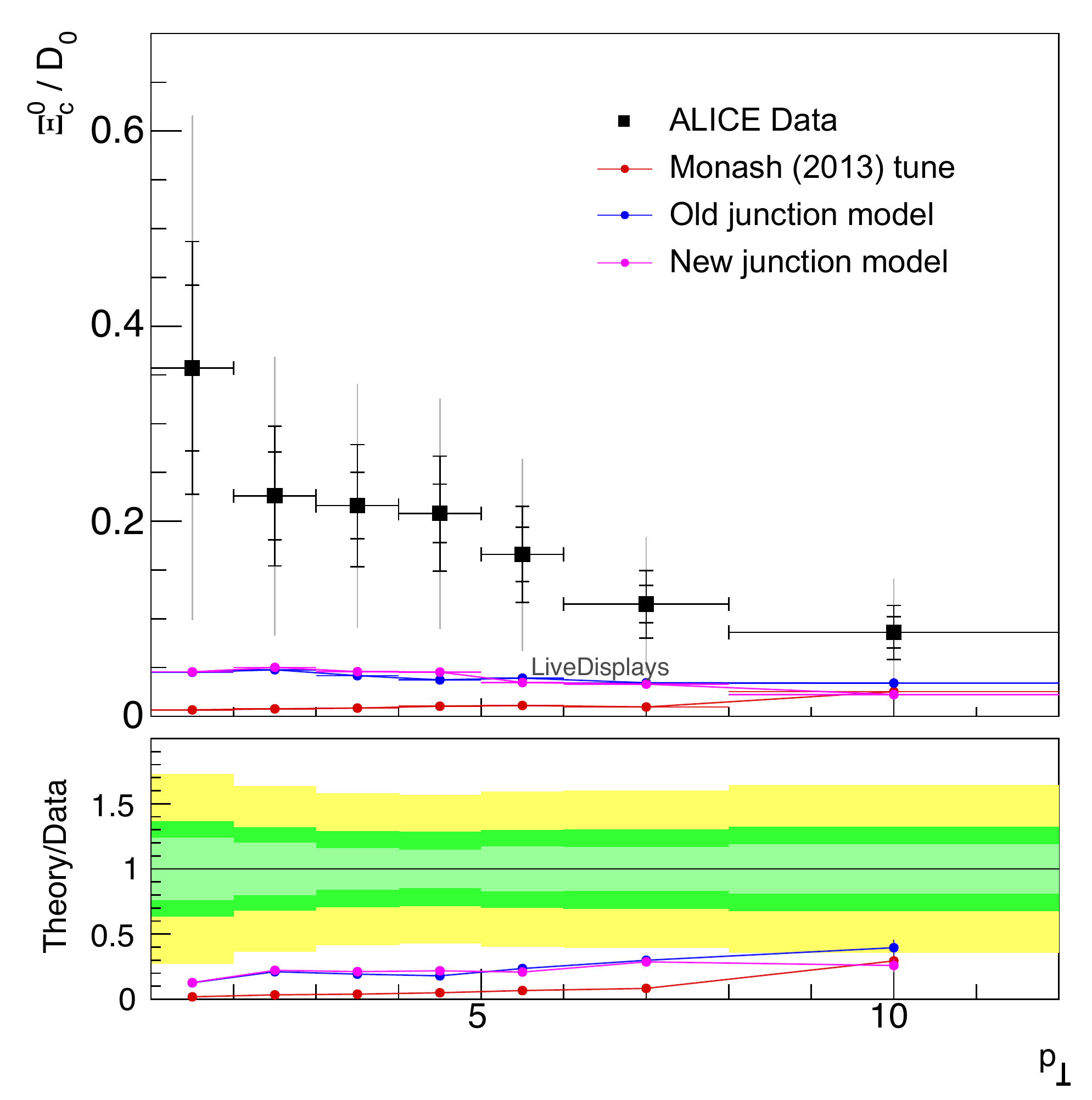}~
    \includegraphics[width = 0.45\textwidth]{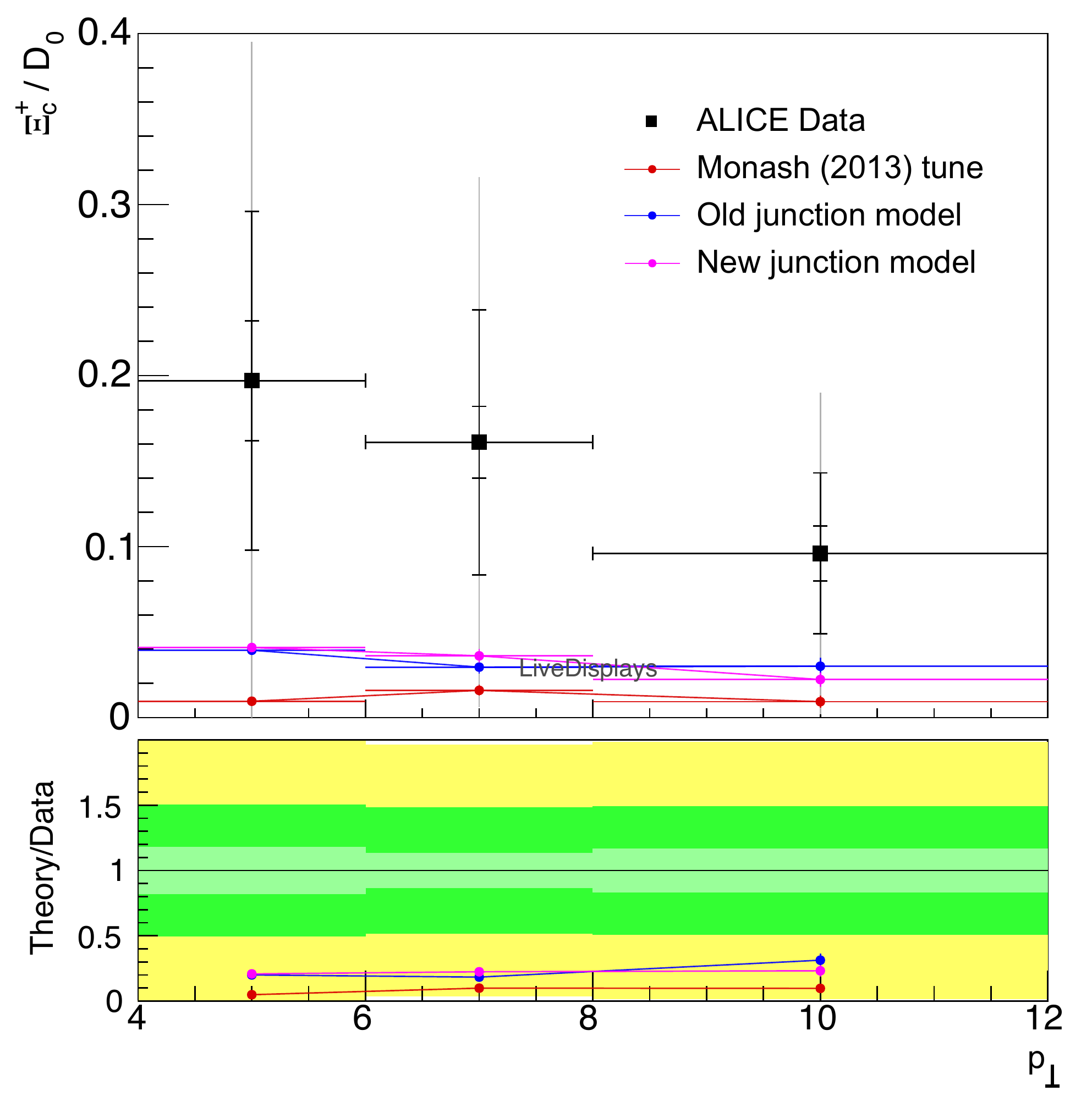}
    \includegraphics[width = 0.45\textwidth]{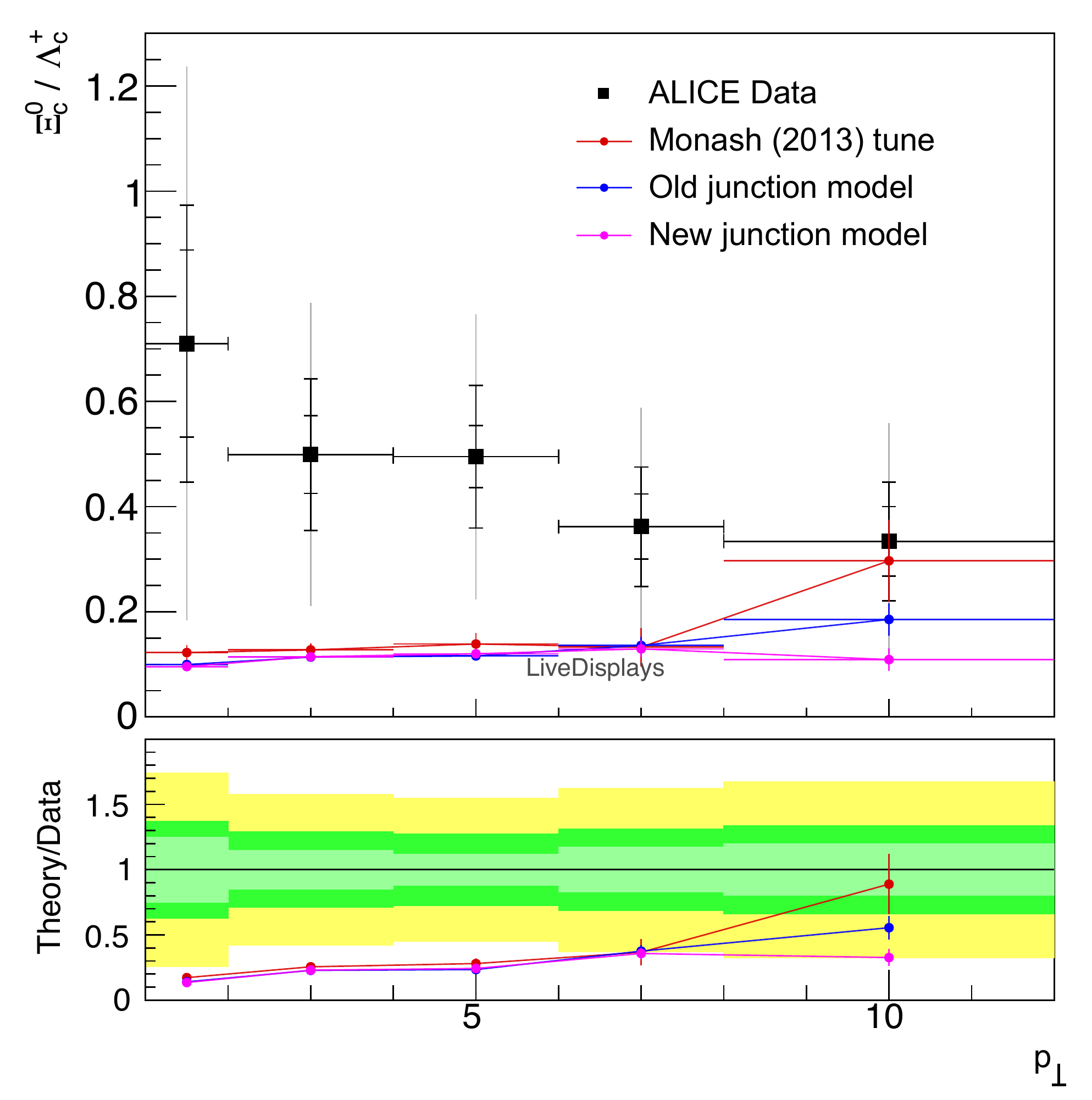}~
    \includegraphics[width = 0.45\textwidth]{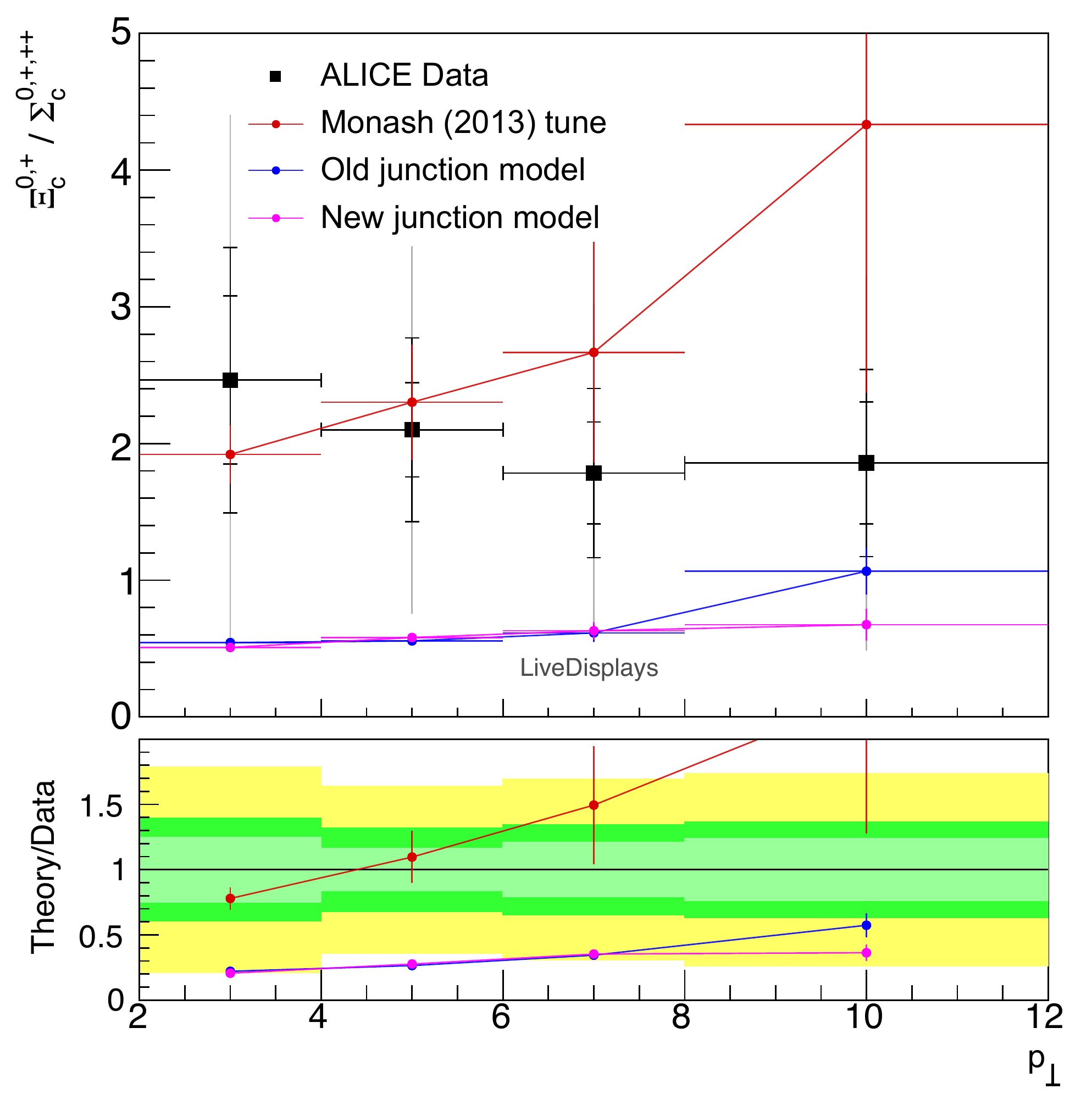}
    \end{center}
    \caption{$\Xi_c$-to-hadron ratios as a function of transverse momentum for $\sqrt{s}=7$ TeV events at ALICE~\cite{ALICE:2021bli} at midrapidity $|y|<0.5$. Top left: $\Xi_c^0/D^0$. Top right: $\Xi_c^+/D^0$. Bottom left: $\Xi_c^0/\Lambda_c^+$. Bottom right: $\Xi_c^{0,+}/\Sigma_c^{0,+,++}$.}
    \label{fig:xiRatios}
\end{figure}

Though junction formation provides enhancement of low-$p_\perp$ baryon production, as mentioned above in relation to the $\Xi/\Lambda$ ratio fig.~\ref{fig:CMSratios}, it is insufficient to describe strange baryon production. 
This is further evident given the distributions shown in fig.~\ref{fig:xiRatios}, which depicts ratios 
involving the single strange $\Xi_c$ baryon to non-strange hadrons. 
One can clearly observe that \Py underpredicts these sets of data by a significant amount, regardless of  which form of junction modelling is used, demonstrating that additional baryon enhancement from junctions is insufficient to describe strange baryon production rates. 
Alongside other strange-hadron data such as the strange hadron-to-pion ratios from ALICE \cite{ALICE:2016fzo}, there is a clear indication for a need for an additional strangeness enhancement mechanism.
An example of such a strangeness-enhancement model which is already implemented in \Py would be the Rope hadronization mode \cite{Bierlich:2016faw, Bierlich:2017sxk}. Though we will not elaborate on strangeness further in this paper, we include these distributions here to emphasise that CR and the enhanced baryon production from junctions alone are not sufficient to describe strange-particle distributions.

\section{Conclusions and Outlook}

We have revisited the equations of motion for string junctions with special attention to limits of soft and/or slow endpoints, for which the previous fragmentation framework in \Py was not robust. Based on our considerations, we have developed a revised fragmentation of string junctions which we have implemented in \Py. A main part of this work has been an updated procedure to find the so-called ``average'' junction rest frame. This was largely motivated by the lack of special treatment of soft-leg junctions as well as the previous procedure resulting in convergence errors for around 10\% of events. 

In the updated modelling, the reliance on an iterative procedure has been removed entirely.
We do so by stepping away from the notion that the average JRF must be some overall Mercedes frame (which does not exist in all cases), and instead map out the junction motion over time and then explicitly average this behaviour. We also ensure to include special treatment of soft junction legs, including the mapping out of oscillations of a soft leg around a junction and the novel pearl-on-a-string treatment. 
Though we do not expect large differences in hadron distributions given the updated modelling, the overall treatment of junctions outlined in this paper results in a more stable and, we believe, reliable junction hadronization procedure that is founded in more rigorous modelling. This allows us to draw stronger conclusions on junction predictions. For future studies a full retuning effort should be undertaken, including both these junction revisions along with all other code updates in the updated version of \Py 8.311.

Further experimental data for comparisons would be of great interest. Most interesting perhaps would be further studies vs.\ multiplicity measures, as well as heavy-flavour hadron ratios, such as $\Lambda_b/\Lambda_c$  and $B^0/D^0$ ratios, and studies of global event properties in events that contain low-$p_\perp$ charm or beauty hadrons. This could assist in finding the source of the discrepancy between the string modelling and experimental results given the current overprediction of the $\Lambda_b/B^0$ ratio. Further studies of baryon species sensitive to the spin-1 vs.\ spin-0 diquark compositions would also be revealing, and e.g.\ extending this to the $b$ sector; as would studies of multiply-heavy-flavoured baryons. 
Baryon correlations would also be very insightful as it may give some indication to the production source of baryons. Diquark-antidiquark pair creation via standard string breaks results in a baryon produced next to an antibaryon, resulting in strong correlations. The popcorn mechanism~\cite{Andersson:1984af, Eden:1996xi} for diquark production allows for meson production between a diquark-antidiquark pair, resulting in more mild baryon-antibaryon correlations. Junction baryons are expected to be even less correlated.

\begin{figure}[t]
    \begin{center}
    \includegraphics[width = 0.5\textwidth]{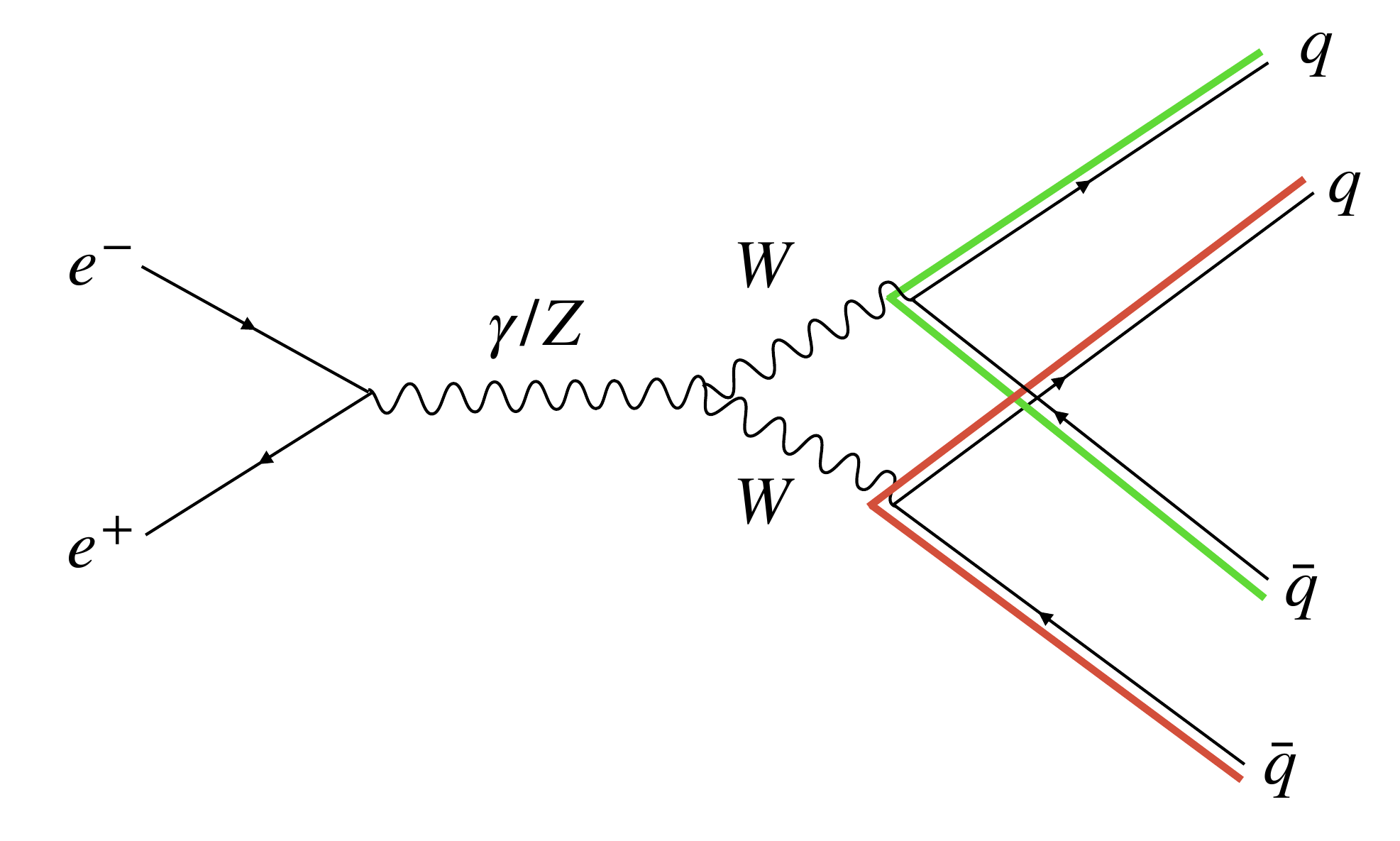}
    \includegraphics[width = 0.8\textwidth]{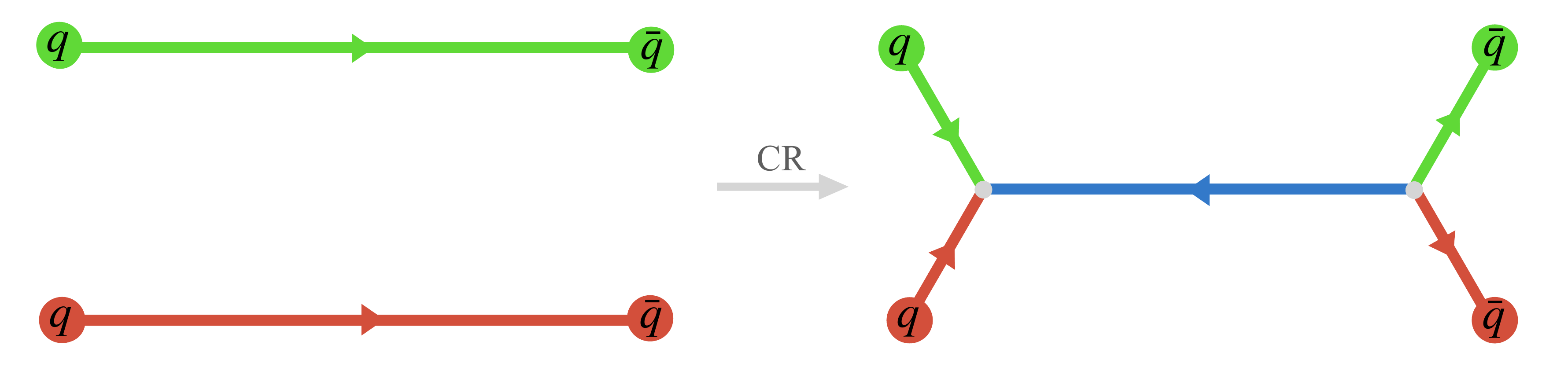}
    \end{center}
    \caption{Top panel: $e^+e^-\rightarrow WW$ process. Both $W$ bosons decay to a $q\bar{q}$ pair with colours green-antigreen and red-antired. In this example the two (anti)quarks move are moving in the same direction. Bottom panel: the left image shows the LC string configuration given the $e^+e^-\rightarrow WW$ process above, and the right image shows a possible junction-antijunction string configuration formed after colour reconnections given the red-green combination results in an antitriplet. }
    \label{fig:eeWW}
\end{figure}

There is also reason to reexamine $e^+e^-\rightarrow WW$ events with junctions in view of the possible formation of junction-antijunction configurations there. An example of such a process is depicted in the top image of fig.~\ref{fig:eeWW}, with the bottom image of fig.~\ref{fig:eeWW} showing the two possible string configurations for such a process; the LC dipole strings and the possible beyond-LC junction-antijunction configuration. These junction-antijunction topologies are formed when there are two quarks moving in one direction, and two antiquarks moving in the opposite direction. As a result, one would expect to find an enhancement of baryons, originating from a junction baryon and antibaryon moving in opposite directions, a behaviour not captured by previous $e^+e^-$ models. Measurements of possible baryon enhancement from these junction-antijunction structures along with the correlation of such baryons would be of particular interest for future studies into junction modelling. Additionally there are many other reasons to further explore $e^+e^-$ collision systems as several interesting experimental observables were not considered during the LEP experiment such as multiplicity dependent measurements, cf., e.g.,~\cite{Hunt-Smith:2020lul}. These could also form targets for fragmentation studies at future $e^+e^-$ colliders, where even relatively rare (e.g., multi-strange, heavy-strange, etc.) baryon species can be probed with high statistics. 

Particularly careful examination of the density of string systems is also required for future studies. As mentioned above, the inclusion of junctions alone is insufficient to describe the strange $\Xi$ and $\Xi_c$ hadron ratios, which along with multiplicity-dependent light strange-hadron-to-pion ratios from ALICE~\cite{ALICE:2017jyt} is indicative of a need for a strangeness-enhancement mechanism that scales with system size. Though existing proposals for such mechanisms such as rope hadronization~\cite{Bierlich:2017sxk} have already shown the ability to describe strange-hadron-to-pion ratios from ALICE, they appear to remain insufficient in describing the enhancement seen in $\Xi_c$. Further, baryon enhancement at low $p_\perp$ does not seem the solution given the well described $\Lambda_c/D^0$ ratio and the overpredicted $\Lambda_b/B^0$ ratio. Hence one must question whether strangeness enhancement for heavy-flavour baryons occurs in some subtly different way than for the light flavours. 
It would also be insightful to examine $\Xi_b$ hadron ratios, particularly comparison of the $\Xi_b/\Lambda_b$ to $\Xi_c/\Lambda_c$ or $\Xi_b/B^0$ to $\Xi_c/D^0$ ratios, which may help pick apart the patterns of baryon- vs.\ strangeness-enhancement mechanisms. 

In fact, our \Py 8.311 implementation also includes the ability to modify the strangeness probability for breaks around junctions themselves, which may in turn result in more heavy strange baryons in particular and perhaps assist in describing the $\Xi_c$ baryon distribution. A detailed examination into such modelling however is left to a future study. 

More generally, further studies into the string environment dependence of fragmentation processes would be interesting, particularly the examination of differences between triplet and octet string fragmentation. Experimentally this could correspond to looking at the tips of gluon jets, LEP data of the so-called hairpin configuration, and perhaps even diffractive LHC events (which may be enriched in gluon fragmentation without significant transverse excitations) may also be able to provide useful insight.

There are of course also many more general open-ended questions within the modelling of strings which one can tackle, including alternatives to the Schwinger mechanism (e.g.,~\cite{Fischer:2016zzs}) and looking beyond the Lund string model. 
The Lund model assumes a constant string tension motivated by the linear portion of the QCD Cornell potential. Lattice support for this is based on assuming static colour charges, and the evidence from hadron spectroscopy also has elements of steady-state evolution. However, in a hadronization model we are concerned with dynamic strings which are expanding at near the speed of light and fragmenting as they do so. This begs the question whether a straightforward model with a constant string tension is sufficient. Few forays into more general models have so far been done (e.g.,~\cite{Hunt-Smith:2020lul}) and more would certainly provide for interesting future extensions to the string modelling in \Py. 

\acknowledgments

We are grateful to G\"osta Gustafson and to Torbj\"orn Sj\"ostrand for
many insightful discussions of the motion and fragmentation of
junction string systems. We also thank the staff and students at the
Rudolf Peierls Centre for Theoretical Physics in Oxford for 
a wonderful working environment during the completion of this work. 
This work was funded by the Australian Research Council via Discovery
Project DP230103014 ``Beautiful strings'', by the Royal Society
Wolfson Visiting Fellowship ``Piercing the precision barrier in
high-energy particle physics'', and by the Monash-Warwick Alliance for
Particle Physics. The work was also supported in part by the European
Union’s Horizon 2020 research and innovation programme under the Marie
Sklodowska-Curie grant agreement No 722105 — MCnetITN3.  

\clearpage\appendix

\section{Model Parameters}
\label{app:parms}
The parameters used in the models in this paper (specifically the ones that differ from the Monash tune and CR Mode 2) are listed in tab.~\ref{tab:models}.
\begin{table}[h]
    \centering
    \begin{tabular}{p{7cm}|lll}
       & \multicolumn{1}{p{2.4cm}}{\centering Monash tune} & \multicolumn{1}{p{2.4cm}}{\centering ``Old'' model \Py 8.310} & \multicolumn{1}{p{2.5cm}}{\centering ``New'' model \Py 8.311 ~[gluon approximation~]} \\
       \hline
       StringPT:sigma & = 0.335 & = 0.335 & = 0.335 \\
       StringZ:aLund & = 0.68 & = 0.36 & = 0.36\\
       StringZ:bLund & = 0.98 & = 0.56 & = 0.56\\
       StringZ:rFactC & = 1.5 & = 1.5 & = 1.5 \\
        \hline
       StringFlav:mesonCvector & = 1.5 & = 1.5 & = 1.5 \\
       StringFlav:probStoUD & = 0.217 & = 0.2 & = 0.2\\
       StringFlav:probQQtoQ & = 0.081 & = 0.078 & = 0.078\\
       StringFlav:probQQ1toQQ0join & = 0.5, & = 0.5 & = 0.5\\
        & ~~~0.7, & ~~~0.7, & ~~~0.7,\\
        & ~~~0.9, & ~~~0.9, & ~~~0.9,\\
        & ~~~1.0 & ~~~1.0 & ~~~1.0\\
        \hline
       MultiPartonInteractions:pT0Ref & = 2.28 & = 2.25 & = 2.25\\
        \hline
       BeamRemnants:remnantMode & = 0 & = 1 & = 1\\
       BeamRemnants:saturation & -- & = 5 & = 5\\
        \hline
       ColourReconnection:mode & = 0 & = 1 & = 1 \\
       ColourReconnection:allowDoubleJunRem & = on & = off & = off \\
       ColourReconnection:m0  & -- & = 0.3 & = 0.3\\
       ColourReconnection:allowJunctions & -- & = on & = on\\
       ColourReconnection:junctionCorrection  & -- & = 1.20 & = 1.20 \\
       ColourReconnection:timeDilationMode & -- & = 2 & = 2 \\
       ColourReconnection:timeDilationPar  & -- & = 0.18 & = 0.18 \\
       ColourReconnection:lambdaForm  & = 0 & = 0 & = 0\\
        \hline
       StringFragmentation:pNormJunction  & = 2 & = 2 & = 2  \\
       StringFragmentation:pearlFragmentation   & = off & = off & = off ~[= on]\\
       StringFragmentation:pearlProbFactor  & --& -- & -- ~[= 4]\\
    \end{tabular}
    \caption{Parameter values used for the model comparisons in this study.}
    \label{tab:models}
\end{table}

\clearpage
\section{Pearl-on-a-string derivation}
\label{app:derivation}

Here we show the derivation of eq.~\eqref{eqn:vHeavy} given the Ariadne frame setup with two back-to-back massless quarks aligned along the z-axis and a massive quark with momentum along the x-axis, with pearl-on-a-string cases corresponding to the massive quark has having initial momentum $p_0$ such that the initial velocity is less than 1/2.

\begin{figure}[tp]\centering
    \includegraphics*[width=0.45\textwidth]{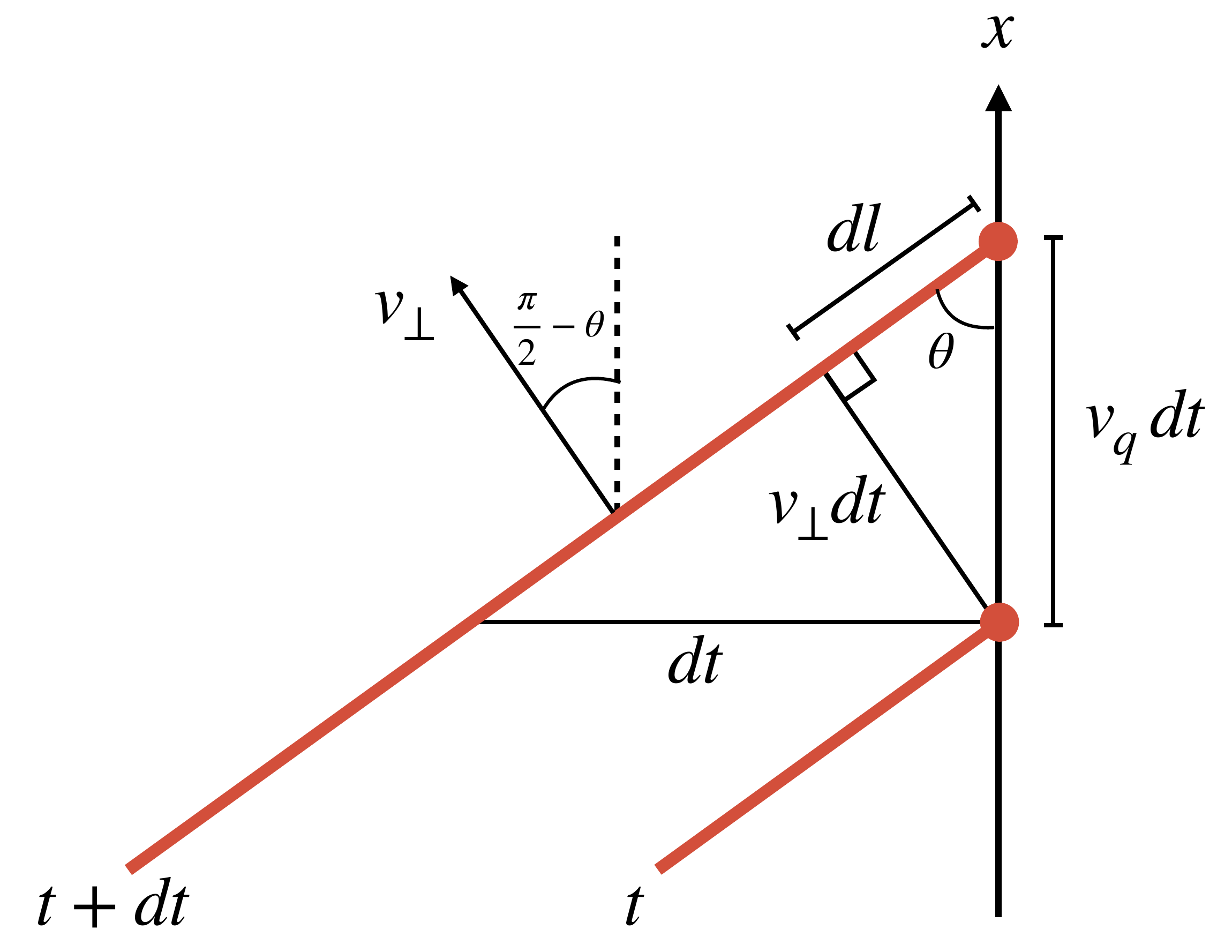}
    \caption{Single string piece near the massive pearl quark at times $t$ and $t+dt$, with transverse velocity $v_\perp$, and string length $dl$ gained due to the motion of the pearl quark.}
    \label{fig:pearlDeriv}
\end{figure}

We study the motion of the massive quark by looking at a string pieces close to the pearl quark at times $t$ and $t+dt$ as depicted in fig.~\ref{fig:pearlDeriv}. By considering the angle $\theta$ the string piece makes with pearl direction of motion, one can easily parameterize the pearl velocity $v_q$, string transverse velocity $v_\perp$, and string length $dl$ in terms of $\theta$,

\begin{subequations}
\begin{equation}
    v_q = \frac{v~dt}{dt}  = \cot\theta ,
\end{equation}
\begin{equation}
    v_\perp = v_q \sin\theta = \cos\theta,
    \label{eqn:vT}
\end{equation}
\begin{equation}
    dl = v_q~dt \cos\theta = dt \frac{\cos^2\theta}{\sin\theta} .
    \label{eqn:dl}
\end{equation}  
\end{subequations}

Given the string tension $\kappa$ represents the string energy per unit length, the energy $dE_s$ gained by the string from string piece $dl$ is the given by $dE_s = \kappa dl \gamma$ given the transverse velocity of the string. Using eq.~\eqref{eqn:vT} and~\eqref{eqn:dl}, this we can rewrite this as 

\begin{equation}
    dE_s = \kappa\frac{dl}{\sqrt{1-v_\perp^2}} = \kappa \left(dt \frac{\cos^2\theta}{\sin\theta}\right)\left(\frac{1}{\sin\theta}\right) = \kappa \cot^2\theta dt .
\end{equation}

As there is no energy transfer along the string, the energy of the string piece $dl$ must be sourced from the reduced energy of the pearl quark $dE_q$, hence 

\begin{equation}
    dE_q = -dE_s = -\kappa \cot^2\theta dt.
    \label{eqn:dEq}
\end{equation}

Then by considering $p_q^2 = E_q^2 - m^2_q$ and eq.~\eqref{eqn:dEq}, the equation of motion of the pearl quark can then be defined,

\begin{subequations}
\begin{equation}
    \frac{dp_q}{dt} = \frac{dp_q}{dE_q}\frac{dE_q}{dt} = \frac{E_q}{\sqrt{E_q^2 - m^2_q}}\frac{dE_q}{dt} = \left( \frac{1}{v_q}\right)\left(-\kappa\cot^2\theta\right) = -\kappa\cot\theta = -\kappa\frac{dx}{dt},
\end{equation}
\begin{equation}
    \frac{dp_q}{dx} = \frac{dp_q}{dt} \frac{dt}{dx} = -\kappa .
\end{equation}
\end{subequations}

Accounting for both string pieces pulling on the massive quark, we get an additional factor of 2 resulting in the momentum loss relation for the pearl,

\begin{equation}
    \frac{dp_q}{dx} = -2\kappa.
    \label{eqn:dpdx}
\end{equation}

Given initial pearl momentum $p_0$, we can then solve eq.~\eqref{eqn:dpdx} for the momentum of the pearl,

\begin{equation}
    p_q(t) = p_0-2\kappa x_q(t) = \frac{mv_q(t)}{\sqrt{1-v^2_q(t)}},
\end{equation}

with the right-hand side of the equation coming from the relation $p=mv\gamma$. Therefore this can be simply rearranged to get 

\begin{equation}
    v_q(t) = \frac{1}{\sqrt{1+\frac{m^2}{p_q^2(t)}}} = \frac{1}{\sqrt{1+\frac{m^2}{(p_0-2\kappa x_q(t))^2}}},
\end{equation}

which is precisely eq.~\eqref{eqn:vHeavy}.



\bibliographystyle{JHEP}
\bibliography{bibliography.bib}

\providecommand{\href}[2]{#2}\begingroup\raggedright\begin{thebibliography}{10}

\bibitem{CMS:2011fsn}
{\scshape CMS} collaboration, \emph{{Measurement of Strange Particle Production
  in Underlying Events in proton-proton collisions at $\sqrt{s}$ = 7 TeV}},
  2011.
\newblock CMS-PAS-QCD-11-010.

\bibitem{ATLAS:2011xhu}
{\scshape ATLAS} collaboration, \emph{{Kshort and $\Lambda$ production in $pp$
  interactions at $\sqrt{s}=0.9$ and 7 TeV measured with the ATLAS detector at
  the LHC}}, \href{https://doi.org/10.1103/PhysRevD.85.012001}{\emph{Phys. Rev.
  D} {\bfseries 85} (2012) 012001}
  [\href{https://arxiv.org/abs/1111.1297}{{\ttfamily 1111.1297}}].

\bibitem{CMS:2013zgf}
{\scshape CMS} collaboration, \emph{{Measurement of Neutral Strange Particle
  Production in the Underlying Event in Proton-Proton Collisions at $\sqrt{s}$
  = 7 TeV}}, \href{https://doi.org/10.1103/PhysRevD.88.052001}{\emph{Phys. Rev.
  D} {\bfseries 88} (2013) 052001}
  [\href{https://arxiv.org/abs/1305.6016}{{\ttfamily 1305.6016}}].

\bibitem{CDF:2013kip}
{\scshape CDF} collaboration, \emph{{Production of $K_S^0, K^{*\pm}(892)$ and
  $\phi^0(1020)$ in Minimum Bias Events and $K_S^0$ and $\Lambda^0$ in Jets in
  $p\bar p$ Collisions at $\sqrt s=1.96 TeV$}},
  \href{https://doi.org/10.1103/PhysRevD.88.092005}{\emph{Phys. Rev. D}
  {\bfseries 88} (2013) 092005}
  [\href{https://arxiv.org/abs/1308.3371}{{\ttfamily 1308.3371}}].

\bibitem{ALICE:2016fzo}
{\scshape ALICE} collaboration, \emph{{Enhanced production of multi-strange
  hadrons in high-multiplicity proton-proton collisions}},
  \href{https://doi.org/10.1038/nphys4111}{\emph{Nature Phys.} {\bfseries 13}
  (2017) 535} [\href{https://arxiv.org/abs/1606.07424}{{\ttfamily
  1606.07424}}].

\bibitem{ALICE:2018pal}
{\scshape ALICE} collaboration, \emph{{Multiplicity dependence of light-flavor
  hadron production in pp collisions at $\sqrt{s}$ = 7 TeV}},
  \href{https://doi.org/10.1103/PhysRevC.99.024906}{\emph{Phys. Rev. C}
  {\bfseries 99} (2019) 024906}
  [\href{https://arxiv.org/abs/1807.11321}{{\ttfamily 1807.11321}}].

\bibitem{ALICE:2023egx}
{\scshape ALICE} collaboration, \emph{{Multiplicity-dependent production of
  $\Sigma(1385)^{\pm}$ and $\Xi(1530)^{0}$ in pp collisions at $\sqrt{s}=13$
  TeV}},  \href{https://arxiv.org/abs/2308.16116}{{\ttfamily 2308.16116}}.

\bibitem{ANDERSSON198331}
B.~Andersson, G.~Gustafson, G.~Ingelman and T.~Sjöstrand, \emph{Parton
  fragmentation and string dynamics},
  \href{https://doi.org/https://doi.org/10.1016/0370-1573(83)90080-7}{\emph{Physics
  Reports} {\bfseries 97} (1983) 31}.

\bibitem{Andersson:2001yu}
B.~Andersson, S.~Mohanty and F.~S{\"o}derberg, \emph{{The Lund fragmentation
  process for a multigluon string according to the area law}},
  \href{https://doi.org/10.1007/s100520100757}{\emph{Eur. Phys. J. C}
  {\bfseries 21} (2001) 631}
  [\href{https://arxiv.org/abs/hep-ph/0106185}{{\ttfamily hep-ph/0106185}}].

\bibitem{ARTRU1983147}
X.~Artru, \emph{Classical string phenomenology. how strings work},
  \href{https://doi.org/https://doi.org/10.1016/0370-1573(83)90081-9}{\emph{Physics
  Reports} {\bfseries 97} (1983) 147}.

\bibitem{Andersson:1998tv}
B.~Andersson, \emph{{The Lund model}}, vol.~7, Cambridge University Press (7,
  2005).

\bibitem{Bierlich:2022pfr}
C.~Bierlich et~al., \emph{{A comprehensive guide to the physics and usage of
  PYTHIA 8.3}},
  \href{https://doi.org/10.21468/SciPostPhysCodeb.8}{\emph{SciPost Phys.
  Codeb.} {\bfseries 2022} (2022) 8}
  [\href{https://arxiv.org/abs/2203.11601}{{\ttfamily 2203.11601}}].

\bibitem{Sjostrand:1993hi}
T.~Sj{\"o}strand and V.A.~Khoze, \emph{{On Color rearrangement in hadronic W+
  W- events}}, \href{https://doi.org/10.1007/BF01560244}{\emph{Z. Phys. C}
  {\bfseries 62} (1994) 281}
  [\href{https://arxiv.org/abs/hep-ph/9310242}{{\ttfamily hep-ph/9310242}}].

\bibitem{L3:2003ohc}
{\scshape L3} collaboration, \emph{{Search for color singlet and color
  reconnection effects in hadronic $Z$ decays at LEP}},
  \href{https://doi.org/10.1016/j.physletb.2003.12.003}{\emph{Phys. Lett. B}
  {\bfseries 581} (2004) 19}
  [\href{https://arxiv.org/abs/hep-ex/0312026}{{\ttfamily hep-ex/0312026}}].

\bibitem{L3:2003oci}
{\scshape L3} collaboration, \emph{{Search for color reconnection effects in
  $e^{+} e^{-} \to W^{+} W^{-} \to$ hadrons through particle flow studies at
  LEP}}, \href{https://doi.org/10.1016/S0370-2693(03)00490-8}{\emph{Phys. Lett.
  B} {\bfseries 561} (2003) 202}
  [\href{https://arxiv.org/abs/hep-ex/0303042}{{\ttfamily hep-ex/0303042}}].

\bibitem{OPAL:2003njc}
{\scshape OPAL} collaboration, \emph{{Tests of models of color reconnection and
  a search for glueballs using gluon jets with a rapidity gap}},
  \href{https://doi.org/10.1140/epjc/s2004-01809-2}{\emph{Eur. Phys. J. C}
  {\bfseries 35} (2004) 293}
  [\href{https://arxiv.org/abs/hep-ex/0306021}{{\ttfamily hep-ex/0306021}}].

\bibitem{Siebel:2005uw}
M.~Siebel, \emph{{A Study of the charge of leading hadrons in gluon and quark
  fragmentation}},  in \emph{{40th Rencontres de Moriond on QCD and High Energy
  Hadronic Interactions}}, pp.~209--213, 5, 2005
  [\href{https://arxiv.org/abs/hep-ex/0505080}{{\ttfamily hep-ex/0505080}}].

\bibitem{OPAL:2005cdr}
{\scshape OPAL} collaboration, \emph{{Colour reconnection in e+ e-
  ---\ensuremath{>} W+ W- at s**(1/2) = 189-GeV - 209-GeV}},
  \href{https://doi.org/10.1140/epjc/s2005-02439-x}{\emph{Eur. Phys. J. C}
  {\bfseries 45} (2006) 291}
  [\href{https://arxiv.org/abs/hep-ex/0508062}{{\ttfamily hep-ex/0508062}}].

\bibitem{ALEPH:2006cdc}
{\scshape ALEPH} collaboration, \emph{{Measurement of the $W$ boson mass and
  width in $e^{+} e^{-}$ collisions at LEP}},
  \href{https://doi.org/10.1140/epjc/s2006-02576-8}{\emph{Eur. Phys. J. C}
  {\bfseries 47} (2006) 309}
  [\href{https://arxiv.org/abs/hep-ex/0605011}{{\ttfamily hep-ex/0605011}}].

\bibitem{DELPHI:2006tie}
{\scshape DELPHI} collaboration, \emph{{Investigation of colour reconnection in
  WW events with the DELPHI detector at LEP-2}},
  \href{https://doi.org/10.1140/epjc/s10052-007-0304-9}{\emph{Eur. Phys. J. C}
  {\bfseries 51} (2007) 249} [\href{https://arxiv.org/abs/0704.0597}{{\ttfamily
  0704.0597}}].

\bibitem{ALEPH:2006jcu}
{\scshape ALEPH} collaboration, \emph{{Test of Colour Reconnection Models using
  Three-Jet Events in Hadronic Z Decays}},
  \href{https://doi.org/10.1140/epjc/s10052-006-0017-5}{\emph{Eur. Phys. J. C}
  {\bfseries 48} (2006) 685}
  [\href{https://arxiv.org/abs/hep-ex/0604042}{{\ttfamily hep-ex/0604042}}].

\bibitem{ALEPH:2013dgf}
{\scshape ALEPH, DELPHI, L3, OPAL, LEP Electroweak} collaboration,
  \emph{{Electroweak Measurements in Electron-Positron Collisions at
  W-Boson-Pair Energies at LEP}},
  \href{https://doi.org/10.1016/j.physrep.2013.07.004}{\emph{Phys. Rept.}
  {\bfseries 532} (2013) 119}
  [\href{https://arxiv.org/abs/1302.3415}{{\ttfamily 1302.3415}}].

\bibitem{Sjostrand:1987su}
T.~Sj{\"o}strand and M.~van Zijl, \emph{{A Multiple Interaction Model for the
  Event Structure in Hadron Collisions}},
  \href{https://doi.org/10.1103/PhysRevD.36.2019}{\emph{Phys. Rev. D}
  {\bfseries 36} (1987) 2019}.

\bibitem{Field:2005sa}
{\scshape CDF} collaboration, \emph{{PYTHIA tune A, HERWIG, and JIMMY in Run 2
  at CDF}},  \href{https://arxiv.org/abs/hep-ph/0510198}{{\ttfamily
  hep-ph/0510198}}.

\bibitem{Skands:2010ak}
P.Z.~Skands, \emph{{Tuning Monte Carlo Generators: The Perugia Tunes}},
  \href{https://doi.org/10.1103/PhysRevD.82.074018}{\emph{Phys. Rev. D}
  {\bfseries 82} (2010) 074018}
  [\href{https://arxiv.org/abs/1005.3457}{{\ttfamily 1005.3457}}].

\bibitem{Rathsman:1998tp}
J.~Rathsman, \emph{{A Generalized area law for hadronic string
  re-interactions}},
  \href{https://doi.org/10.1016/S0370-2693(99)00291-9}{\emph{Phys. Lett. B}
  {\bfseries 452} (1999) 364}
  [\href{https://arxiv.org/abs/hep-ph/9812423}{{\ttfamily hep-ph/9812423}}].

\bibitem{Sandhoff:2005jh}
M.~Sandhoff and P.Z.~Skands, \emph{{Colour annealing - a toy model of colour
  reconnections}},  in \emph{{4th Les Houches Workshop on Physics at TeV
  Colliders}}, 12, 2005, FERMILAB-CONF-05-518-T.

\bibitem{Buttar:2006zd}
C.~Buttar et~al., \emph{{Les houches physics at TeV colliders 2005, standard
  model and Higgs working group: Summary report}},  in \emph{{4th Les Houches
  Workshop on Physics at TeV Colliders}}, 4, 2006
  [\href{https://arxiv.org/abs/hep-ph/0604120}{{\ttfamily hep-ph/0604120}}].

\bibitem{Skands:2007zg}
P.Z.~Skands and D.~Wicke, \emph{{Non-perturbative QCD effects and the top mass
  at the Tevatron}},
  \href{https://doi.org/10.1140/epjc/s10052-007-0352-1}{\emph{Eur. Phys. J. C}
  {\bfseries 52} (2007) 133}
  [\href{https://arxiv.org/abs/hep-ph/0703081}{{\ttfamily hep-ph/0703081}}].

\bibitem{Gieseke:2012ft}
S.~Gieseke, C.~Rohr and A.~Siodmok, \emph{{Colour reconnections in Herwig++}},
  \href{https://doi.org/10.1140/epjc/s10052-012-2225-5}{\emph{Eur. Phys. J. C}
  {\bfseries 72} (2012) 2225}
  [\href{https://arxiv.org/abs/1206.0041}{{\ttfamily 1206.0041}}].

\bibitem{Argyropoulos:2014zoa}
S.~Argyropoulos and T.~Sj\"ostrand, \emph{{Effects of color reconnection on
  $t\bar{t}$ final states at the LHC}},
  \href{https://doi.org/10.1007/JHEP11(2014)043}{\emph{JHEP} {\bfseries 11}
  (2014) 043} [\href{https://arxiv.org/abs/1407.6653}{{\ttfamily 1407.6653}}].

\bibitem{Christiansen:2015yqa}
J.R.~Christiansen and P.Z.~Skands, \emph{{String Formation Beyond Leading
  Colour}}, \href{https://doi.org/10.1007/JHEP08(2015)003}{\emph{JHEP}
  {\bfseries 08} (2015) 003}
  [\href{https://arxiv.org/abs/1505.01681}{{\ttfamily 1505.01681}}].

\bibitem{Bellm:2019wrh}
J.~Bellm, C.B.~Duncan, S.~Gieseke, M.~Myska and A.~Si\'odmok, \emph{{Spacetime
  colour reconnection in Herwig 7}},
  \href{https://doi.org/10.1140/epjc/s10052-019-7533-6}{\emph{Eur. Phys. J. C}
  {\bfseries 79} (2019) 1003}
  [\href{https://arxiv.org/abs/1909.08850}{{\ttfamily 1909.08850}}].

\bibitem{Sjostrand:2002ip}
T.~Sj{\"o}strand and P.Z.~Skands, \emph{{Baryon number violation and string
  topologies}},
  \href{https://doi.org/10.1016/S0550-3213(03)00193-7}{\emph{Nucl. Phys. B}
  {\bfseries 659} (2003) 243}
  [\href{https://arxiv.org/abs/hep-ph/0212264}{{\ttfamily hep-ph/0212264}}].

\bibitem{LHCb:2017bdt}
{\scshape LHCb} collaboration, \emph{{Measurement of $B^0$, $B^0_s$, $B^+$ and
  $\Lambda^0_b$ production asymmetries in 7 and 8 TeV proton-proton
  collisions}},
  \href{https://doi.org/10.1016/j.physletb.2017.09.023}{\emph{Phys. Lett. B}
  {\bfseries 774} (2017) 139}
  [\href{https://arxiv.org/abs/1703.08464}{{\ttfamily 1703.08464}}].

\bibitem{ALICE:2020wla}
{\scshape ALICE} collaboration, \emph{{$\Lambda^+_c$ production in $pp$ and in
  $p$-Pb collisions at $\sqrt {s_{NN}}$=5.02 TeV}},
  \href{https://doi.org/10.1103/PhysRevC.104.054905}{\emph{Phys. Rev. C}
  {\bfseries 104} (2021) 054905}
  [\href{https://arxiv.org/abs/2011.06079}{{\ttfamily 2011.06079}}].

\bibitem{ALICE:2021bli}
{\scshape ALICE} collaboration, \emph{{Measurement of the cross sections of
  $\Xi^0_{\rm c}$ and $\Xi^+_{\rm c}$ baryons and branching-fraction ratio
  BR($\Xi^0_{\rm c} \rightarrow \Xi^-{\rm e}^+\nu_{\rm e}$)/BR($\Xi^0_{\rm c}
  \rightarrow \Xi^-\pi^+$) in pp collisions at 13 TeV}},
  \href{https://arxiv.org/abs/2105.05187}{{\ttfamily 2105.05187}}.

\bibitem{ALICE:2021rzj}
{\scshape ALICE} collaboration, \emph{{Measurement of Prompt D$^{0}$,
  $\Lambda_{c}^{+}$, and $\Sigma_{c}^{0,++}$(2455) Production in
  Proton\textendash{}Proton Collisions at $\sqrt s$ = 13\,\,TeV}},
  \href{https://doi.org/10.1103/PhysRevLett.128.012001}{\emph{Phys. Rev. Lett.}
  {\bfseries 128} (2022) 012001}
  [\href{https://arxiv.org/abs/2106.08278}{{\ttfamily 2106.08278}}].

\bibitem{LHCb:2021xyh}
{\scshape LHCb} collaboration, \emph{{Observation of a
  $\Lambda_b^0-\overline{\Lambda}_b^0$ production asymmetry in proton-proton
  collisions at $\sqrt{s} = 7 \textrm{ and } 8\,\textrm{TeV}$}},
  \href{https://doi.org/10.1007/JHEP10(2021)060}{\emph{JHEP} {\bfseries 10}
  (2021) 060} [\href{https://arxiv.org/abs/2107.09593}{{\ttfamily
  2107.09593}}].

\bibitem{ALICE:2022cop}
{\scshape ALICE} collaboration, \emph{{First measurement of
  \ensuremath{\Omega_c^0} production in pp collisions at
  \ensuremath{\sqrt{s}=13} TeV}},
  \href{https://doi.org/10.1016/j.physletb.2022.137625}{\emph{Phys. Lett. B}
  {\bfseries 846} (2023) 137625}
  [\href{https://arxiv.org/abs/2205.13993}{{\ttfamily 2205.13993}}].

\bibitem{ALICE:2022exq}
{\scshape ALICE} collaboration, \emph{{First measurement of
  \ensuremath{\Lambda}c+ production down to pT=0 in pp and p-Pb collisions at
  sNN=5.02 TeV}},
  \href{https://doi.org/10.1103/PhysRevC.107.064901}{\emph{Phys. Rev. C}
  {\bfseries 107} (2023) 064901}
  [\href{https://arxiv.org/abs/2211.14032}{{\ttfamily 2211.14032}}].

\bibitem{ALICE:2023jgm}
{\scshape ALICE} collaboration, \emph{{Exploring the non-universality of charm
  hadronisation through the measurement of the fraction of jet longitudinal
  momentum carried by $\Lambda_{\rm c}^+$ baryons in pp collisions}},
  \href{https://arxiv.org/abs/2301.13798}{{\ttfamily 2301.13798}}.

\bibitem{ALICE:2023xiu}
{\scshape ALICE} collaboration, \emph{{Inclusive and multiplicity dependent
  production of electrons from heavy-flavour hadron decays in pp and p-Pb
  collisions}}, \href{https://doi.org/10.1007/JHEP08(2023)006}{\emph{JHEP}
  {\bfseries 08} (2023) 006}
  [\href{https://arxiv.org/abs/2303.13349}{{\ttfamily 2303.13349}}].

\bibitem{ALICE:2023wbx}
{\scshape ALICE} collaboration, \emph{{Study of flavor dependence of the
  baryon-to-meson ratio in proton-proton collisions at
  \ensuremath{\sqrt{s}=13}\,\,TeV}},
  \href{https://doi.org/10.1103/PhysRevD.108.112003}{\emph{Phys. Rev. D}
  {\bfseries 108} (2023) 112003}
  [\href{https://arxiv.org/abs/2308.04873}{{\ttfamily 2308.04873}}].

\bibitem{ALICE:2023sgl}
{\scshape ALICE} collaboration, \emph{{Charm production and fragmentation
  fractions at midrapidity in pp collisions at $ \sqrt{\textrm{s}} $ = 13
  TeV}}, \href{https://doi.org/10.1007/JHEP12(2023)086}{\emph{JHEP} {\bfseries
  12} (2023) 086} [\href{https://arxiv.org/abs/2308.04877}{{\ttfamily
  2308.04877}}].

\bibitem{LHCb:2023wbo}
{\scshape LHCb} collaboration, \emph{{Enhanced Production of
  \ensuremath{\Lambda}b0 Baryons in High-Multiplicity pp Collisions at
  \ensuremath{\sqrt{s}=13}\,\,TeV}},
  \href{https://doi.org/10.1103/PhysRevLett.132.081901}{\emph{Phys. Rev. Lett.}
  {\bfseries 132} (2024) 081901}
  [\href{https://arxiv.org/abs/2310.12278}{{\ttfamily 2310.12278}}].

\bibitem{Eichten:1978tg}
E.~Eichten, K.~Gottfried, T.~Kinoshita, K.D.~Lane and T.-M.~Yan,
  \emph{{Charmonium: The Model}},
  \href{https://doi.org/10.1103/PhysRevD.17.3090}{\emph{Phys. Rev. D}
  {\bfseries 17} (1978) 3090}.

\bibitem{Bali:1992ab}
G.S.~Bali and K.~Schilling, \emph{{Static quark - anti-quark potential: Scaling
  behavior and finite size effects in SU(3) lattice gauge theory}},
  \href{https://doi.org/10.1103/PhysRevD.46.2636}{\emph{Phys. Rev. D}
  {\bfseries 46} (1992) 2636}.

\bibitem{Gustafson:1986db}
G.~Gustafson, \emph{{Dual Description of a Confined Color Field}},
  \href{https://doi.org/10.1016/0370-2693(86)90622-2}{\emph{Phys. Lett. B}
  {\bfseries 175} (1986) 453}.

\bibitem{Bierlich:2023okq}
C.~Bierlich, G.~Gustafson, L.~L\"onnblad and H.~Shah, \emph{{The dynamic
  hadronization of charm quarks in heavy-ion collisions}},
  \href{https://arxiv.org/abs/2309.12452}{{\ttfamily 2309.12452}}.

\bibitem{Andersson:1984af}
B.~Andersson, G.~Gustafson and T.~Sj{\"o}strand, \emph{{Baryon Production in
  Jet Fragmentation and $\Upsilon$ Decay}},
  \href{https://doi.org/10.1088/0031-8949/32/6/003}{\emph{Phys. Scripta}
  {\bfseries 32} (1985) 574}.

\bibitem{PhysRev.82.664}
J.~Schwinger, \emph{On gauge invariance and vacuum polarization},
  \href{https://doi.org/10.1103/PhysRev.82.664}{\emph{Phys. Rev.} {\bfseries
  82} (1951) 664}.

\bibitem{Fischer:2016zzs}
N.~Fischer and T.~Sj\"ostrand, \emph{{Thermodynamical String Fragmentation}},
  \href{https://doi.org/10.1007/JHEP01(2017)140}{\emph{JHEP} {\bfseries 01}
  (2017) 140} [\href{https://arxiv.org/abs/1610.09818}{{\ttfamily
  1610.09818}}].

\bibitem{CMS:2011jlm}
{\scshape CMS} collaboration, \emph{{Strange Particle Production in $pp$
  Collisions at $\sqrt{s}=0.9$ and 7 TeV}},
  \href{https://doi.org/10.1007/JHEP05(2011)064}{\emph{JHEP} {\bfseries 05}
  (2011) 064} [\href{https://arxiv.org/abs/1102.4282}{{\ttfamily 1102.4282}}].

\bibitem{Skands:2014pea}
P.~Skands, S.~Carrazza and J.~Rojo, \emph{{Tuning PYTHIA 8.1: the Monash 2013
  Tune}}, \href{https://doi.org/10.1140/epjc/s10052-014-3024-y}{\emph{Eur.
  Phys. J. C} {\bfseries 74} (2014) 3024}
  [\href{https://arxiv.org/abs/1404.5630}{{\ttfamily 1404.5630}}].

\bibitem{Bierlich:2016faw}
C.~Bierlich, \emph{{Hadronisation Models and Colour Reconnection}},
  \href{https://doi.org/10.22323/1.265.0051}{\emph{PoS} {\bfseries DIS2016}
  (2016) 051} [\href{https://arxiv.org/abs/1606.09456}{{\ttfamily
  1606.09456}}].

\bibitem{Bierlich:2017sxk}
C.~Bierlich, \emph{{Rope Hadronization and Strange Particle Production}},
  \href{https://doi.org/10.1051/epjconf/201817114003}{\emph{EPJ Web Conf.}
  {\bfseries 171} (2018) 14003}
  [\href{https://arxiv.org/abs/1710.04464}{{\ttfamily 1710.04464}}].

\bibitem{Sjostrand:1984ic}
T.~Sj{\"o}strand, \emph{{Jet Fragmentation of Nearby Partons}},
  \href{https://doi.org/10.1016/0550-3213(84)90607-2}{\emph{Nucl. Phys. B}
  {\bfseries 248} (1984) 469}.

\bibitem{Sjostrand:1984iu}
T.~Sj{\"o}strand, \emph{{The Merging of Jets}},
  \href{https://doi.org/10.1016/0370-2693(84)91354-6}{\emph{Phys. Lett. B}
  {\bfseries 142} (1984) 420}.

\bibitem{Norrbin:2000zc}
E.~Norrbin and T.~Sj{\"o}strand, \emph{{Production and hadronization of heavy
  quarks}}, \href{https://doi.org/10.1007/s100520000460}{\emph{Eur. Phys. J. C}
  {\bfseries 17} (2000) 137}
  [\href{https://arxiv.org/abs/hep-ph/0005110}{{\ttfamily hep-ph/0005110}}].

\bibitem{CMS:2010tjh}
{\scshape CMS} collaboration, \emph{{Transverse-momentum and pseudorapidity
  distributions of charged hadrons in $pp$ collisions at $\sqrt{s}=7$ TeV}},
  \href{https://doi.org/10.1103/PhysRevLett.105.022002}{\emph{Phys. Rev. Lett.}
  {\bfseries 105} (2010) 022002}
  [\href{https://arxiv.org/abs/1005.3299}{{\ttfamily 1005.3299}}].

\bibitem{CMS:2010qvf}
{\scshape CMS} collaboration, \emph{{Charged Particle Multiplicities in $pp$
  Interactions at $\sqrt{s}=0.9$, 2.36, and 7 TeV}},
  \href{https://doi.org/10.1007/JHEP01(2011)079}{\emph{JHEP} {\bfseries 01}
  (2011) 079} [\href{https://arxiv.org/abs/1011.5531}{{\ttfamily 1011.5531}}].

\bibitem{Workman:2022ynf}
{\scshape Particle Data Group} collaboration, \emph{{Review of Particle
  Physics}}, \href{https://doi.org/10.1093/ptep/ptac097}{\emph{PTEP} {\bfseries
  2022} (2022) 083C01}.

\bibitem{Eden:1996xi}
P.~Eden and G.~Gustafson, \emph{{Baryon production in the string fragmentation
  picture}}, \href{https://doi.org/10.1007/s002880050445}{\emph{Z. Phys. C}
  {\bfseries 75} (1997) 41}
  [\href{https://arxiv.org/abs/hep-ph/9606454}{{\ttfamily hep-ph/9606454}}].

\bibitem{Hunt-Smith:2020lul}
N.~Hunt-Smith and P.~Skands, \emph{{String fragmentation with a time-dependent
  tension}}, \href{https://doi.org/10.1140/epjc/s10052-020-08654-9}{\emph{Eur.
  Phys. J. C} {\bfseries 80} (2020) 1073}
  [\href{https://arxiv.org/abs/2005.06219}{{\ttfamily 2005.06219}}].

\bibitem{ALICE:2017jyt}
{\scshape ALICE} collaboration, \emph{{Enhanced production of multi-strange
  hadrons in high-multiplicity proton-proton collisions}},
  \href{https://doi.org/10.1038/nphys4111}{\emph{Nature Phys.} {\bfseries 13}
  (2017) 535} [\href{https://arxiv.org/abs/1606.07424}{{\ttfamily
  1606.07424}}].

\end{thebibliography}\endgroup

\end{document}